\documentclass[aps,prb,twocolumn,amsmath,amssymb,floatfix,superscriptaddress,eqsecnum]{revtex4-2}
\usepackage{graphicx}
\usepackage{dcolumn}
\usepackage{bm}
\usepackage{hyperref}
\usepackage{braket}
\usepackage{shorthand}

\usepackage[normalem]{ulem}

\newcommand{\chila}{\chi_{LA}}
\newcommand{\chilb}{\chi_{LB}}
\newcommand{\chira}{\chi_{RA}}
\newcommand{\chirb}{\chi_{RB}}
\newcommand{\phila}{\phi_{LA}}
\newcommand{\philb}{\phi_{LB}}
\newcommand{\phira}{\phi_{RA}}
\newcommand{\phirb}{\phi_{RB}}

\begin{document}

\title{Entanglement Entropy of Generalized Moore-Read Fractional Quantum Hall State Interfaces}

\author{Ramanjit Sohal}
\affiliation{Department of Physics and Institute for Condensed Matter Theory,
University of Illinois at Urbana-Champaign, 1110 West Green Street, Urbana, Illinois 61801, USA}
\author{Bo Han}
\affiliation{T.C.M. Group, Cavendish Laboratory, University of Cambridge,
J.J. Thomson Avenue, Cambridge, CB3 0HE, United Kingdom}
\affiliation{Department of Physics and Institute for Condensed Matter Theory,
University of Illinois at Urbana-Champaign, 1110 West Green Street, Urbana, Illinois 61801, USA}
\author{Luiz H. Santos}
\affiliation{Department of Physics, Emory University, Atlanta, Georgia 30322, USA}
\author{Jeffrey C. Y. Teo}
\affiliation{Department of Physics, University of Virginia, Charlottesville, VA22904, USA}

\date{\today}

\begin{abstract}
Topologically ordered phases of matter can be characterized by the presence of a universal, constant contribution to the entanglement entropy known as the topological entanglement entropy (TEE). The TEE can been calculated for Abelian phases via a ``cut-and-glue" approach by treating the entanglement cut as a physical cut, coupling the resulting gapless edges with explicit tunneling terms, and computing the entanglement between the two edges. We provide a first step towards extending this methodology to non-Abelian topological phases, focusing on the generalized Moore-Read (MR) fractional quantum Hall states at filling fractions $\nu=1/n$. We consider interfaces between different MR states, write down explicit gapping interactions, which we motivate using an anyon condensation picture, and compute the entanglement entropy for an entanglement cut lying along the interface. Our work provides new insight towards understanding the connections between anyon condensation, gapped interfaces of non-Abelian phases, and TEE.
\end{abstract}

\maketitle

\tableofcontents

\section{Introduction \label{sec:introduction}}

Entanglement has become an indispensable tool in the characterization of quantum many-body systems, particularly topologically ordered phases of matter, which cannot be identified through local order parameters. The most elementary measure of entanglement is provided by the entanglement entropy (EE). Given a state $\ket{\psi}$ and a bipartition of the Hilbert space $\cH = \cH_A \otimes \cH_B$, the EE is given by
\begin{align}
S = -\mathrm{Tr}_A(\rho_A \ln \rho_A)
\end{align}
where $\rho_A = \mathrm{Tr}_B \ket{\psi}\bra{\psi}$ is the reduced density matrix of $A$. Specializing to $2+1$-dimensional systems, if $\ket{\psi}$ is the ground state of a local Hamiltonian and we choose a spatial bipartitioning of the Hilbert space, then the EE satisfies
\begin{align}
S = \alpha L - \gamma
\end{align}
in the thermodynamic limit, where $L$ is the length of the entanglement cut separating regions $A$ and $B$. The first term in this expression is known as the area law, where $\alpha$ is a non-universal constant. In contrast, $\gamma$ is a universal quantity known as the topological entanglement entropy (TEE) and is non-zero for topologically ordered systems  \cite{Kitaev2006,Levin2006}. If $A$ has the topology of a smooth disc, then $\gamma = \ln \mathcal{D}$, where $\mathcal{D}$ is the total quantum dimension, a quantity which characterizes the anyon content of a topological order.

As a single number, the TEE provides a rather coarse grained description of a gapped state. A more descriptive object is provided by the entanglement spectrum (ES) \cite{Li2008}, which is defined by first formally writing the reduced density matrix for region $A$ in the form of a thermal density matrix,
\begin{align}
\rho_A \propto e^{-\cH_e}. 
\end{align}
The ES is then given by the spectrum of the operator $\cH_e$, which is known as the entanglement Hamiltonian. Remarkably, for (chiral) topological phases, the low-lying part \cite{Thomale2010} of the ES for a spatial entanglement cut corresponds to the physical spectrum of the conformal field theory (CFT) describing the edge of the topological order. This was first demonstrated numerically in fractional quantum Hall systems \cite{Li2008}, while analytic arguments for the correspondence appeared shortly thereafter \cite{Fidkowski2010,Chandran2011,Qi2012,Swingle2012,Dubail2012}. 

Of particular interest to us is the work of Qi, Katsura, and Ludwig \cite{Qi2012}, which employed a ``cut-and-glue" approach to calculate the ES. These authors argued that one can compute the ES by physically cutting the system along the entanglement cut between $A$ and $B$ and turning on an interaction between the resulting gapless edge states. Since the correlation length vanishes in the bulk, any entanglement between $A$ and $B$ should come from the the coupled edges. Using boundary CFT techniques, Qi et. al. deduced the ground state of the coupled edge system and showed that the ES does indeed match that expected for the bulk topological order. Subsequent works applied this approach to the specific cases of \emph{Abelian} topological phases, whose edges are described by multi-component Luttinger liquids \cite{Wen1995}. In this case, one can write down explicit gapping terms for which the ground state can readily be approximated, without recourse to boundary CFT methods \cite{Lundgren2013,Cano2015} (see also Refs. \cite{Furukawa2011,Chen2013} for related calculations). 

The utility of the cut-and-glue approach was made manifest in the work of Cano et. al. \cite{Cano2015}, in which the TEE for an entanglement cut along a gapped interface between \emph{distinct} Abelian topological phases was computed. The authors demonstrated that the TEE in fact receives (universal) corrections depending on the choice of interactions used to gap out the interface, even for an interface between two regions with the same topological order \footnote{see also Ref. \cite{Zou2016} for related considerations and Ref. \cite{Fliss2017} for a calculation using the bulk Chern-Simons theory}. Gapped interfaces of topological phases are of physical interest, due to the possibility of realizing non-Abelian defects at their endpoints ~\cite{BarkeshliQi-2012,Lindner-2012,Clarke-2013,Cheng-2012,Vaezi-2013,BarkeshliJianQi-2013-a,BarkeshliJianQi-2013-b,Mong-2014,khanteohughes-2014,SantosHughes-2017,May-Mann2019,santos2019}. In fact, it was demonstrated that the aforementioned TEE corrections are directly related to the emergence of 1D SPTs along these interfaces \cite{Santos2018}. Recently, progress has also been made in understanding (gapless) interfaces of topological phases beyond effective field theory constructions through numerical simulations \cite{Crepel2019a,Crepel2019b,Crepel2019c}.

The goal of the present work is to provide a first step towards extending the above story to non-Abelian topologically ordered phases of matter. Namely, we would like to, for some class of non-Abelian states, (1) use the cut-and-glue approach to compute the TEE in all topological sectors. Furthermore, we will aim to (2) identify when a gapped interface can be formed between these states and what interactions can generate these interfaces, as well as (3) compute the TEE for an entanglement cut along such an interface. The second of these issues -- the construction of explicit gapping interactions -- has been extensively studied for Abelian systems \cite{Levin2013,Barkeshli2013,Wang2015}, but is less well understood for non-Abelian phases (although interfaces of non-Abelian states have been studied at an abstract level \cite{Bais2009a,Beigi2011,Kitaev2012,Fuchs2013,Kong2014,Lan2015,Hung2015,Ji2019,Lan2019}).

To these ends, we focus on the generalized Moore-Read (MR) states \cite{Moore1991}, which provide examples of the simplest non-Abelian fractional quantum Hall (FQH) states. These states may be viewed as arising from $p+ip$ pairing of composite fermions \cite{Read2000} and, accordingly, their edge theories are described by a free compactified chiral boson and a free Majorana fermion \cite{Milovanovic1996}. One might then expect the computation of the TEE in the MR state to be an uneventful extension of the Abelian case. However, the choice of the local electron operator, which determines the allowed quasiparticles and provides the origin of the non-Abelian properties of these phases, glues the bosonic and fermionic sectors of the Hilbert space together in a non-trival manner. As we will see, the calculation of the EE requires a careful treatment of this organization of the Hilbert space. Before delving into these calculations, given the length of this paper, we first provide a summary of our results.

\subsection{Summary of Results}

(1) We first demonstrate that the correct ES and TEE is obtained for uniform MR interfaces on a torus in all topological sectors using the cut-and-glue approach. On a torus, the ground state of each topological sector, $a$, is a \emph{minimum entropy state} \cite{Dong2008,Zhang2012} and for an entanglement cut splitting the torus into two cylinders, the TEE in these states is given by
\begin{align}
    \gamma = 2 \ln (\mathcal{D}/d_a),
\end{align}
where $d_a$ is the quantum dimension of the anyon associated to the $a$ topological sector. For the MR state at filling $\nu=1/n$, $\mathcal{D}=\sqrt{4n}$, while the allowed anyons have either $d_a=1$ or $d_a= \sqrt{2}$ \cite{Fendley2007,Dong2008}. The local interaction that gaps the interface corresponds to a single-electron backscattering term. This interaction is given by a sine-Gordon operator coupled to a Majorana mass and simultaneously gaps out the charged, chiral boson and neutral Majorana sectors. As in Refs. \cite{Lundgren2013,Cano2015}, we will take the strong coupling limit and approximate this interaction to quadratic order in fluctuations of the fields about their vacuum expectation values. This approximation violates the requirement of electron locality alluded to above and must be supplemented by a projection into the correct topological sector. 

(2) We investigate interfaces of MR states at filling fractions $\nu_A = 1/pb^2$ and $\nu_B=1/pa^2$, where $p,a,b \in \mathbb{Z}$ and we take $a$ and $b$ to be coprime. Although gapped interfaces of non-Abelian states have been studied in the literature \cite{Bais2009a,Beigi2011,Kitaev2012,Fuchs2013,Kong2014,Lan2015,Hung2015,Ji2019,Lan2019}, a systematic understanding of interactions generating distinct classes of these interfaces is lacking. So, we use anyon condensation \cite{Bais2009,Burnell2018} as a guide to deduce when gapped interfaces should exist and to motivate explicit gapping terms. Interestingly, although we can always gap out an interface between MR states at fillings $\nu_A$ and $\nu_B$, we find that when $a$ and $b$ are both odd, a single interaction term is needed, whereas when one of $a$ and $b$ is even and the other odd, two terms are needed. Moreover, in the latter case, we find that the gapped interface is most easily constructed using an alternative representation of the $\nu=1/n$ MR edge CFT which is topologically equivalent to its standard description in terms of a chiral Majorana and a $U(1)_n$ chiral boson. 
In particular, we will make use of the fact that we can rewrite the Ising CFT as
\begin{align*}
\mathrm{Ising} = \frac{SO(N+1)_1}{SO(N)_1} \sim SO(N+1)_1 \boxtimes \overline{SO(N)}_1,
\end{align*}
where $G_k$ denotes a Wess-Zumino-Witten (WZW) theory with Lie group $G$ at level $k$ and the symbol $\boxtimes$ indicates a tensor product supplemented by the condensation of a particular set of bosons. The nature of the equivalence will be explained in more detail later on. This will allow us to express the MR edge in terms of a chiral boson and multiple chiral and anti-chiral Majorana fermions, which can be used to construct the appropriate gapping interactions.

(3) Combining the above results, it is then straightforward to compute the TEE for an entanglement cut along an interface between MR states at fillings $\nu_A$ and $\nu_B$. In this calculation, we must take into account the additional constraints on the ground states imposed by the specific forms of the gapping interactions, in a manner analogous to that of the calculation for Abelian interfaces \cite{Cano2015}. Again working on the torus, we find the TEE in the vacuum sector to be given by
\begin{align}
	\gamma = 2 \ln (2\sqrt{pa^2b^2})
\end{align}
for $a$ and $b$ both odd while,
\begin{align}
	\gamma = 2 \ln (4\sqrt{pa^2b^2})
\end{align}
for one of $a$ and $b$ odd and the other even. Finally, we discuss the connection between these values of the TEE with the existence of a ``parent" topological phase for the two MR states on either side of the interface.

It should be emphasized that ours is not the first work to investigate the EE of non-Abelian systems through a cut-and-glue type approach. The work of Qi et. al. applies to generic uniform chiral topological orders (both Abelian and non-Abelian) and demonstrated that the ground state of the coupled edge system at the interface should be described by so-called Ishibashi states \cite{Ishibashi1988,Cardy2004}. Wen et. al. \cite{Wen2016} later showed that appropriately regularized Ishibashi states furnish the correct entanglement structure for generic chiral phases and generic bipartitions on manifolds of arbitrary genus (a related, earlier calculation was also performed in Ref. \cite{Das2015}). Interfaces between distinct non-Abelian and/or Abelian orders have also been considered, where the interface was conjectured to be described by an appropriately constructed Ishibashi state \cite{Lou2019}. One of the main contributions of this work is a more microscopic justification of these results, for a specific set of non-Abelian phases, starting from an explicit effective field theory description of the interface.

The remainder of the paper is structured as follows. We begin by reviewing the MR edge theory, placing special emphasis on the interpretation of the distinct topological sectors in the CFT language in Section \ref{sec:mr-review}. Section \ref{sec:cut-and-glue} provides a review of the cut-and-glue approach and our handling of the topological sectors. We proceed to calculate the EE for a uniform MR state in Section \ref{sec:uniform}. In Section \ref{sec:non-uniform} we identify the two distinct classes of interfaces between MR states at different fillings and write down explicit gapping terms. The computation of the EE for each of these interfaces is presented in Section \ref{sec:non-uniform-EE}. We provide a discussion of our results and conclude in Section \ref{sec:discussion}. Finally, the appendixes collect some technical details. 
 
\section{Review of Moore-Read Edge Theory \label{sec:mr-review}}

\begin{figure}
  \centering
    \includegraphics[width=0.45\textwidth]{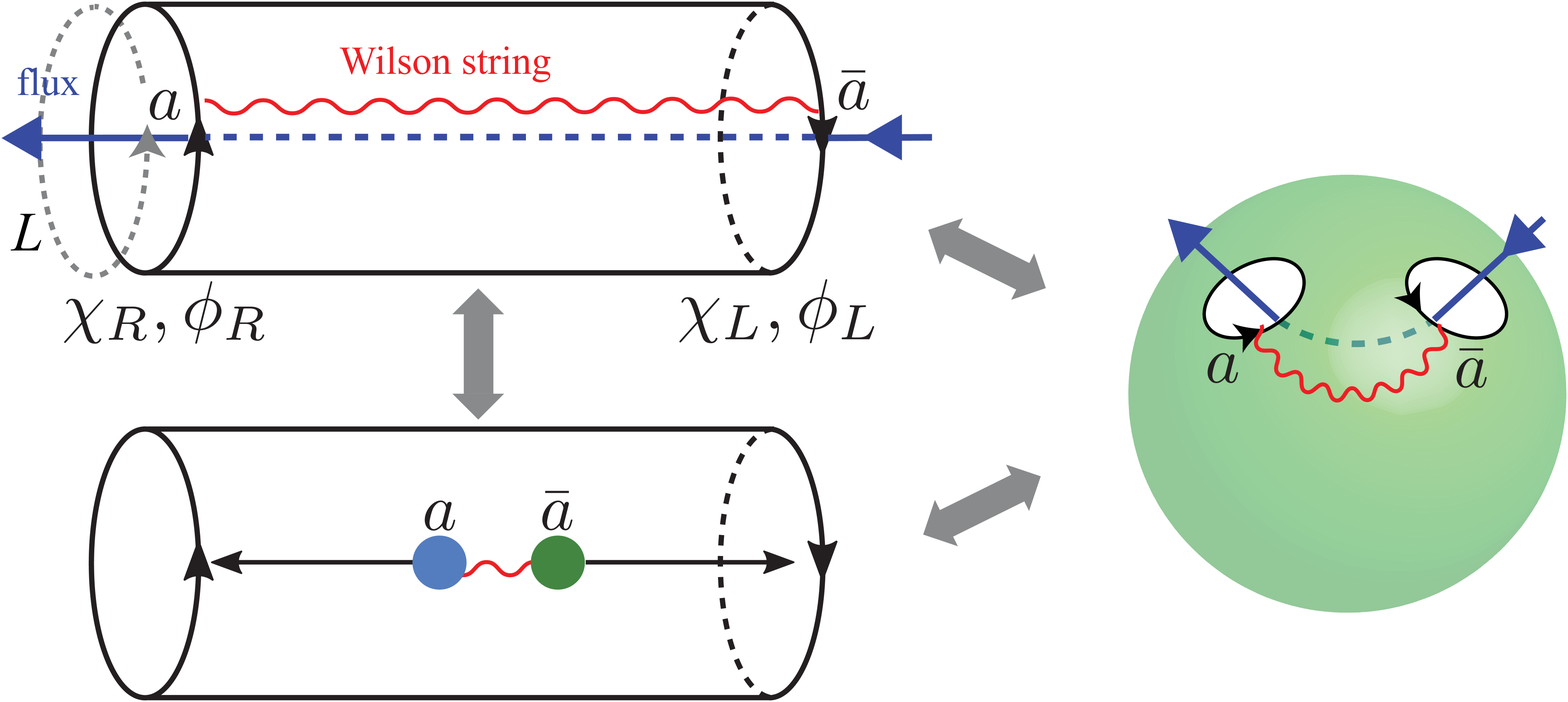}
    \caption{Moore-Read state on a cylinder with chiral edge states. The insertion of an anyon flux $a$ through the cylinder (top) is equivalent to nucleating a conjugate anyon pair in the bulk of the cylinder and dragging them to opposite edges (bottom). The cylinder geometry is homotopic to a sphere with two punctures (right), which bounds the anyon pair.} \label{fig:cylinder}
\end{figure}

We begin by reviewing the edge theory for the MR state at filling fraction $\nu = 1/n$ \cite{Milovanovic1996}. Note that $n$ may be either even or odd. If $n$ is even, we have a MR state of electrons (i.e. fermions) while, if $n$ is odd, we have a MR state of bosons. In the following, we will often refer to the local particles comprising the MR FQH state as electrons, regardless of whether $n$ is even or odd (and hence regardless of whether the local particles are fermions or bosons).

Now, let us consider a MR state defined on a cylinder with circumference $L$. Standard arguments imply that the edges of the cylinder will be described by CFTs of opposite chirality, $\mu=L,R=+,-$, as indicated in Fig. \ref{fig:cylinder}. Specifically, as described in the introduction, the edge theory contains both a neutral Majorana fermion $\chi$ sector and a charged $U(1)$ boson $\phi$ sector. The two edges are formally described by the Lagrangian densities
\begin{align}
\begin{split}
\cL_\mu &= \frac{i}{2}\chi_\mu(\partial_t-\mu v_n\partial_x)\chi_\mu+\frac{n}{4\pi}\partial_x\phi_\mu(\mu\partial_t-v_c\partial_x)\phi_\mu,
\end{split} \label{eqn:outer-edge-lagrangians}
\end{align}
where $v_n>0$ and $v_c>0$ are the velocities of the Majorana and boson, respectively. The Majorana fermion and the $U(1)$ boson are Hermitian: $\chi^\dagger=\chi$, $\phi^\dagger=\phi$. The fields obey the equal-time (anti)commutation relations
\begin{align}
\left[\phi_\mu(x), \partial_y \phi_\mu(y)\right] &= \frac{2\pi i\mu}{n} \delta(x-y) \label{eqn:bosonic-etcr}  \\
\left\{ \chi_\mu(x), \chi_\mu(y) \right\} &= \delta(x-y). \label{eqn:fermionic-etcr}
\end{align}
The bosons are compactified on a circle of radius $R=1$ so that $\phi_{L/R} \equiv \phi_{L/R} + 2\pi$, and the primary fields in the $U(1)$ sector are normal-ordered vertex operators $e^{ir\phi_\mu}$ with integral $r$. The charge densities on the two edges are given by $\rho_{L/R} = \partial_x \phi_{L/R} / (2\pi)$. Note that this means the \emph{winding numbers} of the scalars around the length of the edges, 
\begin{align}
N_{\mu} \equiv \int_0^L \frac{\partial_x \phi_{\mu}(x)}{2\pi} dx = \int_0^L  \rho_{\mu}(x) dx,
\end{align}
count the total charge carried by the edges (in units of $e$ above the ground state) and so can only take values in the set of rational numbers, as determined by the charge of the minimal charge anyon.

At the level of the Lagrangian, it would appear that the charge and neutral sectors are decopuled and hence that the MR edge theory is described by an $\mathrm{Ising}\times U(1)_n$ CFT. This is not the case, as the physical theory is not fully defined until the electronic (i.e. local) operators are specified. This determines the anyon content and hence the Hilbert space topological sectors, as all physical excitations must have trivial braiding statistics with respect to the electron. This constraint of electron locality is ultimately a consequence of the fact that the bulk topological state is constructed from electrons. In the MR edge theory, the charge $e$ operators, 
\begin{align}
\psi_{e,L} = \chi_L e^{i n \phi_L}, \quad \psi_{e,R} = \chi_R e^{-i n \phi_R} \label{eqn:electrons}
\end{align}
are defined to be electronic operators. 

For later use, let us also define the fermion parity operator, $(-1)^F$, which anti-commutes with the fermions of \emph{both} edges:
\begin{align}
(-1)^F \chi_{R/L} = - \chi_{R/L} (-1)^F.
\end{align}
A similar operator for the bosonic sector is given by $(-1)^{N_R+N_L}$ which, using the commutation relations of Eq. (\ref{eqn:bosonic-etcr}), is seen to have the action
\begin{align}
(-1)^{N_R+N_L} e^{in\phi_\mu} = -e^{in\phi_\mu} (-1)^{N_R+N_L}.
\end{align}
Hence, the combined operator,
\begin{align}
G \equiv (-1)^F (-1)^{N_R+N_L}, \label{eqn:Z2-symmetry}
\end{align}
which measures the relative parity between the fermion number and bosonic winding number (i.e.~charge) of both edges, clearly commutes with the electron operators of both edges. 

Having specified the electron operators, we can now enumerate the anyon content of the theory. Explicitly, the MR theory of the $\mu=L,R$ edge carries the following primary fields, 
\begin{align}
e^{ir\phi_{\mu}}, \quad \chi_{\mu} e^{ir\phi_{\mu}}, \quad \sigma_{\mu} e^{i (r+1/2)\phi_{\mu}}, \label{eqn:MR-anyons}
\end{align}
where $r=1,\dots , n$. We can restrict to these values of $r$, as two excitations are considered equivalent if they differ by fusion with an electron operator or a bosonic oscillator mode.
Here, $1,\,\chi$ and $\sigma$ are the primary fields of the neutral Ising sector, where $\chi$ is the Majorana fermion and $\sigma$ represents the non-Abelian Ising twist field. They obey the Ising fusion rules,
\begin{align}
\begin{split}
\chi \times \chi &= 1 \\
\chi \times \sigma &= \sigma \\
\sigma \times \sigma &= 1 + \chi.
\end{split} \label{eqn:ising-fusion-rules}
\end{align}
The vertex operators $e^{ir\phi}$ are charge-carrying Laughlin quasiparticles. In the bulk, the braiding phase between the fields $e^{ir_1\phi}$ and $e^{ir_2\phi}$ is $e^{2\pi ir_1r_2/n}$. In contrast to the Laughlin $U(1)_n$ edge theory, the charge $e$ boson (fermion) $e^{in\phi_\mu}$, for $n$ even (odd), is fractional and is {\em not} a local excitation. This allows for the existence of the non-Abelian twist fields $\sigma e^{i(r+1/2)\phi}$, which exhibit $-1$ braiding with respect to the boson/fermion $e^{in\phi}$ (from the $e^{i\phi/2}$ factor) and the Majorana fermion $\chi$ (from the $\sigma$ factor), but are local with respect to the electronic quasiparticles in Eq. \eqref{eqn:electrons}. In the bulk language, $\sigma e^{i\phi/2}$ corresponds to a non-Abelian half vortex, which traps a Majorana zero-mode (MZM), represented by $\sigma$.  The MZM flips the boundary condition of the Majorana fermion at the edge, since it exhibits a braiding phase of $-1$ with respect to $\chi$. Note that, although the $\mathrm{Ising}\times U(1)_n$ CFT is described by the same Lagrangian as the MR CFT, its anyon content is given by a direct product of that of the Ising and $U(1)_n$ topological orders, as the ``electron operator" is the vertex operator $e^{in\phi}$: $\{1,\sigma,\chi\} \times \{e^{ir\phi}\}_{r=1,\dots,n}$.

The quantum dimension $d_a$ of an anyon $a$ is defined to respect the fusion rules so that $d_ad_b=\sum_cN_{ab}^cd_c$ if $a\times b=\sum_cN_{ab}^cc$. The $2n$ Abelian anyons $e^{ir\phi}$ and $\chi e^{ir\phi}$ have quantum dimension $d=1$ while the remaining $n$ non-Abelian Ising anyons $\sigma e^{i(r+1/2)\phi}$ have quantum dimension $d=\sqrt{2}$. The total quantum dimension $\mathcal{D}$ is defined as
\begin{align}
    \mathcal{D}^2=\sum_c d_a^2.
    \end{align} 
For the MR state, $\mathcal{D}=\sqrt{4n}$. The conjugate $\bar{a}$ of an anyon $a$ is the unique anyon type that annihilates $a$ under fusion $a\times\bar{a}=1+\ldots$, i.e.~$N_{a\bar{a}}^1=1$. For example, $\overline{\sigma_\mu e^{i(r+1/2)\phi_\mu}}\simeq\sigma_\mu e^{i(n-r-1/2)\phi_\mu}$. Note that, in any physical excited state supporting some number of anyons $a_i$, fusing together all the $a_i$'s must yield the vacuum, since any physical state must ultimately be constructed from electrons. 

As noted above, the choice of electron operator glues together the bosonic and fermionic sectors in a non-trivial way not specified at the level of the Lagrangian. In particular, we must restrict the edge CFT Hilbert space to states satisfying $G=1$ [Eq. \eqref{eqn:Z2-symmetry}]. That is, the parity of charge must match the fermion parity (as measured with respect to the ground state). Physically, this is just the statement that all physical states must be constructed out of electrons and acting with an electron operator changes the Majorana fermion parity by the same amount as the winding number parity. Loosely speaking, one may view the invariance of the electron operators under conjugation by $G$ as reflecting a $\mathbb{Z}_2$ gauge symmetry and the constraint $G=1$ as a projection to the gauge-invariant subspace. This rule organizes the states of the theory into topological sectors, which are in one-to-one correspondence with the fundamental anyon excitations. From the bulk perspective, these topological sectors are excited states corresponding to the insertion of Wilson lines connecting the two edges or, equivalently, the process of nucleating of an anyon and its conjugate in the bulk and dragging them to opposite edges, as shown in Fig. \ref{fig:cylinder}. (If one glues the edges together to form a torus as we shall do later, the Wilson line becomes a Wilson loop and the topological sectors now correspond to degenerate ground states.) In the following, we describe how these distinct sectors manifest themselves in the edge CFT.

Let us first consider the ground state of the MR theory on the cylinder (which implies there is no flux through the hole of the cylinder). Clearly, this state has $N_R=N_L=0$ and no fermionic excitations; hence $G=1$ in this state. Acting with the electron operator $\psi_{e,L} = \chi_L e^{in\phi_L}$ on the left edge of the cylinder, we obtain an excited state which is still, by definition, within the same topological sector. Since $\psi_{e,L}$ and $G$ commute, it immediately follows that the application of the electron operator on the ground state can only yield states in which the fermion parity has flipped \emph{and} the bosonic winding has increased by one. All states in this topological sector can be obtained by the application of an arbitrary number of electron operators and $\partial_x \phi$ operators, the latter of which simply create charge density fluctuations without changing the charge or fermion parity. Hence, the states in the identity ($\textbf{1}$) sector are characterized by having their fermion parity \emph{equal} to the bosonic winding parity (equivalently, the parity of charge added above the ground state), individually on each edge. That is to say,
\begin{align}
\mathbf{1}\text{ sector:} \quad (-1)^{N_{R/L}}(-1)^{F_{R/L}} = +1. \label{eqn:1-sector-condition}
\end{align}
Here we have defined individual fermion parities for each edge, $(-1)^{F_{L/R}}$. This is possible because, in the untwisted sector, the fermions obey anti-periodic boundary conditions and so do not possess zero modes. So, acting on a state with, say, a right-moving fermion operator cannot change the left-moving fermion parity.

Let us now consider the states within the $\chi$ sector. Starting from the ground state, we can supply some energy to the bulk to nucleate a pair of neutral $\chi$ anyons and drag them to opposite edges (see Fig. \ref{fig:cylinder}). This defines a state in the $\chi$ sector, in which the fermion parity is odd but the bosonic winding is even (zero). Constructing the remaining states within this topological sector using the $\chi e^{in\phi}$ and $\partial_x \phi$ operators, we see that all states within the $\chi$ sector have fermion parity \emph{opposite} to that of the bosonic winding number parity. In other words,
\begin{align}
\chi\text{ sector:} \quad (-1)^{N_{R/L}}(-1)^{F_{R/L}} = -1.
\end{align}

Distinct topological sectors can also be obtained by inserting $r$ magnetic flux quanta through the hole of the cylinder. This is equivalent to nucleating a Laughlin quasiparticle, $e^{ir\phi}$, and its conjugate in the bulk and dragging them to opposite edges. The Majorana fermions, being electrically neutral, are unaffected by this flux insertion. The winding number parity $(-1)^{N_{L/R}}=e^{i\pi N_{L/R}}$ becomes fractional in this sector. The anyon flux, which in low-energy is represented by the vertex combination $e^{ir\phi_L}e^{ir\phi_R}$ on the two edges, associates a phase $e^{i\pi\mu r/n}$ to the winding number parity because \begin{align}(-1)^{N_\mu}e^{ir\phi_L}e^{ir\phi_R}=e^{i\pi\mu r/n}e^{ir\phi_L}e^{ir\phi_R}(-1)^{N_\mu},\end{align} for $\mu=L,R=+,-$. In other words, the electron operators on both edges pick up a phase of $e^{2\pi ir}$ when transported around the circumference of the cylinder. It is straightforward to see that this implies
\begin{align}
\phi_\mu(x+L) &= \phi_\mu(x)+2\pi N_\mu\nonumber\\
&\equiv \phi_\mu(x) + 2\pi\mu\frac{r}{n} \quad\mbox{modulo }2\pi\mathbb{Z} \label{eqn:bosonic-winding-BCs}
\end{align}
and so the winding numbers are quantized as
\begin{align}
N_{L} = \tilde{N}_{L} + \frac{r}{n}, \quad N_{R} = \tilde{N}_{R} - \frac{r}{n}, \quad \tilde{N}_{L/R} \in \mathbb{Z}. \label{eqn:bosonic-winding}
\end{align}
Hence, in the $e^{ir\phi}$ sector, we have
\begin{align}
e^{ir\phi}\text{ sector:} \quad (-1)^{N_{L}}(-1)^{F_{L}} &= [(-1)^{N_{R}}(-1)^{F_{R}}]^\ast \nonumber\\ &= e^{i\pi r / n}. \label{eqn:eirphi-sector-condition}
\end{align}
Likewise, starting in the $\chi$ sector, we can insert $r$ magnetic flux quanta in addition to the $\chi$ flux to obtain the $\chi e^{ir\phi}$ sectors:
\begin{align}
\chi e^{ir\phi}\text{ sector:} \quad (-1)^{N_{L}}(-1)^{F_{L}} &= [(-1)^{N_{R}}(-1)^{F_{R}}]^\ast \nonumber\\&= -e^{i\pi r / n}. \label{eqn:chieirphi-sector-condition}
\end{align}
Note that in all of these sectors, we still have $G=1$.

Thus far, we have only considered untwisted sectors -- that is, topological sectors in which the Majorana fermions obey anti-periodic boundary conditions. The twisted sectors are obtained by inserting a $\pi$ flux through the cylinder to which only the Majoranas are sensitive, flipping their boundary conditions from anti-periodic to periodic (note that the Majorana fermions, being real, can only see fluxes which are multiples of $\pi$). However, the electron operators, being local objects, cannot have their boundary conditions changed, which implies we must simultaneously insert a magnetic flux of (an odd integer multiple of) $\pi$ to which the chiral bosons are sensitive. This particular flux insertion corresponds precisely to the half-vortex of the bulk theory, represented by $\sigma e^{i\phi/2}$ (or $\sigma e^{i(r+1/2)\phi}$ in general, for $r\in\mathbb{Z}$) in the CFT. 

Now, it is clear that the effect on the chiral bosons is to simply change the quantization of their winding to
\begin{align}
\phi_\mu(x+L) \equiv \phi_\mu(x) + 2\pi\mu \frac{r+1/2}{n},\quad\mbox{modulo } 2\pi\mathbb{Z}, \label{eqn:twsited-bosonic-winding-BCs}
\end{align}
and therefore the winding numbers are quantized as
\begin{align}
N_L = \tilde{N}_{L} + \frac{r+1/2}{n}, \quad N_{R} = \tilde{N}_{R} - \frac{r+1/2}{n}, \quad \tilde{N}_{L/R} \in \mathbb{Z}. \label{eqn:twisted-bosonic-winding}
\end{align}
The $\sigma e^{i(r+1/2)\phi}$ flux through the cylinder can be detected by 
\begin{align}
\sigma e^{i(r+1/2)\phi}\text{ sector:} \quad e^{2\pi iN_L}=e^{-2\pi iN_R}=e^{2\pi i(r+1/2)}.\label{eqn:sigmaeirphi-sector-condition}
\end{align}

The effect on the Majorana fermions, as stated above, is to change their boundary conditions to being periodic. As a result, each edge possesses a Majorana zero mode (MZM), $\chi_L(k=0)=c_0$, $\chi_R(k=0)=\tilde{c}_0$; these must be paired together to form a single, physical complex fermion mode, $f=(c_0+i\tilde{c}_0)/\sqrt{2}$, which may be occupied or unoccupied. This is a reflection of the Ising fusion rules of the $\sigma$ particles, Eq. (\ref{eqn:ising-fusion-rules}). Note that this means we can no longer define separate fermion parities for the two edges, as the MZM operator changes the occupancy of this complex fermion mode, $\{c_0,(-1)^{N_f}\}=\{\tilde{c}_0,(-1)^{N_f}\}=0$ for $(-1)^{N_f}=(-1)^{f^\dagger f}$. In the twisted sector, one can construct a physical state for given windings $N_{R/L}$ in \eqref{eqn:twisted-bosonic-winding} by filling up an arbitrary number of finite momentum Majorana fermion states on either edge, and then choosing the complex fermion zero mode $f$ to be either occupied or unoccupied to satisfy the $G=1$ condition.

Altogether, we see that there are $2n$ untwisted and $n$ twisted sectors, corresponding to the $2n$ Abelian and $n$ non-Abelian anyons of the $\nu =\frac{1}{n}$ MR state. The $2n$ Abelian anyon fluxes $e^{ir\phi}$ and $\chi e^{ir\phi}$ through the cylinder can be distinguished by the local edge combined parity $(-1)^{N_\mu}(-1)^{F_\mu}$, which is identical to a Wilson loop of anyon type $\sigma_\mu e^{i\phi_\mu/2}$ around the cylinder. The phases in \eqref{eqn:eirphi-sector-condition} and \eqref{eqn:chieirphi-sector-condition} are identical to the monodromy braiding phases between $\sigma_\mu e^{i\phi_\mu/2}$ and $e^{ir\phi}$, $\chi e^{ir\phi}$ \begin{align}\mathcal{D}S_{\sigma_\mu e^{i\phi_\mu/2},e^{ir\phi_\mu}}=e^{i\pi\mu r/n}.\end{align} As noticed previously, the remaining $n$ non-Abelian fluxes $\sigma e^{i(r+1/2)\phi}$ through the cylinder cannot be detected by the same local edge combined parities because separate fermion parities for each edge, $(-1)^{F_\mu}$, cannot be defined in these twisted sectors. This is consistent with the trivial modular $S$-matrix entries $S_{\sigma e^{i\phi/2},\sigma e^{i(r+1/2)\phi}}=0$. Instead, the twisted sectors can be distinguished by their $U(1)$ sector according to $e^{2\pi iN_\mu}$, which is identical to a Wilson loop of anyon type $e^{i\phi_\mu}$ around the cylinder. The phases in \eqref{eqn:sigmaeirphi-sector-condition} are identical to the monodromy braiding phases between $e^{i\phi_\mu}$ and $\sigma_\mu e^{i(r+1/2)\phi_\mu}$ \begin{align}\mathcal{D}S_{e^{i\phi_\mu},\sigma e^{i(r+1/2)\phi_\mu}}=e^{2\pi i\mu(r+1/2)/n}.\end{align}  Note that passing from one topological sector to another requires the application of a non-local Wilson line operator. In our computation of the EE using the cut-and-glue approach, we will thus need to ensure that any approximations we make do not mix topological sectors since the ``gluing" will be achieved via local electronic interactions. We describe this calculation and how we handle this subtlety next.

\section{Cut-and-Glue Approach Review and Topological Sector Projection \label{sec:cut-and-glue}}

\begin{figure}
  \centering
    \includegraphics[width=0.3\textwidth]{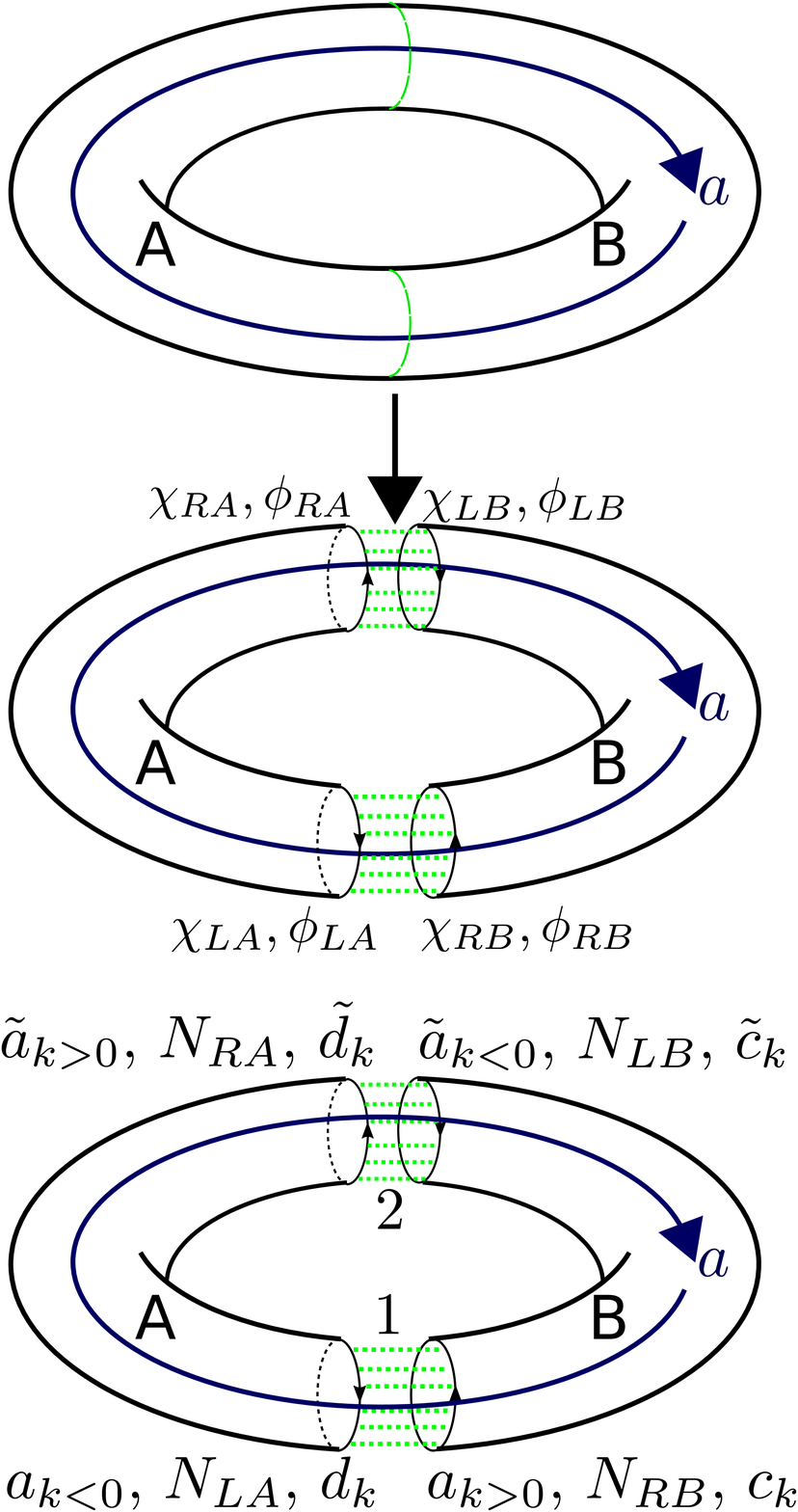}
    \caption{(Top) Moore-Read state on a torus. The arrow passing through the $x$-cycle (i.e. the vertical cycle) of the torus represents an anyon flux $a$. The green dashed lines represent an entanglement cut between regions $A$ and $B$. In Sections \ref{sec:uniform} and \ref{sec:non-uniform}, we will consider the situation in which regions $A$ and $B$ are occupied by MR states with equal and unequal, respectively, filling fractions. (Middle) A cartoon of the cut and glue approach to computing the entanglement entropy. The dotted green lines represent the electron tunneling terms added to glue the edges together. (Bottom) Same as the middle figure, but with each edge at interfaces $1$ and $2$ labelled by which mode operators act on them.} \label{fig:torus}
\end{figure}

As described in the introduction, our EE calculation is based on the cut-and-glue approach \cite{Qi2012} as it is employed in Refs. \cite{Lundgren2013,Cano2015} and which we now review in the context of the MR state. The application of this methodology to non-Abelian states such as the MR state brings with it new subtleties regarding the careful treatment of the edge theory's topological sectors, as noted above. We will discuss these issues below and describe in detail our approach, which is an important new aspect of our work, for addressing them in Sec. \ref{sec:projection}.

Consider a MR state on the torus. We wish to compute the EE associated with the entanglement cut splitting the torus into two cylinders, with the left and right halves labeled as regions $A$ and $B$, respectively, as depicted in Fig. \ref{fig:torus}. The cut-and-glue approach employs the fact that, since the correlation length of the system is vanishingly small in a topological phase, we can approximate the EE as arising purely from entanglement between degrees of freedom near the entanglement cut. To that end, we can treat the entanglement cut as a physical cut and split the torus into two cylinders labeled as $A$ and $B$. Adding electron tunneling interactions will gap out the edges and heal the cut. We can then compute the entanglement between the resulting coupled edge theories. In the case of a torus geometry, we will have two interfaces, as depicted in Fig. \ref{fig:torus}, which we label as the $LA/RB$ and $RA/LB$ interfaces, or interface $1$ and interface $2$, respectively.

The edges at interface $1$, before coupling them through a tunneling term, are described by the Hamiltonians: 
\begin{align}
\begin{split}
H_{\mathrm{dec},1} &= \int_0^L dx \left[ \frac{v_c}{4\pi} (\partial_x \phirb)^2  - v_n \chirb \frac{i}{2}\partial_x \chirb  \right] \\ 
&+  \int_0^L dx \left[ \frac{v_c}{4\pi} (\partial_x \phila)^2  + v_n \chila \frac{i}{2}\partial_x \chila \right]. \end{split}\label{eqn:BareEdgeHam}
\end{align}
The Majorana fields have mode expansions
\begin{align}
\chirb (x) &= \frac{1}{\sqrt{L}} \sum_k e^{ikx} c_k, \quad \chila (x) = \frac{1}{\sqrt{L}} \sum_k e^{ikx} d_k, \label{eqn:majorana-mode-expansion}
\end{align}
with half-integer quantized momenta in the untwisted sectors, $k = \frac{2\pi}{L}(j+1/2)$, $j \in \mathbb{Z}$, and integer quantized momenta in the twisted sectors, $k = \frac{2\pi}{L}j$, $j \in \mathbb{Z}$. The mode operators satisfy
\begin{align}
c^\dagger_k = c_{-k}^{\pdg}, \quad d^\dagger_k = d_{-k}^{\pdg}
\end{align}
and obey the anti-commutation relations
\begin{align}
\{c^\dg_k , c_{k'}^{\pdg} \} = \{d^\dg_k , d_{k'}^{\pdg} \} = \delta_{k,k'}, \quad \{c_k, d_{k'} \} = 0.
\end{align}
The boson fields have mode expansions
\begin{align}
\begin{split}
\phirb &= \phi_{RB,0} + 2\pi N_{RB} \frac{x}{L} + \sum_{k>0} \sqrt{\frac{2\pi}{nL|k|}} (a_k^{\pdg} e^{ikx} + a_k^\dg e^{-ikx}) \\
\phila &= \phi_{LA,0} + 2\pi N_{LA} \frac{x}{L} + \sum_{k<0} \sqrt{\frac{2\pi}{nL|k|}} (a_k^{\pdg} e^{ikx} + a_k^\dg e^{-ikx})
\end{split} \label{eqn:boson-mode-expansion}
\end{align}
with integer quantized momenta in all sectors: $k = \frac{2\pi}{L}j$, $j \in \mathbb{Z}/\ \{0\}$. The mode operators obey the commutation relations:
\begin{align}
&[a_k^\dg, a_{k'}^{\pdg} ] = \delta_{k,k'}, \quad [a_k, a_{k'}] = 0, \\
&[\phi_{RB,0}, N_{RB}] = -[\phi_{LA,0}, N_{LA}] = -\frac{i}{n}.
\end{align}
The quantization of the winding numbers is determined by the topological sector, as detailed in Section \ref{sec:mr-review}.

Likewise, before adding any couplings, interface $2$ is described by 
\begin{align}
\begin{split}
H_{\mathrm{dec},2} &= \int_0^L dx \left[ \frac{v_c}{4\pi} (\partial_x \philb)^2  - v_n \chilb \frac{i}{2}\partial_x \chilb  \right] \\
&+  \int_0^L dx \left[ \frac{v_c}{4\pi} (\partial_x \phira)^2  + v_n \chira \frac{i}{2}\partial_x \chira \right]. \end{split}
\end{align}
We write the mode expansions of the interface $2$ fields as follows:
\begin{align}
\chilb &= \frac{1}{\sqrt{L}} \sum_k e^{ikx} \tilde{c}_k, \quad \chira = \frac{1}{\sqrt{L}} \sum_k e^{ikx} \tilde{d}_k, \label{eqn:majorana-mode-expansion-interface-2} \\
\begin{split}
\phira &= \phi_{RA,0} + 2\pi N_{RA} \frac{x}{L} + \sum_{k>0} \sqrt{\frac{2\pi}{nL|k|}} (\tilde{a}_k^{\pdg} e^{ikx} + \tilde{a}_k^\dg e^{-ikx}) \\
\philb &= \phi_{LB,0} + 2\pi N_{LB} \frac{x}{L} + \sum_{k<0} \sqrt{\frac{2\pi}{nL|k|}} (\tilde{a}_k^{\pdg} e^{ikx} + \tilde{a}_k^\dg e^{-ikx}),
\end{split} \label{eqn:boson-moode-expansion-interface-2}
\end{align}
where the quantization of the momenta and winding numbers are determined in the same way as for the interface $1$ fields.

Now, the (quasi-)electron operators are given by
\begin{align}
\psi_{e,L\alpha} = \chi_{{L \alpha}} e^{i n \phi_{L\alpha}}, \quad \psi_{e,R \alpha} = \chi_{R \alpha} e^{-in \phi_{R\alpha}}.
\end{align}
We also define a $\mathbb{Z}_2$ symmetry operator for \textit{each} cylinder:
\begin{align}
G_{\alpha} = (-1)^{F_\alpha} (-1)^{N_{L\alpha} + N_{R\alpha}}, \quad \alpha = A,B, \label{eqn:g_alpha}
\end{align}
where $(-1)^{F_\alpha}$ is the fermion parity operator on the two edges of cylinder $\alpha$. Since we have physically split the torus into two cylinders, we require separately that $G_{\alpha} = 1$ for $\alpha=A,B$. As before, in the untwisted sectors, we can define separate fermion parities for each edge of either cylinder: $(-1)^{F_{L/R\alpha}}$.

We now wish to glue the two edges together to heal the cut. So, we add in the electron tunnelling terms
\begin{align}
\begin{split}
H_{AB} &= \int_0^L dx \left[\frac{2g}{2\pi}\left(\psi_{e,LA}^\dg \psi_{e,RB} + h.c. \right)\right] \\
&+ \int_0^L dx \left[\frac{2g}{2\pi}\left(\psi_{e,LB}^\dg \psi_{e,RA} + h.c. \right)\right] \\
	&= \int_0^L dx \left[ \frac{2g}{\pi} i \chila \chirb \cos[n(\phirb + \phila)] \right] \\
	&+ \int_0^L dx \left[ \frac{2g}{\pi} i \chilb \chira \cos[n(\phira + \philb)] \right], \end{split} \label{eqn:GappingTerm}
\end{align}
where we take $g > 0$. \footnote{Note that this interaction is irrelevant in the renormalization group sense and so need not open up a gap. This can be remedied by adding in a density-density interaction of the form $U\partial_x \phila \partial_x \phirb$. For a range of $U$, the scaling dimensions of the scalar fields will be renormalized so as to make the tunnelling term relevant. However, in the interest of simplicity, we will not include such terms and simply assume $g$ to be large and the edges are gapped out by the interactions.}

Our task is to approximate the ground state of 
\begin{align}
H = H_{\mathrm{dec},1} + H_{\mathrm{dec},2} + H_{AB}, \label{eqn:full-hamiltonian}
\end{align}
which requires us to approximate $H_{AB}$. 
In the strong coupling limit, the ground state is assumed to give rise to individual expectation values of the bosonic operators $i\chila \chirb$ and $\cos[n(\phirb + \phila)]$.
Without loss of generality, the ground state for $g \rightarrow \infty$ is represented by the expectation values
\begin{equation}
\begin{split}
&\,
\la n(\phirb + \phila) \ra = \la n(\phira + \philb) \ra = \pi
\\
&\,
\la i\chila \chirb \ra , \, \la i\chilb \chira \ra > 0
\,,
\end{split} \label{eqn:expectation-values}
\end{equation}
As such, expanding the fields around their classical expectations values yields a harmonic approximation of the interface interaction
\begin{equation}
\label{eq: H AB harmonic}
\begin{split}
&\,
H_{AB}
\approx \int_0^L dx \big[ \mathrm{const.}
+
v_n \tilde{g}i\chila \chirb + 
v_n \tilde{g}i\chilb \chira
\\
&\,
 +
\frac{v_c \lambda\pi}{2} (\phirb + \phila - \pi)^2
+ 
\frac{v_c \lambda\pi}{2} (\phira + \philb - \pi)^2 \big]
\end{split}    
\end{equation}
where $\tilde{g} = -2g/(v_n \pi) < 0$ and $\lambda>0$. Since we are considering only small fluctuations of $\phira + \philb$ and $\phirb + \phila$ about their pinned values, they cannot have non-zero winding numbers, as this would imply they vary significantly over the length of the system. We thus have the constraint \cite{Lundgren2013}
\begin{align}
    N_{RA} + N_{LB} = N_{LA} + N_{RB} = 0, \label{eqn:uniform-winding-constraint}
\end{align}
in this strong coupling limit.

The harmonic approximation Eq. \eqref{eq: H AB harmonic} plays a key role in this work, for it allows us to calculate the entanglement entropy and spectrum at the interface by analytical means. However, important issues underlying this approximation need to be accounted for. First, the approximated tunnelling Hamiltonian violates both the $\mathbb{Z}_2$ gauge symmetry gluing the fermionic and bosonic sectors together (as discussed above) and the $U(1)$ gauge symmetry associated with independent shifts of the bosonic fields \footnote{Note that, when we say the gauge symmetries are violated, we do not mean to imply that a gauge field is being Higgsed. As we explain, we mean simply that the harmonic approximation of $H_{AB}$, taken at face value, will mix topological sectors (that is, it is a non-local expression).}: $\phi_{L/R} \to \phi_{L/R} + 2\pi P_{L/R}, P_{L/R} \in \mathbb{Z}$. Indeed, under conjugation by $G_A$, we see that $i\chila \chirb \to -i\chila \chirb$. This in turn means that the approximated tunnelling Hamiltonian mixes topological sectors. For instance, consider the identity ($\bm 1$) and $\chi$ sectors of the MR theory. Recall that in the former sector, the fermionic parity matches the bosonic winding number parity on each edge, while these two quantities are opposite in the latter. Now, it is easy to see that the $\tilde{g}i \chila \chirb$ term in the approximated interaction will change the fermionic parity on both edges and so will mix the identity and $\chi$ sectors on each half of the torus. The $(\phirb + \phila - \pi)^2$ term also violates the $U(1)$ symmetry associated with the shift symmetry $\phi_{LA/RB} \to \phi_{LA/RB} + c_{LA/RB}$ and so, in principle, will also mix bosonic winding number sectors corresponding to distinct topological sectors. Hence the ground state of this approximated Hamiltonian \emph{cannot} describe an approximation of the ground state of the interface theory in a definite anyon sector.

Our strategy for dealing with the $\mathbb{Z}_2$ gauge symmetry violation encoded in Eq. \eqref{eq: H AB harmonic}, is to promote the theory to an 
``expanded" Hilbert space in which the gauge symmetries are violated. 
In this expanded Hilbert space, the bosonic and fermionic sectors are genuinely decoupled and so we can compute the ground state of the approximated Hamiltonian using straightforward free field theory methods. Once this is done, we can project the resulting state into the appropriate topological sector of the gauge-invariant subspace. Restoring the $U(1)$ gauge symmetry  amounts to projecting to states with appropriately quantized bosonic winding numbers. Restoring the $\mathbb{Z}_2$ symmetry means projecting to states obeying the appropriate matching of the fermion parity and bosonic winding number parity. We describe this in more detail next.

\subsection{Description of the Projection \label{sec:projection}}

Let us denote the exact ground state of the coupled edge system, as described by the Hamiltonian of Eq. (\ref{eqn:full-hamiltonian}), in topological sector $a$ as
\begin{align}
\ket{\psi_a} = \ket{\psi_{1,a}} \otimes \ket{\psi_{2,a}}.
\end{align}
Here, $\ket{\psi_{1,a}}$ and $\ket{\psi_{2,a}}$ are the ground states of interfaces $1$ and $2$, respectively, in the topological sector $a$. Note that, although we can express the ground state as a tensor product of the two interfaces, the ground states of the interfaces are constrained to lie in the same topological sector. This is a consequence of the fact that the anyon flux $a$ passing through one interface must necessarily pass through the other interface, as shown in Fig. \ref{fig:torus}. In particular, this means that we must have $G_A\ket{\psi_a} = G_B\ket{\psi_a} = \ket{\psi_a}$ -- that is to say, each cylinder must, on its own, lie in the physical MR Hilbert space.

We can write the ground state of the approximated Hamiltonian, given by Eq. \eqref{eqn:full-hamiltonian} with the approximation of $H_{AB}$ by Eq. \eqref{eq: H AB harmonic}, in a similar form
\begin{align}
\ket{\hat{\psi}_a} = \ket{\hat{\psi}_{1,a}} \otimes \ket{\hat{\psi}_{2,a}}.
\end{align}
(Henceforth, symbols with hats will denote objects in the unprojected Hilbert space.) As emphasized above, our approximation of the gapping term violates the $\mathbb{Z}_2$ gauge symmetry, and so both $\ket{\hat{\psi}_{1,a}}$ and $\ket{\hat{\psi}_{2,a}}$ will be superpositions of states from different topological sectors of the MR theory. 
Nevertheless, we have written $\ket{\hat{\psi}_{1,a}}$ and $\ket{\hat{\psi}_{2,a}}$ as having dependence on $a$ because they still retain some information about $a$ through the boundary conditions of both the bosonic and fermionic fields. For instance, if we are working in one of the twisted sectors, our approximation of the interaction term will not change the fact that the Majorana fermions obey periodic boundary conditions.

In order to obtain a state in a definite topological sector of the MR theory, we consider
\begin{align}
\ket{\psi_a} &\approx P_a \ket{\hat{\psi}_a} \equiv P_{a,A} P_{a,B} \ket{\hat{\psi}_a} \label{eqn:projection}
\end{align}
where $P_{a,\alpha}$ projects cylinder $\alpha$ to the topological sector $a$, and we have defined $P_a \equiv P_{a,A} P_{a,B}$. 
We will show in Section \ref{sec:uniform} that the projected state Eq. \eqref{eqn:projection}
correctly describes the universal entanglement properties of the MR state in each topological sector $a$ (and takes the expected form of an Ishibashi state \cite{Qi2012,Wen2016}).

For a general topological sector $a$, the action of the projection is most easily understood when writing $\ket{\hat{\psi}_{a}}$ in terms of a superposition of eigenstates of $N_{\mu\alpha}$ and $(-1)^{F_{\mu\alpha}}$, in which case the projection amounts to removing those states in the sum which do not satisfy the $\mathbb{Z}_2$ constraint appropriate to the topological sector in question. Focusing first on the untwisted sectors, we can, as noted above, define separate fermion parities for each edge:
\begin{align}
    (-1)^{F_\alpha} \equiv (-1)^{F_{L\alpha}} (-1)^{F_{R\alpha}} .
\end{align}
This permits us to define operators which project each edge to specific topological sectors of the untwisted sector. Indeed, we can formally write
\begin{align}
    P_{a,\alpha} \equiv P_{a,L\alpha} P_{a,R\alpha}
\end{align}
as the operator which projects cylinder $\alpha$ to the untwisted sector $a$, where $P_{a,\mu \alpha}$ are operators acting on edges $\mu\alpha$. Specialising momentarily to the sector $a = e^{ir\phi}$ and edge $RB$, we define $P_{e^{ir\phi},RB}$ via its action on a basis of states for the edge. An arbitrary state on edge $RB$ can be written as a superposition of the states
\begin{align}
    \ket{N_{RB}, \{n_{a,k}\}_{k>0} ,  \{n_{c,k}\}_{k>0}},
\end{align}
which are eigenstates of $N_{RB}$, $a^\dg_k a^{\pdg}_k$, and $c^\dg_k c^{\pdg}_k$ with eigenvalues $N_{RB}$, $ \{n_{a,k}\}_{k>0} $, and $\{n_{c,k}\}_{k>0}$, respectively. We then define
\begin{align}
    P_{e^{ir\phi},RB}  \ket{N_{RB}, \{n_{a,k}\} ,  \{n_{c,k}\}}  = \ket{N_{RB}, \{n_{a,k}\} ,  \{n_{c,k}\}}
\end{align}
if  $N_{RB}+\frac{r}{n} \in \mathbb{Z}$ and $(-1)^{N_{RB}+\frac{r}{n}+\sum_{k} n_{c,k}} = 1$, while
\begin{align}
    P_{e^{ir\phi},RB}  \ket{N_{RB}, \{n_{a,k}\} ,  \{n_{c,k}\}}  = 0
\end{align}
otherwise. The first condition enforces that the winding number obey the appropriate quantization for the $a=e^{ir\phi}$ sector on edge $RB$, Eq. (\ref{eqn:bosonic-winding}), while the second condition ensures that the $\mathbb{Z}_2$ constraint for sector $a=e^{ir\phi}$, Eq. (\ref{eqn:eirphi-sector-condition}), is satisfied. In physical terms, this operator ensures that the correct magnetic flux is threaded through the circle defined by the edge and that the fermion parity matches the integer part of bosonic winding on this edge. We similarly define for edge $LB$
\begin{align}
    P_{e^{ir\phi},LB}  \ket{N_{LB}, \{n_{\tilde{a},k}\} ,  \{n_{\tilde{c},k}\}} = \ket{N_{LB}, \{n_{\tilde{a},k}\} ,  \{n_{\tilde{c},k}\}}
\end{align}
if $N_{LB}-\frac{r}{n} \in \mathbb{Z}$ and $(-1)^{N_{LB}-\frac{r}{n}+\sum_{k} n_{\tilde{c},k}} = 1$, while
\begin{align}
    P_{e^{ir\phi},LB}  \ket{N_{LB}, \{n_{\tilde{a},k}\} ,  \{n_{\tilde{c},k}\}} = 0
\end{align}
otherwise. The operators $P_{e^{ir\phi},\mu A}$ are defined in an analogous manner. Likewise, the $P_{\chi e^{ir\phi}, \sigma\alpha}$ operators are defined in a similar way, but by instead enforcing the $\mathbb{Z}_2$ constraint of Eq. (\ref{eqn:chieirphi-sector-condition}) on each edge.

As for the twisted sectors, since we cannot define separate fermion parities for each edge, we cannot write down a projection operator as a product of operators acting on the two edges of the cylinder. Let us first consider cylinder $B$. We define a complex fermion from the Majorana zero modes of each edge (recall that the Majorana fermions obey periodic boundary conditions in the twisted sectors),
\begin{align}
    f_B = \frac{1}{\sqrt{2}}(c_0 + i \tilde{c}_0), \label{eqn:f_B}
\end{align}
which explicitly ties together the $\mu = L,R$ Hilbert spaces of the cylinder. An arbitrary state on cylinder $B$ can then be written as a superposition of states of the form
\begin{align}
\begin{split}
    &\ket{N_{RB}, \{n_{a,k}\}_{k> 0}, \{n_{c,k}\}_{k>0}} \\ &\otimes \ket{N_{LB}, \{n_{\tilde{a},k}\}_{k< 0},  \{n_{\tilde{c},k}\}_{k<0}} \otimes \ket{n_B}
\end{split}
\end{align}
which are eigenstates of $N_{R/L,B}$, $a^\dg_k a^{\pdg}_k$, $\tilde{a}^\dg_k \tilde{a}^{\pdg}_k$, $c^\dg_k c^{\pdg}_k$, $\tilde{c}^\dg_k \tilde{c}^{\pdg}_k$, $f_B^\dg f_B^{\pdg}$  with eigenvalues, $N_{R/L,B}$, $ \{n_{a,k}\}_{k\neq 0} $, $\{n_{c,k}\}_{k>0}$, $\{n_{\tilde{c},k}\}_{k<0}$, and $n_B$ respectively. We then define the operator $P_{a,B}$, which projects cylinder $B$ to the twisted topological sector $a = \sigma e^{i(r+1/2)\phi}$, via its action on these states:
\begin{align}
\begin{split}
    &P_{\sigma e^{i(r+1/2)\phi},B}\ket{N_{\mu B}, \{n_{a/\tilde{a},k}, n_{c/\tilde{c},k}\}, n_B} \\
    &\quad = \ket{N_{\mu B}, \{n_{a/\tilde{a},k}, n_{c/\tilde{c},k}\}, n_B}
\end{split}
\end{align}
if $N_{\mu B}+\mu \frac{r}{n} \in \mathbb{Z}$ and $(-1)^{\sum_\mu N_{\mu B}+\sum_{k} (n_{c,k} + n_{\tilde{c},-k})+n_B} = 1$, while
\begin{align}
    P_{\sigma e^{i(r+1/2)\phi},B} \ket{N_{\mu B}, \{n_{a/\tilde{a},k}, n_{c/\tilde{c},k}\}, n_B} = 0
\end{align}
otherwise. Again, the first constraint ensures that the bosonic winding numbers satisfy the quantization of Eq. (\ref{eqn:twisted-bosonic-winding}) while the second condition enforces the $\mathbb{Z}_2$ constraint $G_B=1$. Physically, $P_{\sigma e^{i(r+1/2,\phi)},B}$ ensures the correct magnetic flux passes through the cylinder and that the total fermion parity across both edges matches the total bosonic winding of the two edges. An analogous operator, $P_{\sigma e^{i(r+1/2)\phi},A}$, for cylinder $A$ can be defined, after forming a complex fermion, $f_A$, defined from the Majorana zero modes of the two edges:
\begin{align}
    f_A = \frac{1}{\sqrt{2}}(d_0 + i \tilde{d}_0). \label{eqn:f_A}
\end{align}

One can write down explicit expressions for the projection operators defined above but, for our purposes, the above operational definitions will prove more convenient. We also note that there is a bit of an ambiguity in defining the projection operators for the twisted sectors in that there is a choice as to whether one defines an occupied $f_{A/B}$ state as corresponding to odd or even fermion parity. We will return to this point in Section \ref{sec:uniform-twisted}, when we calculate the EE in the twisted sectors, and in Appendix \ref{sec:appendix-twisted-GS}, where we present explicit expressions for the twisted sector ground states.

\section{Uniform Interface Entanglement Entropy \label{sec:uniform}}
We are now prepared to move on to the actual computation of the ES and EE of the MR states. We first recall that, for an entanglement cut of the torus of the type we are considering (Fig. \ref{fig:torus}), the TEE in the ground state of topological sector $a$ is given by
\begin{align}
\gamma_a = 2\ln (\mathcal{D}/d_a),
\end{align}
where $\mathcal{D}$ is again the total quantum dimension and $d_a$ is the quantum dimension of the anyon $a$. These states (on the torus) are known as \emph{minimum entropy states}, as they maximize the TEE within the space of degenerate ground states \cite{Dong2008,Zhang2012}. As noted in Section \ref{sec:mr-review}, a MR state at filling $\nu = 1 / n$ has $\cD = 2\sqrt{n}$,
the Abelian anyons $e^{ir\phi}$ and $\chi e^{ir\phi}$ all have $d_a = 1$, and the non-Abelian anyons $\sigma e^{i(r+1/2)\phi}$ have $d_a = \sqrt{2}$. Hence, in the untwisted sectors, we expect to find the TEE
\begin{align}
\gamma_a = 2\ln 2\sqrt{n}, \quad a = e^{ir\phi}, \chi e^{ir\phi}, \label{eqn:untwisted-TEE}
\end{align}
while in the twisted sectors we expect
\begin{align}
\gamma_a = 2\ln \sqrt{2n}, \quad a = \sigma e^{i(r+1/2)\phi} \, . \label{eqn:twisted-TEE}
\end{align}
We can glean some intuition for these results by contrasting them with the TEE for the Abelian system consisting of a $p+ip$ superconductor stacked with (and decoupled from) a $\nu=\frac{1}{n}$ Laughlin state. Such a state has $\mathcal{D} = \sqrt{n}$ and an edge is also described by Eq. \eqref{eqn:outer-edge-lagrangians}, but with local (electronic) operators given by $\chi$ and $e^{in\phi}$. The TEE in, for instance, the trivial sector on the torus of this theory is thus $\gamma_1 = 2\ln \sqrt{n}$, in contrast to $\gamma_1 = 2\ln 2\sqrt{n}$ for the MR state. As we will see explicitly, the factor of two difference in the argument of the logarithm arises precisely from the the projection discussed in Section \ref{sec:projection}. Indeed, when writing the approximated ground state $\ket{\hat{\psi}_a}$, as a superposition of states with definite bosonic winding and fermion occupation numbers, we will find that the projection to the physical MR Hilbert space, Eq. \eqref{eqn:projection}, will remove exactly \emph{half} of the states appearing in the superposition. This increases the TEE by $\ln 2 + \ln 2$, with each interface contributing a single $\ln 2$.

A heuristic understanding of the difference between the TEEs of the untwisted and twisted sectors follows from the fact that a cylinder with a $\sigma e^{i(r+1/2)}$ flux traps a MZM at each edge. Gluing two cylinders together to form a torus, as we do, hybridizes the MZMs on the edges. On tracing out one cylinder to compute the EE, one is, loosely speaking, tracing out half of a qubit for each pair of edges, giving a contribution of $2\ln \sqrt{2}$ to the EE.

In the following subsections, we proceed to compute the entanglement spectrum and TEE of the ground state of the MR theory for the Abelian and non-Abelian topological sectors. We will compute the ground state for the interface $1$ explicitly; the calculations for interface $2$ are identical.

\subsection{Abelian (Untwisted) Sectors \label{sec:uniform-untwisted}}
We begin by considering a MR state on a torus in one of the untwisted topological sectors: $e^{ir\phi}$,  $\chi e^{ir\phi}$. The Majorana fields satisfy anti-periodic boundary conditions while the bosons obey the boundary conditions of Eq. (\ref{eqn:bosonic-winding-BCs}) and hence the winding numbers are quantized as in Eq. (\ref{eqn:bosonic-winding}). Now, using the field mode expansions, the full approximated Hamiltonian describing interface $1$ decouples into fermionic and bosonic terms: 
\begin{align}
H_1 \equiv H_{1,f}^{\mathrm{osc}} + H_{1,b}^{\mathrm{osc}} + H_{1,b}^{\mathrm{zero}}. \label{eqn:H-interface-1-approx-untwisted}
\end{align}
The bosonic zero mode Hamiltonian is given by 
\begin{align}
    H_{1,b}^{\mathrm{zero}} &= \frac{\pi v_c n }{2L} (N_{RB} -N_{LA})^2 + \frac{\pi\lambda v_c L}{2} (\phi_{RB,0} + \phi_{LA,0})^2 , \label{eqn:H-b-zero}
\end{align}
where we have made use of the constraint of Eq. (\ref{eqn:uniform-winding-constraint}). The bosonic oscillator part takes the form
\begin{align}
    H_{1,b}^{\mathrm{osc}} &= \frac{v_c}{2}\sum_{k\neq 0} (a_k^\dg \quad a_{-k} )
		\begin{pmatrix}
			A_k && B_k \\
			B_k && A_k		
		\end{pmatrix}
		\begin{pmatrix}
			a_k \\
			a_{-k}^\dg		
		\end{pmatrix}, \label{eqn:H-b-osc}
\end{align}
where
\begin{align}
    A_k = |k| + \frac{2\lambda \pi^2}{n |k|}, \quad B_k = \frac{2\lambda \pi^2}{n|k|}.
\end{align}
Lastly, the fermion oscillator modes are governed by the Hamiltonian
\begin{align}
H^{\mathrm{osc}}_{1,f} &= v_n \sum_{k>0} (c_k^\dg \,\,\,\, d_{-k} )
		\begin{pmatrix}
			k && -i\tilde{g} \\
			i\tilde{g} && -k		
		\end{pmatrix}
		\begin{pmatrix}
			c_k \\
			d_{-k}^\dg		
		\end{pmatrix} . \label{eqn:H-f-osc} 
\end{align}
Since, within our harmonic approximation, the bosons and fermions decouple, we can compute the ground state of these two sectors separately. However, as emphasized above, this decoupling is a manifestation of the violation of the $\mathbb{Z}_2$ gauge symmetry by our approximation. As discussed in Section \ref{sec:projection}, we will have to perform a projection to obtain a state in a definite untwisted topological sector. Having done so, it will then be straightforward to obtain the reduced density matrix for subregion $B$, as the projected ground state will take a simple Schmidt decomposed form.

\subsubsection{Bosonic Sector Ground State}

In the expanded Hilbert space, the computation of the ground state in the bosonic sector is identical to the calculation carried out by Lundgren et. al. \cite{Lundgren2013} for the Laughlin states at filling $\nu = 1 / n$. For completeness, we briefly review the calculation here.

Starting with the oscillator sector, we can diagonalize Eq. (\ref{eqn:H-b-osc}) via a Bogoliubov transformation,
\begin{align}
    \begin{pmatrix}
    a_k \\
    a_{-k}^\dg
    \end{pmatrix} =
    \begin{pmatrix}
    \cosh \theta_k && \sinh \theta_k \\
    \sinh \theta_k && \cosh \theta_k
    \end{pmatrix}
    \begin{pmatrix}
    b_k \\
    b_{-k}^\dg
    \end{pmatrix},
\end{align}
where $\cosh(2\theta_k) = A_k / \varepsilon_k$, $\sinh(2\theta_k) = -B_k/\varepsilon_k$, and $\varepsilon_k = \sqrt{|k|^2 + 4\lambda\pi^2 / n}$. With these definitions, we can write $H_{1,b}^{\mathrm{osc}} = v_c\sum_{k \neq 0} \varepsilon_k ( b_k^\dg b_k^{\pdg} + \frac{1}{2} )$,
so that the ground state is defined by $b_k \ket{G_{b,\mathrm{osc},1}} = 0$. It is readily checked that the ground state is given by the coherent state
\begin{align}
    \ket{G_{b,\mathrm{osc},1}} = \exp\left( \sum_{k>0} e^{-u_k/2} a_k^\dg a_{-k}^\dg \right) \ket{0}, \label{eqn:b-osc-ground-state-1}
\end{align}
where $u_k = \ln \coth^2(2\theta_k)$ and $\ket{0}$ is the ground state of the decoupled system, satisfying $a_k \ket{0} = 0$ for all $k\neq 0$. For $|k| \ll \lambda$,
\begin{align}
    u_k \approx \frac{2}{\pi} \sqrt{\frac{n}{\lambda}}k \equiv v_e k ,
\end{align}
where we have defined the entanglement velocity $v_e =  \frac{2}{\pi} \sqrt{\frac{n}{\lambda}}$. 

As for the zero-mode sector, on defining $X=n(N_{RB}-N_{LA})/2$ and $P=\phi_{LA,0}+\phi_{RB,0}$ so that $[X,P]=i$, we see that Eq. (\ref{eqn:H-b-zero}) describes a simple harmonic oscillator. In the $L\to \infty$ limit, we can ignore the discretization of $X$ and simply write down the ground state:
\begin{align}
    \ket{G_{b,\mathrm{zero},1}} = \sum_{N \in \mathbb{Z} - \frac{r}{n}} e^{-v_e \pi n N^2 / 2L} \ket{N_{RB}=N,N_{LA}=-N},
\end{align}
where we have again made use of the constraint $N_{RB}+N_{LA}=0$ [Eq. Eq. \eqref{eqn:uniform-winding-constraint}] and enforced the quantization of the winding numbers given in Eq. (\ref{eqn:bosonic-winding-BCs}).

\subsubsection{Majorana Sector Ground State}

Turning next to the Majorana fermions, we can perform a unitary transformation to diagonalize Eq. (\ref{eqn:H-f-osc}). We define $\gamma_k = \cos \varphi_k c_k + i \sin\varphi_k d_{-k}^\dg$,
where $\sin\varphi_k = \tilde{g}/\lambda_k$, $\cos\varphi_k = k/\lambda_k$, and $\lambda_k = \sqrt{k^2 + \tilde{g}^2}$. 
The Hamiltonian, in this basis, becomes
$H^{\mathrm{osc}}_{1,f} = v_n \sum_{k \neq 0} \lambda_k ( \gamma_{k}^\dg \gamma_{k}  - \frac{1}{2} )$.
The ground state is defined by $\gamma_{k} \ket{G_{f,\mathrm{osc},1}} = 0$. Explicitly, we can write the ground state of $H^{\mathrm{osc}}_{1,f}$ in BCS form:
\begin{align}
\begin{split}
        \ket{G_{f,\mathrm{osc},1}} &= \exp\left( \sum_{k>0} i e^{-w_k / 2} d_{-k}^\dagger c_k^\dagger \right)\ket{0}, 
\end{split}\label{eqn:f-osc-ground-state-1}
\end{align} 
where we have defined $w_k$ through $e^{-w_k / 2} = -\tan \varphi_k$ (recalling that $\tilde{g}<0)$ and $\ket{0}$ is the ground state of the decoupled system, satisfying $c_{k} \ket{0} = d_{-k} \ket{0} = 0$ for all $k>0$. For $|k| \ll \tilde{g}$, we have that
\begin{align}
    w_k \approx \frac{2 k}{|\tilde{g}|} \equiv \tilde{v}_e k ,
\end{align}
where we have defined $\tilde{v}_e = 2/|\tilde{g}|$. 

\subsubsection{Projecting to the Physical Hilbert Space}
We can now construct the full ground state of the coupled edge system in the expanded Hilbert space by combining the above results with the analogous results for interface $2$ (i.e. the $RA/LB$ interface). Explicitly,
\begin{align}
\begin{split}
\ket{\hat{\psi}_a} &= \ket{\hat{\psi}_{1,a}}\otimes \ket{\hat{\psi}_{2,a}}, \\
\ket{\hat{\psi}_{i,a}} &= \ket{G_{b,\mathrm{zero},i}} \otimes \ket{G_{b,\mathrm{osc},i}} \otimes \ket{G_{f,\mathrm{osc},i}}, \,\, i=1,2
\end{split}
\end{align}
where,
\begin{align}
    \ket{G_{b,\mathrm{zero},2}} &= \sum_{N \in \mathbb{Z} - \frac{r}{n}} e^{-\frac{v_e \pi n N^2}{2L}} \ket{N_{LB}=-N,N_{RA}=N}, \\
     \ket{G_{b,\mathrm{osc},2}} &= \exp\left( \sum_{k>0} e^{-v_e k / 2 } \tilde{a}_k^\dg \tilde{a}_{-k}^\dg \right) \ket{0}, \label{eqn:b-osc-ground-state-2} \\
    \ket{G_{f,\mathrm{osc},2}} &=  \exp\left( \sum_{k>0} i e^{-\tilde{v}_e k / 2} \tilde{c}_{-k}^\dagger \tilde{d}_k^\dagger \right) \ket{0} . \label{eqn:f-osc-ground-state-2}
\end{align}
Note that in the expressions for the oscillator sector ground states, we have taken the low-energy limit by expanding $u_k$ and $w_k$ to linear order in $k$. This is because the correspondence between the entanglement spectrum and the physical edge CFT spectrum only holds for the low lying entanglement spectrum eigenvalues.

In order to obtain an approximation to the true ground state $\ket{\psi_a}$ ($a=e^{ir\phi}$ or $\chi e^{ir\phi}$), we must apply the projection operator $P_a \equiv P_{a,A} P_{a,B}$ defined in Eq. (\ref{eqn:projection}). Now, since $\ket{\hat{\psi}_a}$ is a superposition of states with winding number and fermion parity eigenvalues satisfying $N_{RB}=-N_{LA}$ and $(-1)^{F_{RB}}=(-1)^{F_{LA}}$ as well as $N_{LB}=-N_{RA}$ and $(-1)^{F_{LB}}=(-1)^{F_{RA}}$, it is straightforward to see that
\begin{align}
    \ket{\psi_a} = P_a \ket{\hat{\psi}_a} = P_{a,A} \ket{\hat{\psi}_a} = P_{a,B} \ket{\hat{\psi}_a}. \label{eqn:untwisted-P=PA=PB}
\end{align}
In more physical terms, this expresses the fact that the electron tunneling term enforces that the two cylinders reside in the same topological sector. 

The explicit form of $\ket{\psi_a} = P_{a,B} \ket{\hat{\psi}_a}$ is rather cumbersome, and so we leave it for Appendix \ref{sec:appendix-untwisted-GS}. However, on expanding out the exponentials in $\ket{G_{b,\mathrm{osc},1/2}}$ and $\ket{G_{f,\mathrm{osc},1/2}}$, it is not too difficult to see that $ \ket{\psi_{a}} = P_{a,B}\ket{\hat{\psi}_{a}}$ is in a Schmidt decomposed form. 
Indeed, we have that
\begin{align}
    \begin{split}
    \ket{\psi_{1,a}} =  &e^{-\cH_e^{RB}/2} \sum_{\substack{N_{RB}, \\ \{n_{a,k},n_{c,k}\}}}  i^{\sum_k n_{c,k}} P_{a,RB} \left[ \ket{N_{RB}=-N_{LA}} \right. \\ &\left.\otimes\ket{\{n_{a,k} = n_{a,-k},  n_{c,k} = n_{d,-k}\}_{k>0}} \right] ,
    \end{split}
\end{align}
\begin{align}
    \begin{split}
    \ket{\psi_{2,a}} =  &e^{-\cH_e^{LB}/2}
    \sum_{\substack{N_{LB}, \\ \{n_{\tilde{a},k},n_{\tilde{c},k}\}}} i^{\sum_k n_{\tilde{c},k}} P_{a,LB} \left[ \ket{N_{LB}=-N_{RA}} \right. \\ &\left.\otimes\ket{\{n_{\tilde{a},-k} = n_{\tilde{a},k},  n_{\tilde{c},-k} = n_{\tilde{d},k}\}_{k>0}}\right]
    \end{split}
\end{align}
where,
\begin{align}
\begin{split}
    \cH_e^{RB} = &v_e \left(\frac{\pi n}{L} N_{RB}^2 + \sum_{k>0} ka_k^\dg a_k^{\pdg} - \frac{\pi}{12L} \right) \\
    & + \tilde{v}_e \left(\sum_{k>0} k c_k^\dg c_k^{\pdg} - \frac{\pi}{24L} \right)
\end{split} \label{eqn:untwisted-He-RB}
\end{align}
and
\begin{align}
\begin{split}
    \cH_e^{LB} = &v_e \left(\frac{\pi n}{L} N_{LB}^2 + \sum_{k<0} |k| \tilde{a}_k^\dg \tilde{a}_k^{\pdg} - \frac{\pi}{12L} \right) \\
    & + \tilde{v}_e \left( \sum_{k<0} |k| \tilde{c}_k^\dg \tilde{c}_k^{\pdg} - \frac{\pi}{24 L} \right).
\end{split} \label{eqn:untwisted-He-LB}
\end{align}
Note that we have multipled $\ket{\psi_{a,i}}$ by unimportant overall constants, $e^{-v_e \pi / 24L}$ and $e^{-\tilde{v}_e \pi / 48L}$, for later convenience. For readers familiar with boundary CFT methods, it should hopefully be clear that $\ket{\psi_{1/2,a}}$ are essentially regularized Ishibashi states for the $a$ topological sectors of the MR CFT \cite{Qi2012,Wen2016} (up to unimportant relative phases). In other words, $\ket{\psi_a} = \ket{\psi_{1,a}}\otimes\ket{\psi_{2,a}}$ is a superposition of all states in the $a$ topological sector, regulated by the operator $\exp[-(\cH_e^{LB}+\cH_e^{RB})/2]$. We can thus deduce that the reduced density matrix for, say, cylinder $B$ is given by
\begin{align}
     \rho_{a,B} = \mathrm{Tr}_A[\ket{\psi_a}\bra{\psi_a}] = \frac{1}{Z_{a,e}} P_{a,B}  e^{-\cH_e^{RB}-\cH_e^{LB}}  P_{a,B},
\end{align}
So, the form of the entanglement Hamiltonian precisely matches that of the physical edge Hamiltonian in the topological sector $a$, as expected. The projection operator $P_{a,B}$ ensures the reduced density matrix only acts on states within the topological sector $a$ of the physical Hilbert space.

\subsubsection{Entanglement Spectrum and Entropy}

At this point in the calculation, we are actually done. Indeed, we have argued that the entanglement spectrum exactly matches the physical edge CFT spectrum (taking into account the projection into the appropriate topological sector), and so we will necessarily obtain the correct TEE. Nevertheless, for completeness, we will show explicitly that we obtain the correct TEE for the $e^{ir\phi}$ sectors.

Introducing the fictitious inverse temperature $\beta=1/T$, we wish to compute
\begin{align}
\begin{split}
Z_{e^{ir\phi},e} &= \mathrm{Tr}_B \left[ P_{e^{ir\phi},B} e^{-\beta (\cH_{e}^{RB} + \cH_{e}^{LB})} P_{e^{ir\phi},B} \right] \\
                &= Z_{e^{ir\phi},e}^{RB} Z_{e^{ir\phi},e}^{LB}
\end{split}
\end{align}
where we have defined \footnote{We emphasize that the trace is taken over all states in the physical MR Hilbert space on cylinder $B$. In particular, this means that we cannot, in general, separate the trace into separate traces over the edges $RB$ and $LB$, since the states appearing in the trace must lie in a definite topological sector. However, the presence of the $P_{e^{ir\phi},\mu B}$ operators within the trace ensures that only states on edge $\mu B$ satisfying the winding number quantization of Eq. (\ref{eqn:bosonic-winding}) contribute, ensuring we do not mix topological sectors. So, in this case, we are justified in splitting the trace over $B$ into two traces over its two edges.}
\begin{align}
    Z_{e^{ir\phi},e}^{RB} &= \mathrm{Tr}_{RB} \left[ P_{e^{ir\phi},RB} e^{-\beta \cH_{e}^{RB}} P_{e^{ir\phi},RB} \right], \label{eqn:Z-untwisted-RB-definition} \\
    Z_{e^{ir\phi},e}^{LB} &= \mathrm{Tr}_{LB} \left[ P_{e^{ir\phi},LB} e^{-\beta\cH_{e}^{LB}} P_{e^{ir\phi},LB} \right]. \label{eqn:Z-untwisted-LB-definition}
\end{align}
In the following, we will focus on the computation of $Z_{e^{ir\phi},e}^{RB}$, as the calculation of $Z_{e^{ir\phi},e}^{LB}$ is virtually identical. First, we define the modular parameters
\begin{align}
\tau = i \tau_2 = i \frac{\beta v_e}{L}, \quad \tilde{\tau} = i \tilde{\tau}_2 = i \frac{\beta \tilde{v}_e}{L}
\end{align}
and the variables
\begin{align}
q = e^{2\pi i \tau}, \quad \tilde{q} = e^{2\pi i \tilde{\tau}}. \label{eqn:modular-variables}
\end{align}
We compute the trace using eigenstates of $N_{RB}$, $a_k^\dg a_k^{\pdg}$, and $c_k^\dg c_k^{\pdg}$. Keeping in mind that that the role of the projection operator $P_{e^{ir\phi},RB}$ is to exclude those states which do not satisfy the constraint of Eq. (\ref{eqn:eirphi-sector-condition}), we compute the entanglement partition function to be
\begin{align}
\begin{split}
Z_{e^{ir\phi},e}^{RB} = &\frac{1}{2}\chi_0^{\text{Ising}}(\tilde{q})[\chi_{r/n}^+(q) + \chi_{r/n}^-(q)] \\ 
		&+ \frac{1}{2}\chi_{1/2}^{\text{Ising}}(\tilde{q})[\chi_{r/n}^+(q) - \chi_{r/n}^-(q)], \label{eqn:Z-untwisted-RB}
\end{split}
\end{align}
where, employing the notation of Ref. \cite{Milovanovic1996}, we have defined
\begin{align}
    \chi_0^{\text{Ising}}(\tilde{q}) &=  \frac{1}{2}\tilde{q}^{-\frac{1}{48}}\left[ \prod_{j=0}^\infty (1+\tilde{q}^{j+1/2}) + \prod_{j=0}^\infty (1-\tilde{q}^{j+1/2}) \right] \label{eqn:MW-0-character} \\
    \chi_{1/2}^{\text{Ising}}(\tilde{q}) &=  \frac{1}{2}\tilde{q}^{-\frac{1}{48}}\left[ \prod_{j=0}^\infty (1+\tilde{q}^{j+1/2}) - \prod_{j=0}^\infty (1-\tilde{q}^{j+1/2}) \right] \label{eqn:MW-1/2-character}
\end{align}
and
\begin{align}
    \chi_{r/n}^{\pm}(q) =  q^{-\frac{1}{24}}\left( \sum_{N\in\mathbb{Z}} (\pm 1)^N q^{n (N-\frac{r}{n})^2/2} \right) \prod_{j=1}^\infty \left( 1-q^j \right)^{-1} . \label{eqn:boson-r/n-character}
\end{align}

Let us take a moment to unpack these expressions. The terms $\chi_0^{\text{Ising}}(\tilde{q})$ and $\chi_{1/2}^{\text{Ising}}(\tilde{q})$ are the contributions from the fermionic sector. Focusing first on $\chi_0^{\text{Ising}}(\tilde{q})$, we note that the first product appearing within the square brackets is simply the partition function for a free Majorana fermion with momenta quantized as $k=2\pi (j+1/2)/L$. The second product is the partition function for a free Majorana, but with each state weighted by its fermion parity, $(-1)^F$. So, when these two products are added together, all terms corresponding to a state with an \emph{odd} number of excited Majorana oscillator modes will cancel out. In other words, $\chi_0^{\text{Ising}}(\tilde{q})$ is the partition function for a free Majorana, with the trace restricted to states with an \emph{even} fermion parity, $(-1)^F = +1$. Likewise, $\chi_{1/2}^{\text{Ising}}(\tilde{q})$ is the partition function for a free Majorana, with the trace restricted to states with an \emph{odd} fermion parity, $(-1)^F = -1$. In more formal terms, $\chi_{0,1/2}^{\text{Ising}}(\tilde{q})$ are the characters of the $1$ and $\chi$ sectors of the Ising CFT, respectively. Similarly, $\chi_{r/n}^+(q)$ are the characters for a $U(1)_n$ boson in the $e^{ir\phi}$ sector. In particular, the term in large rounded brackets in Eq. (\ref{eqn:boson-r/n-character}) results from the trace over the winding number sector, while the product outside the brackets results from the trace over the oscillator modes. The term $\chi_{r/n}^-(q)$ is the character for a $U(1)_n$ boson in the $e^{ir\phi}$ sector, but with each term in the trace weighted by the parity of the integer part of its winding number, $(-1)^N$. Hence, $\chi_{r/n}^+(q) \pm \chi_{r/n}^-(q)$ correspond to the partition functions for $U(1)_n$ bosons in the $e^{ir\phi}$ sector with the trace over the winding numbers restricted to states with the integer part of the winding being even and odd, respectively. Altogether, the first (second) line of Eq. (\ref{eqn:Z-untwisted-RB}) corresponds to a trace of $e^{-\cH_e^{RB}}$ over states with even (odd) fermion number and an even (odd) integer part of the bosonic winding number. This accounts for all states in the $e^{ir\phi}$ topological sector. So, the entanglement partition function of the right-movers of the MR theory in the $e^{ir\phi}$ sector is indeed given by Eq. (\ref{eqn:Z-untwisted-RB}).

Now, we can write Eq. (\ref{eqn:Z-untwisted-RB}) in terms of the Dedekind $\eta$ and Jacobi $\theta$ functions (see Appendix \ref{sec:appendix-modular-functions}):
\begin{align}
\begin{split}
&Z_{e^{ir\phi},e}^{RB} = \\
&\frac{1}{4} \left[ \sqrt{\frac{\theta_0^0(\tilde{\tau})}{\eta(\tilde{\tau})}} + \sqrt{\frac{\theta_{1/2}^0(\tilde{\tau})}{\eta(\tilde{\tau})}}\right]  \frac{\theta_0^{-r/m}(n\tau) + e^{-\frac{i\pi r}{m}}\theta_{1/2}^{-r/m}(n\tau)}{\eta(\tau)} \\
+ &\frac{1}{4} \left[ \sqrt{\frac{\theta_0^0(\tilde{\tau})}{\eta(\tilde{\tau})}} - \sqrt{\frac{\theta_{1/2}^0(\tilde{\tau})}{\eta(\tilde{\tau})}}\right] \frac{\theta_{0}^{-r/m}(n\tau) - e^{-\frac{i\pi r}{m}} \theta_{1/2}^{-r/m}(n\tau)}{\eta(\tau)}.
\end{split}
\end{align}
Using the modular transformation properties of the $\eta$ and $\theta$ functions given in Eqs. (\ref{eqn:eta-S-transformation}) and (\ref{eqn:theta-S-transformation}), as well as their asymptotic behaviour in the limit $L \to \infty$ as given in Eqs. (\ref{eqn:eta-asymptotic}) and (\ref{eqn:theta-asymptotic}), we find
\begin{align}
\lim_{L \to \infty} Z_{e^{ir\phi},e}^{RB} \to  \frac{1}{2\sqrt{n}}e^{\frac{\pi L}{12 \beta}\left( \frac{1}{v_e} + \frac{1}{2\tilde{v}_e} \right)}.
\end{align}
Essentially identical calculations yield $Z_{e^{ir\phi},e}^{RB} = Z_{e^{ir\phi},e}^{LB}$ in this limit. Hence,
\begin{align}
\begin{split}
S_{e^{ir\phi}} &= \left. \frac{\partial [ T \ln Z_{e^{ir\phi},e} ]}{\partial T} \right|_{T=1} \\ 
            &= - 2\ln (2\sqrt{n}) + \frac{\pi L}{3} \left(\frac{1}{v_e} + \frac{1}{2 \tilde{v}_e} \right),
\end{split}
\end{align}
and so we obtain the expected TEE [see Eq. \eqref{eqn:untwisted-TEE}].

\subsection{Non-Abelian (Twisted) Sectors \label{sec:uniform-twisted}}
Next we turn to the twisted sectors, corresponding to the insertion of a $\sigma e^{(r+1/2)\phi}$ anyon flux through the torus. The mode expansions of the fields have the same form as that in Eq. (\ref{eqn:majorana-mode-expansion}) and Eq. (\ref{eqn:boson-mode-expansion}), except that the quantization of the quantum numbers has changed. The Majorana fields are now periodic and so have integer-quantized momenta $k = \frac{2\pi}{L} j$, $j \in \mathbb{Z}$. As for the bosons, the momenta will still be quantized as $k=\frac{2\pi j}{L}$, $j \in \mathbb{Z}$. The winding numbers, however, now obey the quantization of Eq. (\ref{eqn:twisted-bosonic-winding}). 

Let us again first focus on interface $1$. The full approximate Hamiltonian takes the form 
\begin{align}
H_1 \equiv H_{1,f}^{\mathrm{osc}} + H_{1,f}^{\mathrm{zero}} + H_{1,b}^{\mathrm{osc}} + H_{1,b}^{\mathrm{zero}}. \label{eqn:H-interface-1-approx-twisted}
\end{align}
Here, $H_{1,f}^{\mathrm{osc}}$, $H_{1,b}^{\mathrm{osc}}$, and $H_{1,b}^{\mathrm{zero}}$ are again given by Eqs. (\ref{eqn:H-b-zero})-(\ref{eqn:H-f-osc}), with appropriate changes to the quantization of the momenta and winding numbers. The new addition is a contribution from the Majorana zero modes
\begin{align}
H^{\text{zero}}_{1,f} = i \tilde{g} d_0 c_0 .
\end{align}
We now proceed to derive the reduced density matrices for each sector, following the same methodology as was employed for the untwisted sectors.

\subsubsection{Bosonic Sector Ground State}

Aside from the change in the quantization of the winding modes, the calculation of the bosonic sector ground state proceeds as before. Hence, we can immediately write the zero mode ground state as
\begin{align}
    \ket{G_{b,\mathrm{zero},1}} = \sum_{N \in \mathbb{Z} - \frac{r+1/2}{n}} e^{-\frac{v_e \pi n N^2}{2L}} \ket{N_{RB}=N,N_{LA}=-N},
\end{align}
with the only change being the quantization of $N_{RB}$. Similarly, the oscillator mode ground state is again given by Eq. \eqref{eqn:b-osc-ground-state-1}. 

\subsubsection{Majorana Sector Ground State}

Likewise, the ground state for the Majorana oscillator mode sector is again given by Eq. \eqref{eqn:f-osc-ground-state-1}, where now $k = \frac{2\pi}{L}j$, $j \in \mathbb{Z}$. The new aspect of the calculation in the twisted sector is the presence of the Majorana zero modes. Constructing complex fermion operators as 
\begin{align}
    f= \frac{1}{\sqrt{2}}(c_0 + id_0), \quad \tilde{f} = \frac{1}{\sqrt{2}}( \tilde{d}_0 + i \tilde{c}_0)
\end{align}
the Hamiltonian describing the zero modes of interfaces $1$ and $2$ can be expressed as
\begin{align}
H^{\mathrm{zero}}_{1,f} + H^{\mathrm{zero}}_{2,f} &= i\tilde{g}d_0 c_0 + i\tilde{g}\tilde{c}_0 \tilde{d}_0 = -\tilde{g}(f^\dg f + \tilde{f}^\dg \tilde{f} - 1)
\end{align}
where $\tilde{g}<0$. Now, a complete basis for the zero-mode Hilbert space is given by $\ket{n,\tilde{n}}$ where $n$ ($\tilde{n}$) denotes the occupation of the $f$ ($\tilde{f}$) fermion. The ground state is then given by $\ket{G_{f,\mathrm{zero}}} = \ket{0,\tilde{0}}$.

We can also form a different pair of complex fermions from the above Majorana zero modes, localized in the two halves of the torus, as defined in Eq. (\ref{eqn:f_B}) and Eq. (\ref{eqn:f_A}):
\begin{align*}
f_A = \frac{1}{\sqrt{2}}(d_0 + i\tilde{d}_0), \quad f_B = \frac{1}{\sqrt{2}}(c_0 + i \tilde{c}_0) .
\end{align*}
Calculating the reduced density matrix for cylinder $B$ will require us to trace out the $f_A$ degree of freedom from the state $\ket{0,\tilde{0}}$, and so we must express $\ket{G_{f,\mathrm{zero}}}$ in terms of the basis states $\ket{n_A,n_B}$, where $n_{A/B}$ denotes the occupation of the $f_{A/B}$ fermion:
\begin{align}
\ket{G_{f,\mathrm{zero}}} = \frac{1}{\sqrt{2}}(\ket{0_A,0_B} + i \ket{1_A,1_B} ).
\end{align}

\subsubsection{Projecting to the Physical Hilbert Space}
Putting everything together, we can write the ground state of the approximated Hamiltonian, Eq. (\ref{eqn:H-interface-1-approx-twisted}), as
\begin{align}
    \begin{split}
    \ket{\hat{\psi}_a} = &\ket{G_{b,\mathrm{osc},1}} \otimes \ket{G_{b,\mathrm{zero},1}} \otimes \ket{G_{f,\mathrm{osc},1}} \\
                        &  \ket{G_{b,\mathrm{osc},2}} \otimes \ket{G_{b,\mathrm{zero},2}} \otimes \ket{G_{f,\mathrm{osc},2}} 
                        \otimes \ket{G_{f,\mathrm{zero}}} \label{eqn:approx-gs-twisted}
    \end{split}
\end{align}
where the explicit forms of $\ket{G_{b,\mathrm{zero},1}}$, $\ket{G_{b,\mathrm{osc},1}}$, $\ket{G_{f,\mathrm{osc},1}}$, and $\ket{G_{f,\mathrm{zero}}}$ are given above, while
\begin{align}
    \ket{G_{b,\mathrm{zero},2}} &= \sum_{N \in \mathbb{Z} - \frac{r+1/2}{n}} e^{-\frac{v_e \pi n N^2}{2L}} \ket{N_{LB}=-N,N_{RA}=N},
\end{align}
$\ket{G_{b,\mathrm{osc},2}}$ is again given by Eq. \eqref{eqn:b-osc-ground-state-2}, and $\ket{G_{f,\mathrm{osc},2}}$ is given by Eq. \eqref{eqn:f-osc-ground-state-2} with $k=\frac{2\pi}{L}j$, $j \in \mathbb{Z}$.

We now obtain an approximation to the physical ground state, $\ket{\psi_a}$, through the projection $P_a=P_{a,A} P_{a,B}$ defined in Section \ref{sec:projection}, with $a=\sigma e^{i(r+1/2)\phi}$. As in the untwisted sector problem, it suffices to apply only one of the projection operators acting on one of the cylinders, say, $P_{a,B}$, due to the form of $\ket{\hat{\psi}_a}$. Indeed, from its explicit form, we see that every state appearing in $\ket{\hat{\psi}_a}$ has $(-1)^{N_{RB}+N_{LB}} = (-1)^{N_{RA}+N_{LA}}$ and $(-1)^{F_{B}}=(-1)^{F_{A}}$. Hence, following the same reasoning given in the untwisted sector calculation, we have that 
\begin{align}
\ket{\psi_a} = P_a\ket{\hat{\psi}_a} = P_{a,A} P_{a,B} \ket{\hat{\psi}_a}= P_{a,B} \ket{\hat{\psi}_a} \label{eqn:twisted-P=PA=PB}
\end{align}

Again, we reserve the explicit form of $\ket{\psi_a}$ for Appendix \ref{sec:appendix-twisted-GS}. We also discuss, in Appendix \ref{sec:appendix-twisted-GS}, an important subtlety regarding the definition of the fermion parity of the complex fermion zero mode. Now, as we did in the untwisted sector problem, we can make use of the fact that $\ket{\psi_a} = P_a \ket{\hat{\psi}_a} = P_{a,B} \ket{\hat{\psi}_{a}}$ is in a Schmidt decomposed form to deduce the form of the reduced density matrix for, say, cylinder $B$. Explicitly,
\begin{align}
\rho_{a,B} &= \frac{1}{Z_{\sigma e^{i(r+1/2)\phi , e}}} P_{a,B} \rho_{\mathrm{zero},B} e^{-\cH_e^{RB}-\cH_e^{LB}}P_{a,B},
\end{align}
where,
\begin{align}
\begin{split}
    \cH_e^{RB} = &v_e \left(\frac{\pi n}{L} N_{RB}^2 + \sum_{k>0} ka_k^\dg a_k^{\pdg} - \frac{\pi}{12L} \right) \\
    & + \tilde{v}_e \left(\sum_{k>0} k c_k^\dg c_k^{\pdg} + \frac{\pi}{12L} \right),
\end{split}
\end{align}
\begin{align}
\begin{split}
    \cH_e^{LB} = &v_e \left(\frac{\pi n}{L} N_{LB}^2 + \sum_{k<0} |k| \tilde{a}_k^\dg \tilde{a}_k^{\pdg} - \frac{\pi}{12L} \right) \\
    & + \tilde{v}_e \left( \sum_{k<0} |k| \tilde{c}_k^\dg \tilde{c}_k^{\pdg} + \frac{\pi}{12 L} \right),
\end{split}
\end{align}
\begin{align}
    \rho_{\mathrm{zero},B} &= \ket{0_B}\bra{0_B} + \ket{1_B}\bra{1_B}.
\end{align}
We have again shifted the entanglement spectrum by a constant for convenience.

\subsubsection{Entanglement Spectrum and Entropy}
Now, introducing the fictitious inverse temperature $\beta=1/T$, we wish to compute (for $a=\sigma e^{i(r+1/2)\phi}$)
\begin{align}
Z_{a,e} = \mathrm{Tr}_B \left[ P_{a,B}\rho_{\mathrm{zero},B} e^{-\beta(\cH_e^{RB}+\cH_e^{LB})} P_{a,B} \right] .
\end{align}
When computing the trace, the presence of the $P_{a,B}$ projection operators requires that we only sum over states in the $a= \sigma e^{i(r+1/2)\phi}$ sector. Now, consider a state $\ket{\beta}$ which obeys the correct quantization of winding numbers for the $\sigma e^{i(r+1/2)\phi}$ sector, but has a fermion parity such that $(-1)^{F_B} \neq (-1)^{N_{RB}+N_{LB}}$, implying it does not lie in the physical MR Hilbert space and so will not contribute to the trace. It follows that by applying either $f_B$ or $f^\dagger_B$ (recall that these are the zero-mode operators on cylinder $B$) to $\ket{\beta}$ will yield a state that \emph{does} satisfy the parity selection rule $(-1)^{F_B} = (-1)^{N_{RB}+N_{LB}}$. Moreover, whichever of $f_B\ket{\beta}$ or $f_B^\dagger \ket{\beta}$ is non-zero will have the same eigenvalue as $\ket{\beta}$ under $\rho_{\mathrm{zero},B} e^{-\beta(\cH_e^{RB}+\cH_e^{LB})}$, since $\rho_{\mathrm{zero},B}$ is simply the identity operator in the zero-mode sector. It is then not too difficult to see that we obtain
\begin{align}
Z_{\sigma e^{i(r+1/2)\phi},e} = Z_{\sigma e^{i(r+1/2)\phi},e}^{RB} Z_{\sigma e^{i(r+1/2)\phi},e}^{LB} 
\end{align}
where, focusing on edge $RB$ and recalling the definitions of Eq. (\ref{eqn:modular-variables}),
\begin{align}
    Z_{\sigma e^{i(r+1/2)\phi},e}^{RB} = \chi^{\text{Ising}}_{1/16}(\tilde{q})\chi_{(r+1/2)/n}^+(q).
\end{align}
Here,
\begin{align}
    \chi^{\text{Ising}}_{1/16}(\tilde{q}) = \tilde{q}^{\frac{1}{24}}\prod_{j=1}^\infty (1 + \tilde{q}^j) \label{eqn:MW-1/16-character}
\end{align}
results from the trace over the (anti-periodic) Majorana oscillator modes and is the character of the Ising CFT in the twisted sector. The quantity $\chi^+_m(q)$ was defined in Eq. (\ref{eqn:boson-r/n-character}). It should be emphasized that the entanglement partition function can be expressed as a product of traces over edges $RB$ and $LB$ because the the Majorana zero modes have been traced over; the Hilbert spaces of edges $RB$ and $LB$ are not genuinely decoupled. 

This expression for the entanglement partition function matches the character of the appropriate topological sector in the MR CFT \cite{Milovanovic1996}, and so it follows immediately that we will obtain the correct EE. Indeed, as usual, we can express the entanglement partition function in terms of modular functions:
\begin{align}
    Z_{\sigma e^{i(r+1/2)\phi},e}^{RB} =  \sqrt{\frac{\theta_0^{1/2}(\tilde{\tau})}{2\eta(\tilde{\tau})}}  \frac{\theta^{-(r+1/2)/n}_0(n\tau)}{\eta(\tau)}.
\end{align}
Making use of the modular transformation and asymptotic properties of the $\theta$ and $\eta$ functions (see Appendix \ref{sec:appendix-modular-functions}), we obtain, in the $L \to \infty$ limit,
\begin{align}
\lim_{L\rightarrow \infty} Z_{\sigma e^{i(r+1/2)\phi},e}^{RB} \approx  \frac{1}{\sqrt{2n}} e^{\frac{\pi}{24 \tilde{\tau}_2}} e^{\frac{\pi}{12 \tau_2}}  .
\end{align}
One finds that $ Z_{\sigma e^{i(r+1/2)\phi},e}^{LB}$ is given by the same expression in this limit. So,
\begin{align}
\begin{split}
S_{\sigma e^{i(r+1/2)\phi}} &= \left.  \lim_{L\rightarrow \infty}\frac{\partial [ T \ln Z_{\sigma e^{i(r+1/2)\phi},e} (\beta) ]}{\partial T} \right|_{T=1} \\
			&= -2\ln (\sqrt{2n}) + \frac{\pi L}{3} \left(\frac{1}{v_e} + \frac{1}{2 \tilde{v}_e} \right),
\end{split}
\end{align}
as required [see Eq. \eqref{eqn:twisted-TEE}].

\section{Non-Uniform Moore-Read Gapped Interfaces \label{sec:non-uniform}}

Thus far, we have demonstrated that the cut-and-glue approach can be extended to the computation of the EE in all topological sectors of the MR theory. However, the utility of this approach is that it may be used to compute the EE for an entanglement cut lying along the interface between two \emph{different} topological phases. This was demonstrated for interfaces of arbitrary Abelian phases in Ref. \cite{Cano2015}. The focus of the remainder of the present work is to conduct a similar analysis of interfaces of MR states at different filling fractions.

As a prerequisite to computing the EE for non-uniform interfaces, it is necessary to first deduce which pairs of MR states actually admit gapped interfaces and what interaction terms can generate such a gap. The corresponding question for arbitrary Abelian states has been studied in great detail \cite{Kapustin2011,Levin2013,Barkeshli2013,Wang2015}. It is now well established that an interface between Abelian topological orders $\mathcal{A}$ and $\mathcal{B}$ can be gapped if and only if (i) $\mathcal{A}$ and $\mathcal{B}$ have identical chiral central charge $c(\mathcal{A})=c(\mathcal{B})$, which is related to the thermal Hall conductance~\cite{KaneFisher97,Cappelli01,Kitaev06} by $\kappa=dI_{\mathrm{energy}}/dT=c\frac{\pi^2k_B^2}{3h}T$, and (ii) the topological order $\mathcal{A}\times\bar{\mathcal{B}}$ (where the overbar indicates time-reversal) possesses a \textit{Lagrangian subgroup}, a maximal set of mutually local bosons which, when condensed, confine all other anyons. Such subgroups, when they exist, are related to the so-called null vectors \cite{Haldane1995}, which label sine-Gordon interactions corresponding to tunneling of integer numbers of electrons.

Interfaces of non-Abelian states have also been studied intensively \cite{Bais2009a,Beigi2011,Kitaev2012,Fuchs2013,Kong2014,Lan2015,Hung2015,Ji2019,Lan2019}, although many open questions still remain.
Indeed, in contrast to Abelian edge theories, which are described by multi-component Luttinger liquids \cite{Wen1995}, non-Abelian edge theories are described by generic CFTs \cite{Moore1991}, whose primary fields need not have free-field representations. As such, a comprehensive approach to classifying gapped interfaces via explicit gapping interactions seems difficult to develop (although specific examples have been considered before, such as those in Ref. \cite{Cappelli2015}). Our goal in this section is to use anyon condensation, which we will briefly review, to understand when interfaces between MR states can be gapped, and then to use this picture to propose explicit gapping interactions.

\subsection{Anyon Condensation Picture of Gapped Interfaces}

Suppose we wish to determine whether one can form a gapped interface between topological phases $\mathcal{A}$ and $\mathcal{B}$, assuming they have identical chiral central charges. This is equivalent to asking whether one can gap out an interface between the phase $\mathcal{A}\times \bar{\mathcal{B}}$ and the vacuum by the folding trick \cite{khanteohughes-2014,Burnell2018}. In the case where $\mathcal{A}$ and $\mathcal{B}$ are both Abelian, the necessary and sufficient criterion for the existence of such an interface is the existence of a Lagrangian subgroup, $\mathcal{L} \subset \mathcal{A}\times \bar{\mathcal{B}}$. If $\mathcal{A}\times \bar{\mathcal{B}}$ is a bosonic topological order (i.e. the local ``electron" operators have bosonic statistics), then a Lagrangian subgroup is a set of anyons defined by the requirements that (1) for all $a \in \cL$, $e^{i\theta_a} = 1$, where $\theta_a$ is the spin of $a$, (2) for all $a,b \in \cL$, $e^{i\theta_{a,b}} = 1$, where $\theta_{a,b}$ is the braiding phase between $a$ and $b$, and (3) for any $b \notin \mathcal{L}$, there exists some $a \in \mathcal{L}$ such that $e^{i \theta_{a,b}} \neq 1$. Now, in the anyon condensation picture of Bais and Slingerland \cite{Bais2009}, if one condenses all anyons in $\mathcal{L}$, all other anyons in the theory will become confined. If $\mathcal{A}\times\bar{\mathcal{B}}$ is fermionic, then condition (1) is relaxed to the constraint $e^{i\theta_a} = \pm 1$ -- that is, the anyons in $\cL$ can have bosonic or fermionic self-statistics. This is because a fermionic anyon $a \in \cL$ can be fused with a local fermion (an electron) to obtain a bosonic quasiparticle which can be condensed. In either case, $\mathcal{A} \times \bar{\mathcal{B}}$ can be reduced to the vacuum or a trivial state \emph{without} the closing of a gap, implying the existence of a gapped interface between $\mathcal{A}$ and $\mathcal{B}$. 

It is believed that a similar anyon condensation criterion can be used to identify gapped interfaces of non-Abelian states \cite{Hung2015,Lan2015}. In this case, the picture is a bit more subtle as non-Abelian anyons may ``split" under condensation, and so the maximal set of condensable anyons may not be closed under fusion. For this reason, we will call such a set of anyons a Lagrangian subset, as opposed to a subgroup. Although, to the best of our knowledge, there is no rigorous proof of connection between the existence of a Lagrangian subset and the gappability of a non-Abelian interface, we can use this picture as motivation for writing down explicit gapping terms for the Moore-Read states. After first reviewing gapped Laughlin interfaces, this will be the next order of business.

\subsubsection{Review of Laughlin Interfaces}

Let us consider an interface between Laughlin states at fillings $\nu_1 = 1/k_1$ and $\nu_2=1/k_2$, as studied in Ref.~\cite{SantosHughes-2017}. The free part of the Lagrangian describing the interface is given by
\begin{align}
\cL_0 = \frac{k_1}{4\pi} \partial_x \phi_L (\partial_t - \partial_x ) \phi_L + \frac{k_2}{4\pi} \partial_x \phi_R (-\partial_t - \partial_x ) \phi_R.
\end{align}
The interaction term we add in to gap out the interface must be constructed from local degrees of freedom (i.e. electron operators). It will be sufficient to restrict our attention to an electron tunneling term:
\begin{align}
\cL_{\mathrm{int}} = (\psi_L^\dg)^a \psi_R^b + \mathrm{H.c.} = \cos (ak_1 \phi_L + b k_2 \phi_R), \label{eqn:laughlin-non-uniform-tunneling}
\end{align}
where $\psi_L = e^{-i k_1 \phi_L}$ and $\psi_R = e^{i k_2 \phi_2}$ are the local electron operators.
Here, $\Lambda = (a,b)$ must satisfy Haldane's null vector criterion \cite{Haldane1995}
\begin{align}
    \begin{pmatrix}
        a &
        b
    \end{pmatrix}
    \begin{pmatrix}
        k_1 && 0 \\
        0 && -k_2
    \end{pmatrix}
    \begin{pmatrix}
        a \\ b
    \end{pmatrix} = 0.
\end{align}
This ensures the argument of the cosine argument behaves as a classical variable and so can obtain an expectation value in the strongly interacting limit, gapping out the scalar fields. In the present case, this means
\begin{align}
a^2 k_1 - b^2 k_2 = 0.
\end{align}
We also require $\Lambda$ to be primitive \cite{Levin2012}, so as to not introduce a spurious ground state degeneracy, meaning that $a$ and $b$ must be co-prime. These two requirements can be shown to constrain the fillings to be \cite{SantosHughes-2017}
\begin{align}
\nu_1 =  k_1^{-1} = \frac{1}{pb^2}, \quad \nu_2 = k_2^{-1} = \frac{1}{pa^2}.
\end{align}
Hence, there exists a gapped interface between Laughlin states $\mathcal{A}$ and $\mathcal{B}$ at the filling fractions:
\begin{align}
(\mathcal{A}) \quad \nu = \frac{1}{pb^2} \qquad | \qquad (\mathcal{B}) \quad \nu = \frac{1}{pa^2}.
\end{align}

Let us now confirm that there indeed exists a Lagrangian subgroup for $\mathcal{A} \times \bar{\mathcal{B}}$, which is condensed by Eq. (\ref{eqn:laughlin-non-uniform-tunneling}). The anyon content of $\mathcal{A} \times \bar{\mathcal{B}}$ is
\begin{align}
\mathcal{A} \times \bar{\mathcal{B}} = \{ e^{ir\phi_L} \}_{r=1,\dots , pb^2} \times \{ e^{is\phi_R} \}_{s=1,\dots , pa^2}.
\end{align}
For concreteness, $r$ and $s$ will henceforth always index the $\mathcal{A}$ and $\bar{\mathcal{B}}$ factors, respectively. These anyons have spin
\begin{align}
h_{r,s} = \frac{1}{2} \left( \frac{r^2}{pb^2} - \frac{s^2}{pa^2} \right).
\end{align}
Hence, anyons of the form $(r,s) = l(b,a)$ have trivial spin; it is also straightforward to see that they have trivial braiding statistics with each other and non-trivial statistics with respect to all other anyons. So, the anyons
\begin{align}
\cL = \{ e^{ilb\phi_L} e^{ila\phi_R}\}_{l=1,\dots , pab}
\end{align}
form a Lagrangian subgroup and their condensation fully gaps the interface. Note, in particular, that \begin{align}
    (e^{ib\phi_L} e^{ia\phi_R})^{pab} = e^{ipab^2 \phi_L} e^{ipa^2 b \phi_R} 
\end{align}
corresponds to the composite electron operator $\psi_L^a \psi_R^b$ appearing in Eq. (\ref{eqn:laughlin-non-uniform-tunneling}) and will obtain an expectation value when the argument of the cosine is pinned, resulting in the condensation of all anyons in $\mathcal{L}$. This makes explicit the connection between Lagrangian subgroups and electron tunneling terms.

\subsubsection{Extension to Moore-Read Interfaces}

We would now like to identify gapped interfaces between generalized MR states at different filling fractions. Absent a correspondence between gapping terms and Lagrangian subsets, as exists in the Abelian case, we can at best use the anyon condensation picture as a source of intuition for identifying candidate gapping terms. As a first step, however, we can restrict which filling fractions to consider by focusing on gapping terms that correspond to tunneling of electrons. Indeed, if we consider an interface between MR states at filling fractions $\nu_1 = 1/k_1$ and $\nu_2 = 1/k_2$, the most general electron tunneling term we can write down is given by
\begin{align}
\cL_{\mathrm{int}} = (\psi_L^\dg)^a \psi_R^b + \mathrm{H.c.} = i\chi_L^a \chi_R^b  \cos (ak_1 \phi_L + b k_2 \phi_R). \label{eqn:MR-naive-e-tunneling}
\end{align}
We will analyze this interaction term in more detail in the following subsection. For now, we emphasize that our implementation of the cut-and-glue approach required that the Majorana and bosonic parts of the interaction term were separately bosonic and so separately obtained expectation values in the strongly interacting limit [see the discussion around Eq. \eqref{eqn:expectation-values}]. Using our analysis of Laughlin interfaces above, we see that this is only possible if $k_1=pb$ and $k_2=pa$, with $a$ and $b$ co-prime \footnote{The condition that $a$ and $b$ be co-prime arose in the Abelian case by requiring primitivity of the gapping term. We do not have a systematic understanding of what constitutes a primitive gapping term in the MR case, but we can at least justify requiring $a$ and $b$ being co-prime by noticing that any tunneling term of the form $(\psi_L^\dg)^{qa} \psi_R^{qb} + H.c.$ with $q$ integer will necessarily be less relevant (in the renormalization group sense) than Eq. (\ref{eqn:MR-naive-e-tunneling}).}. So, we will restrict our attention to gapped interfaces (GIs) between two MR phases, $\mathcal{A}$ and $\mathcal{B}$, at filling fractions
\begin{align}
(\mathcal{A}) \quad \nu = \frac{1}{pb^2} \qquad | \qquad (\mathcal{B}) \quad \nu = \frac{1}{pa^2}.
\end{align}
This is not to say that GIs cannot be formed between MR states at other filling fractions, only that these GIs are those most obviously amenable to our cut-and-glue approach to the calculation of the EE.

In this case, the anyon content of $\mathcal{A} \times \bar{\mathcal{B}}$ is 
\begin{align}
\begin{split}
\mathcal{A} \otimes \bar{\mathcal{B}} = &\{ e^{ir\phi_L}, \chi_L  e^{ir\phi_L}, \sigma_L  e^{i(r+1/2)\phi_L} \}_{r=1,\dots , pb^2} \\
&\otimes \{ e^{is\phi_R}, \chi_R e^{is\phi_R}, \sigma_R e^{i(s+1/2)\phi_R} \}_{s=1,\dots , pa^2}.
\end{split}
\end{align}
Again, our goal is to condense a set of bosonic anyons such that all other anyons will be confined. Our strategy is as follows: we will first condense all possible Abelian anyons. This will yield a new topological order in which all of the non-Abelian anyons will have, hopefully, either become confined or have split into Abelian ones. It will then be straightforward to see whether that order can be reduced to a trivial one.

Motivated by our analysis of the Laughlin problem, we start by condensing the following set of Abelian anyons:
\begin{align}
\cL_0 = \{ e^{ilb\phi_L} e^{ila\phi_R}\}_{l=1,\dots , pab} \times \{ 1_L 1_R, \chi_L \chi_R \}
\label{eqn:abelian-ls}
\end{align}
It follows immediately that all anyons of the form $e^{ir\phi_L} e^{is\phi_R}$ and $\chi_L e^{ir\phi_L} \chi_R e^{is\phi_R}$ not lying in $\cL_0$ will be confined. The condensation pattern of the remaining anyons depends on whether $a$ and $b$ are odd or even. Since we have assumed $a$ and $b$ to be coprime, there are only two cases to consider: (i) one of $a$ and $b$ even, the other odd and (ii) both $a$ and $b$ odd.

\vspace{5mm}

\noindent \textbf{Case (i):} One of $a$ and $b$ even, the other odd

Without loss of generality, let us take $a$ to be even and $b$ odd. In this case, the anyons in the set
\begin{align}
\{ \chi_L e^{ilb\phi_L} e^{ila\phi_R} , e^{ilb\phi_L} \chi_R e^{ila\phi_R}\}_{l=1,\dots , pab},
\end{align}
despite being fermionic, can be condensed after combining them with (fermionic) electrons. In fact, they are all \emph{equivalent} to products of anyons in $\cL_0$, up to fusion with electron operators. For instance,
\begin{align}
\begin{split}
1_L\chi_R &\sim 1_L\chi_R \times (\chi_L e^{ipb^2 \phi_L})^a \times (\chi_R e^{i p a^2 \phi_R})^b \\ 
&= e^{i pab^2 \phi_L} e^{i p a^2 b \phi_R },
\end{split}
\end{align}
where the tilde indicates an equivalence up to fusion with electrons. Here we made use of the fact that $\chi^a \sim 1$, since $a\in 2\mathbb{Z}$. Thus, we should extend the Lagrangian subset from $\cL_0$ to
\begin{align}
\cL &= \{ e^{ilb\phi_L} e^{ila\phi_R}\}_{l=1,\dots , 2pab}  \times \{ 1_L 1_R, \chi_L \chi_R \}. \label{eqn:LS-opp-parity}
\end{align}
It immediately follows that all of the non-Abelian anyons will be confined. Indeed, any anyons of the form 
\begin{align}
\sigma_L e^{i(r+1/2)\phi_L} e^{is\phi_R} \sim \sigma_L e^{i(r+1/2)\phi_L} \chi_R e^{is\phi_R}
\end{align}
and 
\begin{align}
e^{ir\phi_L} \sigma_R e^{i(s+1/2)\phi_R} \sim \chi_L e^{ir\phi_L} \sigma_R e^{i(s+1/2)\phi_R}
\end{align}
will be confined, since they all possess non-trivial braiding with $\chi_L \chi_R$. As for anyons of the form,
\begin{align}
\sigma_{L,r}\sigma_{R,s} \equiv \sigma_L e^{i(r+1/2)\phi_L} \sigma_R e^{i(s+1/2)\phi_R}, \label{eqn:sigr-sigs-anyon}
\end{align}
we can compute their braiding with $e^{ilb\phi_L} e^{ila\phi_R}$ and $\chi_L e^{ilb\phi_L} \chi_R e^{ila\phi_R}$ to be
\begin{align}
e^{i\theta_{(r,s),l}} = \exp \left( 2\pi i\frac{l}{pab} \left[ ar - bs + \frac{1}{2}(a-b) \right] \right) . \label{eqn:sigr-sigs-L0-braiding}
\end{align}
Since $a$ is even while $b$ is odd, this phase can never be trivial. Hence all anyons of the form $\sigma_{L,r}\sigma_{R,s}$ will be confined. Thus, we obtain a gapped interface, but one which is opaque to non-Abelian anyons since they are all confined. 

\vspace{5mm}

\noindent \textbf{Case (ii):} $a, \, b$ both odd

As a first step, we again condense $\mathcal{L}_0$. Upon doing so, the anyon
\begin{align}
f \equiv \chi_L e^{ilb\phi_L} e^{ila\phi_R} \sim \chi_L 1_R \sim 1_L \chi_R \sim e^{ilb\phi_L} \chi_R e^{ila\phi_R}
\end{align}
remains deconfined, where the equivalences come from fusion with elements of $\cL_0$. However, any other anyon of the form $\chi_L e^{ir\phi_L} e^{is\phi_R}$ or $e^{ir\phi_L} \chi_R e^{is\phi_R}$ will clearly be confined, as the chiral boson factors will yield non-trivial braiding with the elements of $\cL_0$. It is also straightforward to see that any anyons of the form $\sigma_L e^{i(r+1/2)\phi_L} e^{is\phi_R} \sim \sigma_L e^{i(r+1/2)\phi_L} \chi_R e^{is\phi_R}$ and $e^{ir\phi_L} \sigma_R e^{i(s+1/2)\phi_R} \sim \chi_L e^{ir\phi_L} \sigma_R e^{i(s+1/2)\phi_R}$ will be confined, since they all possess non-trivial braiding with $\chi_L \chi_R$.

This leaves us with the anyons of Eq. (\ref{eqn:sigr-sigs-anyon}).
Their braiding with $e^{ilb\phi_L} e^{ila\phi_R}$ and $\chi_L e^{ilb\phi_L} \chi_R e^{ila\phi_R}$ is again given by Eq. (\ref{eqn:sigr-sigs-L0-braiding}). Since $a$ and $b$ are both odd, it follows that $a-b \in 2\mathbb{Z}$ and so this phase can be trivial for an appropriate choice of $r$ and $s$.  Specifically, we need to look for $r,s \in \mathbb{Z}$ satisfying the Diophantine equation
\begin{align}
ar-bs + \frac{1}{2}(a-b) = pab t, \quad t \in \mathbb{Z} \label{eqn:anyon-diophantine}
\end{align}
in order to identify the deconfined non-Abelian anyons. One can show that solutions to this equation for arbitrary $t$ are equivalent to those for $t=0$, up to fusion with electrons. It is easy to see that, for the $t=0$ case, one solution to the Diophantine equation is given by
\begin{align}
r_0 = \frac{b-1}{2}, \quad s_0 = \frac{a-1}{2}.
\end{align}
All other solutions can be parameterized as
\begin{align}
r_u = r_0 + u b, \quad s_u = s_0 + u a \label{eqn:diophantine-solution-parameterization}
\end{align}
and correspond to fusing $\sigma_{L,r_0} \sigma_{R,s_0}$ with a condensed anyon in $\cL_0$. Hence, after condensing $\cL_0$, $\sigma_{L,r_0} \sigma_{R,s_0}$ is the only non-Abelian anyon (up to fusion with electrons and condensed anyons) which is not confined.

In order to understand the fate of $\sigma_{L,r_0} \sigma_{R,s_0}$ after condensing the anyons in $\cL_0$, let us check the fusion of $\sigma_{L,r_0} \sigma_{R,s_0}$ with itself. We have that
\begin{align}
\begin{split}
& \sigma_{L,r_0} \sigma_{R,s_0} \times \sigma_{L,r_0} \sigma_{R,s_0}  \\
&= (1+ \chi_L 1_R + 1_L \chi_R + \chi_L \chi_R ) e^{ib \phi_L} e^{ia \phi_R} \\
			&\to 2 \times 1 + 2 \times f
\end{split}
\end{align}
where, in the last step, we applied the identifications arising from condensing $\mathcal{L}_0$. Since the vacuum appears twice in this fusion rule, $\sigma_{L,r_0} \sigma_{R,s_0}$ must split \cite{Bais2009} into two Abelian anyons: $\sigma_{L,r_0} \sigma_{R,s_0} \to e + m$, with the fusion rules $e^2=m^2=f^2=1$ and $e\times m=f$. So, after condensing the Abelian anyons in $\cL_0$, we are left with the Abelian anyons $\{1,e,m,f\}$. Now, since $\sigma_{L,r_0} \sigma_{R,s_0}$ has bosonic self-statistics,
\begin{align}
e^{i\theta_{(r_0,s_0)}} &= \exp \left( \pi i\left[\frac{(r_0+1/2)^2}{pb^2} - \frac{(s_0+1/2)^2}{pa^2}\right] \right) = 1,
\end{align}
it follows that the daughter $e$ and $m$ anyons must also be self-bosons. Additionally, the monodromy associated with braiding $f$ around $\sigma_{L,r_0}\sigma_{R,s_0}$, and hence also around either $e$ or $m$, is $-1$. So, this condensation pattern is essentially that of the $\mathrm{Ising}\times\bar{\mathrm{Ising}} \to \text{Toric code}$ transition.  We can then condense either $e$ or $m$ to fully gap out the interface. In contrast to the previous case, however, a subset of non-Abelian anyons \emph{can} pass through this interface. 

We thus conclude that we can always form a GI between Moore-Read states at filling fractions $\nu^{-1} = pa^2$ and $\nu^{-1} = pb^2$, although the nature of the interface depends on whether or not $a-b \in 2 \mathbb{Z}$.

\subsection{Gapping Terms for \texorpdfstring{$\nu^{-1}_1 = pb^2$}{1/nu=pb2} and \texorpdfstring{$\nu^{-1}_2 = pa^2$}{1/nu=pb2} MR Interfaces \label{sec:nonuniform-gapping-terms}}

We now turn to the problem of constructing explicit interactions which can gap out these interfaces by drawing some intuition from the above anyon condensation pictures.

\subsubsection{Equal Parity Interface: \texorpdfstring{$a, \, b \in 2\mathbb{Z}+1$}{a,b in 2Z+1}}

Let us first focus on the interface between $\nu^{-1}_1 = pb^2$ and $\nu^{-1}_2 = pa^2$ with $a$ and $b$ both odd. In this case, the na\"ive electron tunneling term of Eq. (\ref{eqn:MR-naive-e-tunneling}) takes the form
\begin{align}
    \cL_{\mathrm{int}} = i \chi_L \chi_R \cos(pab^2 \phi_L + pa^2 b \phi_R), \label{eqn:MR-same-parity-gap-int}
\end{align}
where we used the fusion rule $\chi^2 = 1$. [To be more careful about this, one should point-split Eq. (\ref{eqn:MR-naive-e-tunneling}) and perform an operator product expansion to obtain Eq. (\ref{eqn:MR-same-parity-gap-int})]. It is straightforward to see that, in the strongly interacting limit, this interaction term will gap out both the scalar fields and Majorana fermions.

How does this interaction term connect with the anyon condensation picture described above? As a start, one may ask what anyon (or anyons) generate the set of anyons, $\cL_0$, of Eq. (\ref{eqn:abelian-ls}). First, we note that up to the electronic combinations $\chi_L e^{-ipb^2 \phi_L}$ and $\chi_R e^{-ipa^2 \phi_R}$,
\begin{align}
\begin{split}
& e^{i pab^2 \phi_L} e^{i p a^2 b \phi_R } \\
&\sim (\chi_L e^{-ipb^2 \phi_L})^a \times (\chi_R e^{-i p a^2 \phi_R})^b \times e^{i pab^2 \phi_L} e^{i p a^2 b \phi_R } \\
		&= \chi_L \chi_R.
\end{split}
\end{align}
So, all anyons in $\cL_0$ can be obtained by fusing the anyon $e^{ib\phi_L} e^{ia\phi_R}$ with itself some number of times, which is to say, $\cL_0$ is generated by a single anyon. Additionally, we observed above that
\begin{align}
\begin{split}
& \sigma_{L,r_0} \sigma_{R,s_0} \times \sigma_{L,r_0} \sigma_{R,s_0}  \\
&= (1+ \chi_L 1_R + 1_L \chi_R + \chi_L \chi_R ) e^{ib \phi_L} e^{ia \phi_R}, \label{eqn:sigRsigL-fusion}
\end{split}
\end{align}
which means the elements of $\mathcal{L}_0$, and hence the full Lagrangian subset, can all be generated from this single non-Abelian anyon. This suggests that the corresponding gapped edge can be obtained using a single gapping term, namely that given by Eq. (\ref{eqn:MR-same-parity-gap-int}). Indeed, in the strong coupling limit, the argument of the cosine will be pinned and $\chi_L \chi_R$ will obtain an expectation value, corresponding to the condensation of $\chi_L\chi_R$ and all anyons of the form $e^{ilb\phi}e^{ila\bar{\phi}}$, as suggested by the Lagrangian subset picture. That, roughly speaking, $\sigma_{L,r_0}\sigma_{R,s_0}$ is condensed can be inferred from Eq. \eqref{eqn:sigRsigL-fusion}, since $e^{ib\phi}e^{ia\bar{\phi}}$ is also condensed, or by analogy with the standard Ising model, in which the condensation of $\chi_L \chi_R$ implies a gap for the full theory.

\subsubsection{Opposite Parity Interface: \texorpdfstring{$a \in 2\mathbb{Z}, \, b \in 2\mathbb{Z}+1$}{a in 2Z, b in 2Z+1}}

In contast to the previous case, the naive tunneling term of Eq. (\ref{eqn:MR-naive-e-tunneling}) will not serve to gap out the interface. Indeed, since $a$ is even and $b$ is odd, we have that $\psi_L^a \psi_R^b$ is fermionic and so cannot obtain a non-zero expectation value. In order to identify an appropriate gapping interaction, let us try to draw some intuition from the above anyon condensation picture. In particular, we may ask which anyons generate the set $\mathcal{L}$ of Eq. (\ref{eqn:LS-opp-parity}). By inspection, we see that $\mathcal{L}$ has the group structure $\mathbb{Z}_{2pab} \times \mathbb{Z}_2$. (Note that $\chi_L\chi_R$ is \emph{not} equivalent up to fusion with electrons with $e^{ipab^2\phi_L}e^{ipa^2b\phi_R}$ when one of $a$ and $b$ is even and the other odd.) In particular, $\cL$ is generated by $e^{ib\phi_L}e^{ia \phi_R}$ and $\chi_L\chi_R$. This suggests that we will need two distinct tunneling terms to condense the anyons in each of the $\mathbb{Z}_{2pab}$ and $\mathbb{Z}_{2}$ factors and hence fully gap the interface.

Motivated by this observation, we can write down what is effectively the square of the na\"ive electron tunneling operator of Eq. (\ref{eqn:MR-naive-e-tunneling}):
\begin{align}
    \cL_{c} = (\psi_L^\dg)^{2a}\psi_R^{2b} + H.c. = \cos(2pab^2 \phi_L + 2pa^2 b \phi_R),
\end{align}
where we again used the fusion rule $\chi^2 = 1$. It is clear that this interaction can gap out the charged sector (i.e. the scalar fields) and the pinning of the argument of the cosine will correspond to the condensation of the anyons $e^{ilb\phi_L} e^{ila\phi_R}$ in Eq. (\ref{eqn:LS-opp-parity}).

We are thus left with the task of gapping out the neutral degrees of freedom, namely the Majorana fermions. The na\"ive expectation, on inspection of Eq. (\ref{eqn:LS-opp-parity}), is that the neutral sector should be gapped out by a term of the form $(\chi_L \chi_R)^2$, since $\chi_L^2=1$ and $\chi_R^2=1$ are local quasi-particles and $(\chi_L \chi_R)^2$ obtaining an expectation value would correspond to the condensation of $\chi_L \chi_R$. But, it is precisely due to these fusion rules that  $(\chi_L \chi_R)^2 \sim 1$ cannot introduce a gap. More precisely, on point-splitting the interaction, one finds $(\chi_L \chi_R)^2 \sim \chi_L \partial \chi_L \chi_R \partial \chi_R$, which is an irrelevant interaction (in the RG sense) and cannot perturbatively introduce a gap \footnote{In the Ising model, this interaction induces a flow from the tricritical to the critical Ising CFT, all along which the fermions remain massless \cite{Zamoldchikov1991}. Beyond the tricritial Ising CFT fixed point, this interaction does open a gap. Although a similar situation may arise here, we are interested in writing down relevant interactions, which we know will perturbatively introduce a gap}. Evidently, we must employ a more indirect approach to fully gap out the interface.

Indeed, we will make use of an alternative representation of the Ising CFT
\begin{align}
\mathrm{Ising} = \frac{SO(N+1)_1}{SO(N)_1} \sim SO(N+1)_1 \boxtimes \overline{SO(N)}_1,
\end{align}
where $N=2r$ is an even number with $r>1$, $SO(N)_1$ denotes the $SO(N)$ Kac-Moody algbera at level one, and the tensor product $\boxtimes$ denotes a usual tensor product combined with the condensation of a particular set of bosonic anyons to tie the two factors together. The details of this representation are reviewed in Appendix \ref{sec:appendix-ising}. This representation allows us to to re-express the Majorana sector of the MR theory in terms of $N+1$ left-moving and $N$ right-moving Majorana fermions. The topological data of theory (i.e. the anyon content) will remain the same in this alternative reprsentation due to the choice of condensed operators encoded in the $\boxtimes$ notation. In particular, all $2N+1$ Majorana operators belong to a single topological sector. So, we expect to obtain the correct TEE in our entanglement calculation. However, the total central charge will change and will alter the area law term in the entanglement entropy. This, of course, is not distressing since the coefficient of the area law term is a non-universal quantity. The upshot of this alternative representation is that we can write down current-current backscattering interactions which are manifestly local and marginally relevant, which means they \emph{can} induce a gap. 

Explicitly, in this alternative reprsentation, we can write the free part of the $\nu = 1 / n$ MR edge theory as 
\begin{align}
\begin{split}
\cL = &\frac{n}{4\pi} \partial_x \phi (\partial_t - \partial_x) \phi + \frac{1}{4\pi} \sum_{j=1}^{r} \partial_x \phi^j (\partial_t - \partial_x) \phi^j \\ 
&+ \frac{1}{4\pi} \sum_{j=1}^r \partial_x \overline{\phi}^j (-\partial_t - \partial_x) \overline{\phi}^j + \chi \frac{i}{2} (\partial_t - \partial_x)\chi, \end{split} \label{eqn:MR-coset-description}
\end{align}
with the local operators being the electron operator
\begin{align}
\psi_e = \chi e^{in \phi},
\end{align}
the $SO$ currents of Eqs. \eqref{eqn:SO(N+1)-currents-1}-\eqref{eqn:SO(N)-currents}, as well as the condensed operators of Eqs. \eqref{eqn:condensed-currents-1}-\eqref{eqn:condensed-currents-2}. As usual, it is important to understand the organization of the Hilbert space. To that end, let us place this MR phase on a cylinder so that we have chiral and anti-chiral copies on the left ($L$) and right ($R$) edges of the cylinder. We then define the operator 
\begin{align}
\begin{split}
G' &= G (-1)^{\sum_j (N_R^j + N_L^j)} (-1)^{\sum_j (\overline{N}_R^j + \overline{N}_L^j)} \\
   &= (-1)^{N_R+N_L} (-1)^F (-1)^{\sum_j (N_R^j + N_L^j)} (-1)^{\sum_j (\overline{N}_R^j + \overline{N}_L^j)},
   \end{split} \label{eqn:alternate-MR-CFT-Z2-symmetry}
\end{align}
where $N_\mu$, $N_\mu^j$, and $\bar{N}_\mu^j$ are the winding modes of $\phi_\mu$, $\phi_\mu^j$, and $\bar{\phi}^j_\mu$, respectively. One can check that $G'$ commutes with all the local-electronic operators in this theory. Hence, similar to the conventional MR edge theory, the physical Hilbert space is defined by the constraint $G'=1$. This simply states that the charge (i.e. the winding number parity of the $\phi$ field) must match the combined fermion number parity and neutral boson winding number parity. In particular, in the $\mathbf{1}$ sector we can define separate fermion parities for each edge. As such, we can define the operators
\begin{align}
    G'_{\mu} = (-1)^{N_{\mu}} & (-1)^{F_{\mu}} (-1)^{\sum_j N_{\mu}^j} \times (-1)^{\sum_j \overline{N}_{\mu}^j}. \label{eqn:alternate-MR-CFT-Z2-symmetry-edge}
\end{align}
The $\mathbf{1}$ sector is then defined by the constraint $G'_{\mu}=1$. For later convenience, we can define the operator
\begin{align}
    P_{\mathbf{1}} = P_{\mathbf{1},R} P_{\mathbf{1},L} \label{eqn:alternate-MR-1-projection}
\end{align}
which projects states the cylinder to the $\mathbf{1}$ sector of the MR edge theory. Here, $P_{\mathbf{1},\mu}$ acts on edge $\mu$ of the cylinder and for $\ket{\psi}$ an eigenstate of $(-1)^{F_{\mu}}$, $(-1)^{N_{\mu}}$, $(-1)^{N^j_{\mu}}$, and $(-1)^{N^j_{\mu}}$, we have that, schematically,
\begin{align}
    P_{\mathbf{1},\mu} \ket{\psi} = \ket{\psi} 
\end{align}
if $G'_{\mu} \ket{\psi} = \ket{\psi}$ and
\begin{align}
    P_{\mathbf{1},{\mu}} \ket{\psi} = 0
\end{align}
otherwise.

Returning to the non-uniform interface, we can now employ the current-current interactions described in Appendix \ref{sec:appendix-ising} to gap out the neutral modes \cite{Gross1974}:
\begin{align}
\begin{split}
\cL_n = &u \sum_{j_1 \neq j_2} \cos(2\Theta^{j_1}) \cos(2\Theta^{j_2}) + u \sum_{j=1}^{r} \cos(2\Theta^j) i \chi_L \chi_R \\ 
&+u \sum_{j_1 \neq j_2} \cos(2\overline{\Theta}^{j_1}) \cos(2\overline{\Theta}^{j_2})
\end{split}
\end{align}
where we have defined
\begin{align}
2 \Theta^j \equiv \phi^j_R - \phi^j_L , \quad 2\overline{\Theta}^j = \overline{\phi}^j_R - \overline{\phi}^j_L .
\end{align}
In its fermionized form, as presented in Eq. (\ref{eqn:ising-fermionized-gapping-int}) of Appendix \ref{sec:appendix-ising}, we see that $\cL_n$ does indeed, heuristically, represent a $(\chi_L\chi_R)^2$ interaction, in line with our intuition from the anyon condensation picture. It is clear that, taken together, the charge sector and neutral sector interaction terms,
\begin{align}
\cL_{\mathrm{gap}} \equiv \cL_c + \cL_n, \label{eqn:charge-neutral-gapping-term}
\end{align}
will fully gap the interface. 

\section{Non-Uniform Interface Entanglement Entropy \label{sec:non-uniform-EE}}

Having established which interfaces of MR states can be gapped and which explicit interactions can induce these gaps, we can proceed to apply the cut-and-glue approach to the calculation of the EE for these interface systems. We again consider the geometry of Fig. \ref{fig:torus} except, now, region $A$ ($B$) will be occupied by a $\nu^{-1} = pb^2$ ($\nu^{-1} = pa^2$) MR state. The entanglement cut thus lies on the interface between these two distinct topological orders. We will consider the two classes of interfaces discussed in the previous section in turn. Our analysis will parallel that of Ref. \cite{Cano2015}, in that we will first illustrate how the gapping interactions place constraints on the ground state. Aside from these constraints, the actual computation of the ground state and the EE then proceeds in essentially the same way as for the uniform interfaces. We will focus, for simplicity, on the trivial ($\mathbf{1}$) sector. 

\subsection{Equal Parity Interface \label{sec:equal-parity}}

We begin by considering the case where both $a$ and $b$ are odd. As in the uniform interface calculation, we will focus on the $LA/RB$ interface (i.e. interface $1$). For ease of access, we restate here the free Lagrangian,
\begin{align}
\begin{split}
 \cL_{\mathrm{dec},1}& = \chila \frac{i}{2}(\partial_t - v_n \partial_x) \chila + \frac{pb^2}{4\pi} \partial_x \phila (\partial_t - v_c \partial_x ) \phila \\
		 &+ \chirb \frac{i}{2}(\partial_t + v_n \partial_x) \chirb + \frac{pa^2}{4\pi} \partial_x \phirb (-\partial_t - v_c \partial_x ) \phirb ,
\end{split} \label{eqn:BareEdge-LA-RB-EqualParity} 
\end{align}
and the gapping interaction,
\begin{align}
    \cL_{\mathrm{gap},1} &= -\frac{2g}{\pi}  i \chi_{LA} \chi_{RB} \cos(pab^2 \phi_{LA} + pa^2 b \phi_{RB}). \label{eqn:MR-same-parity-gap-int-LA-RB}
\end{align}

\subsubsection{Gapping Term Constraints}
As in the uniform interface problem, we will take the strongly interacting limit and approximate
\begin{align}
\begin{split}
H_{\mathrm{gap},1} \approx & \int_0^L dx \big[ \mathrm{const.}  + v_n \tilde{g}i\chila \chirb \\
& + \frac{v_c \lambda\pi}{2} (b \phila + a \phirb - \pi/pab)^2 \big] dx ,
\end{split}
\end{align}
where $\tilde{g} = -2g/(v_n \pi) < 0$ and we have expanded about the vacuum
\begin{align}
\begin{split}
\la pab^2\phila + pa^2b\phirb \ra &= \pi \\
\la i\chila \chirb \ra &> 0.
\end{split}
\end{align}
We perform a similar approximation for interface $2$. As before, this violates the $\mathbb{Z}_2$ gauge symmetries generated by the $G_\alpha$ operators [Eq. \eqref{eqn:g_alpha}], and so the ground state to the approximated Hamiltonian will need to be projected to the $G_\alpha = 1$ subspace. However, following Ref. \cite{Cano2015}, an additional constraint is imposed by the gapping interaction.

Indeed, as in the case of the uniform interface problem, the pinning of the cosine term implies the linear combination of the scalar fields $b\phi_{LA} + a\phi_{RB}$ cannot fluctuate significantly from its vacuum expectation value over the length of the system. In particular, it cannot have a non-zero winding, which requires that
\begin{align}
    b N_{LA} + a N_{RB} = 0.
\end{align}
Since $a$ and $b$ are coprime, this relation fixes the quantization of the winding numbers to be
\begin{align}
N_{LA} = a z, \quad N_{RB} = -b z, \quad z \in \mathbb{Z}. \label{eqn:equal-parity-winding-quantization}
\end{align}
The physical content of this restriction is clear in view of the form of the gapping interaction, which involves scattering $a$ electrons from edge $LA$ with $b$ holes from edge $RB$. The ground state of the interface will then naturally consist of a superposition of states consisting of multiples of $(\psi_{LA}^\dg)^a \psi_{RB}^b$ particle-hole pairs. This is precisely what is expressed by the above constraint, once we also enforce the $\mathbb{Z}_2$ gauge symmetry constraint, which ties the bosonic winding to the fermionic parity.

\subsubsection{Entanglement Entropy Calculation}

The calculation of the EE is nearly identical to that of the uniform interface case, with the primary difference being that we must take into account the above constraints on the winding numbers. The approximated Hamiltonian again takes the decoupled form
\begin{align}
H_1 \equiv H_{1,f}^{\mathrm{osc}} + H_{1,b}^{\mathrm{osc}} + H_{1,b}^{\mathrm{zero}}. \label{eqn:H-interface-1-approx-equal-parity}
\end{align}
The fermionic part of the approximated Hamiltonian, $H_{1,f}^{\mathrm{osc}}$, is identical to that for the uniform interface problem, Eq. (\ref{eqn:H-f-osc}), and so the ground state of the fermionic sector will again be given by Eq. (\ref{eqn:f-osc-ground-state-1}).  The bosonic parts of the Hamiltonian are now given by: 
\begin{align}
    H_{1,b}^{\mathrm{zero}} &= \frac{\pi v_c p }{2L} (aN_{RB} -bN_{LA})^2 + \frac{\pi\lambda v_c L}{2} (a\phi_{RB,0} + b\phi_{LA,0})^2 , \\
    H_{1,b}^{\mathrm{osc}} &= \frac{v_c}{2}\sum_{k\neq 0} (a_k^\dg \quad a_{-k} )
		\begin{pmatrix}
			A_k && B_k \\
			B_k && A_k		
		\end{pmatrix}
		\begin{pmatrix}
			a_k \\
			a_{-k}^\dg		
		\end{pmatrix},
\end{align}
where
\begin{align}
    A_k = |k| + \frac{2\lambda \pi^2}{p|k|}, \quad B_k = \frac{2\lambda \pi^2}{p|k|}.
\end{align}
Dispensing with the details, we simply jump to writing down the ground state for the approximated Hamiltonian (including both interfaces):
\begin{align}
\ket{\hat{\psi}_{\mathbf{1}}} &= \ket{\hat{\psi}_{1,{\mathbf{1}}}}\otimes \ket{\hat{\psi}_{2,{\mathbf{1}}}}, \\
\ket{\hat{\psi}_{1/2,{\mathbf{1}}}} &= \ket{G_{b,\mathrm{zero},1/2}} \otimes \ket{G_{b,\mathrm{osc},1/2}} \otimes \ket{G_{f,\mathrm{osc},1/2}}
\end{align}
where
\begin{align}
\begin{split}
	\ket{G_{b,\mathrm{zero},1}} &= \sum_{N \in \mathbb{Z}} e^{-\frac{v_e \pi pa^2b^2 N^2}{2L}} \ket{N_{RB}=bN,N_{LA}=-aN} , \label{eqn:b-zero-ground-state-nonuniform-1} \\
	\ket{G_{b,\mathrm{zero},2}} &= \sum_{N \in \mathbb{Z}} e^{-\frac{v_e \pi pa^2b^2 N^2}{2L}} \ket{N_{LB}=-bN,N_{RA}=aN}, 
\end{split}
\end{align}
while $\ket{G_{b,\mathrm{osc},1/2}}$ and $\ket{G_{f,\mathrm{osc},1/2}}$ are again given by Equations (\ref{eqn:b-osc-ground-state-1}), (\ref{eqn:b-osc-ground-state-2}) and (\ref{eqn:f-osc-ground-state-1}), (\ref{eqn:f-osc-ground-state-2}), respectively. The constraint imposed by the gapping interaction manifests itself in the sums over the winding mode states. The entanglement velocities are given by
\begin{align}
    v_e = \frac{2}{\pi}\sqrt{\frac{p}{\lambda}}, \qquad \tilde{v}_e = \frac{2}{|\tilde{g}|}.
\end{align}

Following the now standard procedure, we must apply the projection operator $P_{{\mathbf{1}}} \equiv P_{{\mathbf{1}},A} P_{{\mathbf{1}},B}$ defined in Eq. (\ref{eqn:projection}) to obtain a physical state in the MR Hilbert space. As in the uniform interface case, we again have that $P_{{\mathbf{1}}} \ket{\hat{\psi}_{{\mathbf{1}}}} = P_{{\mathbf{1}},A}\ket{\hat{\psi}_{{\mathbf{1}}}} = P_{{\mathbf{1}},B}\ket{\hat{\psi}_{{\mathbf{1}}}}$. Indeed, we see that every state appearing in $\ket{\hat{\psi}_{{\mathbf{1}}}}$ has $(-1)^{F_{RB}}=(-1)^{F_{LA}}$ and $(-1)^{F_{LB}}=(-1)^{F_{RA}}$. Additionally, since \emph{both} $a$ and $b$ are \emph{odd}, we have that $(-1)^{bN}=(-1)^{aN}$, and so the states also satisfy $(-1)^{N_{RB}}=(-1)^{N_{LA}}$, as well as $(-1)^{N_{LB}}=(-1)^{N_{RA}}$. It then readily follows that 
\begin{align}
    \ket{\psi_{\mathbf{1}}} = P_{\mathbf{1}} \ket{\hat{\psi}_{\mathbf{1}}} = P_{{\mathbf{1}},A} \ket{\hat{\psi}_{\mathbf{1}}} = P_{{\mathbf{1}},B} \ket{\hat{\psi}_{\mathbf{1}}}.
\end{align}
As in the uniform interface problem, $P_{{\mathbf{1}},B} \ket{\hat{\psi}_{\mathbf{1}}}$ is again in a Schmidt decomposed form, and so we can directly read off the entanglement spectrum and hence the reduced density matrix for $B$ (the only difference with the uniform interface calculation is the winding mode sector. We have that
\begin{align}
    \rho_{\mathbf{1},B} =  \frac{1}{Z_{e^{ir\phi},e}} P_{\mathbf{1},B} P_{b} e^{-\cH_e^{RB}-\cH_e^{LB}} P_{b} P_{\mathbf{1},B}, \label{eqn:approx-rho-equal-parity}
\end{align}
where $\cH_e^{RB}$ and $\cH_e^{LB}$ are given by Equations (\ref{eqn:untwisted-He-RB}) and (\ref{eqn:untwisted-He-LB}), respectively, with the substitution $n=pa^2$. The operator $P_b$ enforces the constraint of Eq. (\ref{eqn:equal-parity-winding-quantization}):
\begin{align}
    P_b \ket{N_{RB},N_{LB}} = \delta_{N_{RB},0\,\text{mod} b} \delta_{N_{LB},0\,\text{mod} b} \ket{N_{RB},N_{LB}}.
\end{align}

It is now a straightforward matter to derive the entanglement partition function. As before, we can write $Z_{\mathbf{1},e}$ as a product of contributions from the right and left edges:
\begin{align}
    Z_{\mathbf{1},e} = Z_{\mathbf{1},e}^{RB}Z_{\mathbf{1},e}^{LB}.
\end{align}
Explicitly,
\begin{align}
\begin{split}
Z_{\mathbf{1},e}^{RB} &=  \chi_0^{\text{Ising}}(\tilde{q}) \left( \sum_{N\in\mathrm{even}} q^{pa^2 (bN)^2/2} \right) q^{-\frac{1}{24}} \prod_{j=1}^\infty \left( 1-q^j \right)^{-1} \\
&+  \chi_{1/2}^{\text{Ising}}(\tilde{q}) \left( \sum_{N\in\mathrm{odd}} q^{pa^2 (bN)^2/2} \right) q^{-\frac{1}{24}} \prod_{j=1}^\infty \left( 1-q^j \right)^{-1} ,
\end{split} \label{eqn:same-parity-Z}
\end{align}
where $\chi_0^{\text{Ising}}(\tilde{q})$ and $\chi_{1/2}^{\text{Ising}}(\tilde{q})$ were defined in Eqs. (\ref{eqn:MW-0-character}) and (\ref{eqn:MW-1/2-character}), respectively, and $Z_{\mathbf{1},e}^{LB}$ is given by a similar expression. As in the entanglement partition function for the untwisted sectors of the uniform interface problem, the first (second) line of Eq. (\ref{eqn:same-parity-Z}) arises from the states in the trace which have both an even (odd) fermion parity and winding number parity. It is immediate to see that Eq. (\ref{eqn:same-parity-Z}) is formally equivalent to Eq. (\ref{eqn:Z-untwisted-RB}) with the substitutions $n \to pa^2 b^2$ and $r \to 0$. This implies that Eq. (\ref{eqn:same-parity-Z}) is in fact the partition function in the trivial sector for a MR state at inverse filling $\nu^{-1} = pa^2b^2$. We will have more to say on this point later in this section but, for now, this observation allows us to immediately deduce the EE in the present non-uniform interface problem to be,
\begin{align}
S_{\mathbf{1}}= -2\ln(2 \sqrt{pa^2 b^2}) + \frac{\pi L}{3} \left(\frac{1}{v_e} + \frac{1}{2 \tilde{v}_e} \right).
\end{align}
We thus find the TEE for this nonuniform interface on the torus (in the vacuum sector) is given by
\begin{align}
\gamma_1 = 2\ln(2 \sqrt{pa^2 b^2}),    \label{eqn:tee-opp-parity}
\end{align}
which is one of the main results of this paper.

\subsection{Opposite Parity Interface \label{sec:opposite-parity}}
We now turn to the class of interfaces in which one of $a$ and $b$ is even and the other odd. Without loss of generality, we will again take $a$ to be even and $b$ to be odd. We will also employ the topologically equivalent representation of the MR CFT, as discussed in Section \ref{sec:nonuniform-gapping-terms} and detailed in Appendix \ref{sec:appendix-ising}. Again focusing on interface $1$, the free part of the Lagrangian is given by,
\begin{align}
\begin{split}
&\cL_{\mathrm{dec},1} = \sum_{\mu}\left[ \frac{k_{\mu}}{4\pi} \partial_x \phi_\mu (\mu \partial_t - v_c\partial_x) \phi_\mu + \chi_\mu \frac{i}{2} (\partial_t - \mu v_n \partial_x)\chi_\mu \right. \\ 
& \left. + \frac{1}{4\pi} \sum_{j=1}^r \left\{\partial_x \overline{\phi}^j_\mu (-\mu \partial_t - v_n \partial_x) \overline{\phi}^j_\mu +  \partial_x \phi^j_\mu (\mu \partial_t - v_n \partial_x) \phi^j_\mu \right\} \right], \end{split}
\end{align}
where, in the interest of compactness, we have abused our earlier notation by temporarily redefining $\mu=LA/RB=+/-$. We have also set,
\begin{align}
    k_{LA} = pa^2, \quad k_{RB} = pb^2.
\end{align}
The gapping interaction is given by
\begin{align}
\cL_{\mathrm{gap},1} &= \cL_{c,1} + \cL_{n,1}, \\
\cL_{c,1} &= -\frac{2g}{\pi} \cos(2pab^2 \phi_{LA} + 2pa^2 b \phi_{RB}) \label{eqn:L-charge-gap-1} \\
\begin{split}
\cL_{n,1} &= u \sum_{j_1 \neq j_2}\left[ \cos(2\Theta^{j_1}_1) \cos(2\Theta^{j_2}_1) +  \cos(2\overline{\Theta}^{j_1}_1) \cos(2\overline{\Theta}^{j_2}_1)\right] \\
        &+ u \sum_{j=1}^{r} \cos(2\Theta^j_1) i \chi_{LA} \chi_{RB},
\end{split}
\end{align}
where,
\begin{align}
2 \Theta^j_1 \equiv \phi^j_{RB} - \phi^j_{LA} , \quad 2\overline{\Theta}^j_1 = \overline{\phi}^j_{RB} - \overline{\phi}^j_{LA},
\end{align}
and we take $u,g>0$. We will also require the mode expansions
\begin{align}
\begin{split}
\phi_\mu^j &= \phi_{\mu,0}^j + 2\pi N_{\mu}^j \frac{x}{L} + \sum_{\mu k<0} \sqrt{\frac{2\pi}{L|k|}} \left[a_k^j e^{ikx} + (a_k^j)^\dg e^{-ikx}\right] \\
\bar{\phi}_\mu &= \bar{\phi}_{\mu,0} + 2\pi \bar{N}_{\mu}^j \frac{x}{L} + \sum_{\mu k>0} \sqrt{\frac{2\pi}{L|k|}} \left[\bar{a}_k^j e^{ikx} + (\bar{a}_k^j)^\dg e^{-ikx}\right]
\end{split}
\end{align}
where
\begin{align}
&[(a_k^i)^\dg, a_{k'}^{j} ] = [(\bar{a}_k^i)^\dg, \bar{a}_{k'}^{j} ] =\delta_{k,k'}\delta_{i,j}, \\
&[\phi_{\mu,0}^i, N_{RB}^j] = -[\bar{\phi}^i_{\mu,0}, \bar{N}^j_{\mu}] = -i \delta_{i,j},
\end{align}
and we have temporarily set $\mu=LA/RB=+/-$.

\subsubsection{Gapping Term Constraints}
We now take the strong coupling limit. Without loss of generality, we expand about the vacuum defined by the expectation values
\begin{align}
\begin{split}
\la 2 pab^2\phila + 2 pa^2b\phirb \ra &= \pi \\
    \la 2\Theta_1^j \ra = \la 2\Theta_1^j \ra &= 0 \\
    \la i\chi_{LA} \chi_{RB} \ra &< 0,
\end{split}
\end{align}
so that
\begin{align}
\begin{split}
H_{\mathrm{gap},1} \approx & \int_0^L \Bigg[ \frac{\lambda\pi}{2} \sum_j \left[ (2\Theta^{j} )^2 + (2\overline{\Theta}^{j} )^2 \right] + \tilde{g} i \chi_{LA} \chi_{RB} \\ 
& + \frac{v_c \tilde{\lambda}\pi}{2} (b \phila + a \phirb - \pi /(2pab))^2 \Bigg] dx.
\end{split}
\end{align}
Here, $\lambda,\tilde{\lambda}>0$ and $\tilde{g} = -ru < 0$. As in the equal parity interface problem, the pinning of $b \phila + a \phirb$ enforces the constraint Eq. (\ref{eqn:equal-parity-winding-quantization}), while the pinning of the $2\Theta_1^j$ and $2\bar{\Theta}_1^j$ fields enforces the constraints
\begin{align}
    N_{LA}^j = N_{RB}^j \in \mathbb{Z} , \quad \bar{N}_{LA}^j = \bar{N}_{RB}^j \in \mathbb{Z}.
\end{align}
Note that, at this level of our approximation, the factor of two in the argument of $\mathcal{L}_{c,1}$, which reflects the fact that we must tunnel an even number of electrons, does not play any role. This will be accounted for once we project to the physical Hilbert space. 

\subsubsection{Entanglement Entropy Calculation}

We see that, in the approximated Hamiltonian, the Majorana fermion, neutral boson, and charged boson sectors all decouple. In particular, the Hamiltonians for each of these sectors have already appeared in our calculations for the equal-parity interface in Eq. (\ref{eqn:H-interface-1-approx-equal-parity}). Hence, we will skip the details of the computation and simply jump to writing down the ground state of the approximated Hamiltonian:
\begin{align}
    \begin{split}
    \ket{\hat{\psi}_{{\mathbf{1},1}}} &= \ket{G_{b,\mathrm{zero},1}} \otimes \ket{G_{b,\mathrm{osc},1}} \otimes \ket{G_{f,\mathrm{osc},1}} \\
    & \otimes \prod_{j=1}^r\ket{G_{n,\mathrm{zero},1}^j} \otimes \prod_{j=1}^r \ket{G_{n,\mathrm{osc},1}^j},
    \end{split}
\end{align}
where, $\ket{G_{b,\mathrm{osc},1}}$, $\ket{G_{f,\mathrm{osc},1}}$, and $\ket{G_{b,\mathrm{zero},1}}$ are again given by Equations (\ref{eqn:b-osc-ground-state-1}), (\ref{eqn:f-osc-ground-state-1}), and (\ref{eqn:b-zero-ground-state-nonuniform-1}), respectively, while the ground states for the neutral boson oscillator and zero-mode sectors of interface $1$, respectively, take the form
\begin{align}
\ket{G_{n,\mathrm{osc},1}^j} &= \exp\left( \sum_{k>0} e^{-\frac{v_{e,n} k}{2}} [(a_k^j)^\dg (a_{-k}^j)^\dg + (\bar{a}_k^j)^\dg (\bar{a}_{-k}^j)^\dg] \right) \ket{0} \label{eqn:groundstate-neutral-boson-osc}
\end{align}
\begin{align}
\begin{split}
\ket{G_{n,\mathrm{zero},1}^j} &=  \left( \sum_{\bar{N}^j} e^{-\frac{v_{e,n} \pi (\bar{N}^j)^2}{2L}} \ket{\bar{N}_{RB}^j = \bar{N}^j,\bar{N}_{LA}^j = \bar{N}^j}\right) \\ 
        &\otimes  \left( \sum_{N^j} e^{-\frac{v_{e,n} \pi (N^j)^2}{2L}} \ket{N_{RB}^j = N^j,N_{LA}^j = N^j}\right),
\end{split} \label{eqn:groundstate-neutral-boson-zero}
\end{align}
where the non-universal entanglement velocity $v_{e,n}$ depends on the field expectation values in an unimportant way. The corresponding state for interface $2$, $\ket{\hat{\psi}_{\mathbf{1},2}}$, is given by a similar expression.

As usual, we obtain an approximation to the physical ground state of the unapproximated gapping Hamiltonian in the $\mathbf{1}$ sector by applying a projection to $\ket{\hat{\psi}_{\mathbf{1}}}$. Defining $\mathbb{Z}_2$ symmetry operators, Eq. (\ref{eqn:alternate-MR-CFT-Z2-symmetry-edge}), for each cylinder, $G'_{\mu\alpha}$ (where $\mu=L,R$, $\alpha=A,B$), the $\mathbf{1}$ sector is defined by the constraint $G'_{\mu\alpha}=1$. Likewise, we define copies of the projection operators, Eq. (\ref{eqn:alternate-MR-1-projection}), for each cylinder: $P_{\mathbf{1},\alpha} = P_{\mathbf{1},L\alpha}P_{\mathbf{1},R\alpha}$. We thus obtain an approximation to the ground state in the physical Hilbert space via the projection
\begin{align}
    \ket{\psi_{\mathbf{1}}} = P_{\mathbf{1}}\ket{\hat{\psi}_{\mathbf{1}}} = P_{\mathbf{1},A}P_{\mathbf{1},B}\ket{\hat{\psi}_{\mathbf{1}}}.
\end{align}
In contrast to our earlier calculations, however, the projection requires a bit more care, since $a$ is even while $b$ is odd, and so $(-1)^{aN} \neq (-1)^{bN}$ for $N$ odd. Explicitly, we have that
\begin{align}
\begin{split}
    (-1)^{N_{LA}}\ket{\hat{\psi}_{1,\mathbf{1}}} = \ket{\hat{\psi}_{1,\mathbf{1}}},
\end{split}
\end{align}
since each state appearing in $\ket{\hat{\psi}_{1,\mathbf{1}}}$ is an eigenstate of $N_{LA}$ with eigenvalue $aN$ and $(-1)^{aN} = 1$. So, $P_{\mathbf{1},LA}$ will project out all states in $\ket{\hat{\psi}_{1,\mathbf{1}}}$ with
\begin{align}
(-1)^{F_{LA}} (-1)^{\sum_j N_{LA}^j} (-1)^{\sum_j \overline{N}_{LA}^j} = -1,
\end{align}
that is, those states whose fermion parity does not match the neutral boson winding parity. However, we can see from the explicit form of $\ket{\hat{\psi}_{1,\mathbf{1}}}$ that
\begin{align}
(-1)^{F_{RB}+\sum_j N_{RB}^j+\sum_j \overline{N}_{RB}^j} = (-1)^{F_{LA}+\sum_j N_{LA}^j+\sum_j \overline{N}_{LA}^j}
\end{align}
for each state appearing in $\ket{\hat{\psi}_{1,\mathbf{1}}}$. Now, when we apply $P_{\mathbf{1},RB}$ to $P_{\mathbf{1},LA}\ket{\psi_{1,\mathbf{1}}}$, we must project out those states with $(-1)^{N_{RB}} = -1$, since all the remaining states have $(-1)^{F_{RB}+\sum_j N_{RB}^j+\sum_j \overline{N}_{RB}^j} = +1$. But, each state in $\ket{\hat{\psi}_{1,\mathbf{1}}}$ has $N_L = b N$, with $b$ odd, and $(-1)^{bN} =(-1)^{N}$. Thus, the only states remaining in the sum after projection will have $N \in 2\mathbb{Z}$ -- i.e. $N_{RB} = 2 b z$ and $N_{LA} = -2 a z$, with $z=N/2$. Physically, this reflects the fact that we are scattering an even number of electrons and holes, as manifested by the factor of two in the argument of $\mathcal{L}_{c,1}$ [Eq. \eqref{eqn:L-charge-gap-1}].

It is now a simple matter to deduce the entanglement spectrum and hence the entanglement partition function for, say, cylinder $B$. Taking into account the constraints on the fermion parity and bosonic winding number quantum numbers imposed by the projections, we can read off the entangelement spectrum from the explicit forms of $\ket{\psi_{1,\mathbf{1}}}$ and $\ket{\psi_{2,\mathbf{1}}}$, which are in Schmidt-decomposed form. Indeed, we find for the entanglement partition function,
\begin{align}
\begin{split}
Z_{\mathbf{1},e}^{RB} &= \left( \sum_{N\in\mathrm{even}} q^{pb^2 (aN)^2/2} \right) q^{-\frac{1}{24}} \prod_{j=1}^\infty \left( 1-q^j \right)^{-1} \times  \\
& \left( \chi_0^{\text{Ising}}(\tilde{q}) \sum_{ \substack{ \{N_i\} \\ \sum_i N_i \in \mathrm{even}} } q_n^{\sum_i N_i^2 / 2} \left[ q_n^{-\frac{1}{24}} \prod_{j=1}^\infty (1-q_n^j)^{-1} \right]^{2r} \right. \\
  + & \left. \chi_{1/2}^{\text{Ising}}(\tilde{q}) \sum_{ \substack{ \{N_i\} \\ \sum_i N_i \in \mathrm{odd}} } q_n^{\sum_i N_i^2 / 2} \left[ q_n^{-\frac{1}{24}} \prod_{j=1}^\infty (1-q_n^j)^{-1} \right]^{2r} \right),
\end{split} \label{eqn:Z-opposite-parity-RB}
\end{align}
where $q$ and $\tilde{q}$ are again take forms given by Eq. \eqref{eqn:modular-variables} and we have defined $q_n \equiv \exp(2\pi i \tau_n)$, with $\tau_n \equiv i\beta v_{e,n}/L$ [$v_{e,n}$ is defined implicitly in Eqs. \eqref{eqn:groundstate-neutral-boson-osc}, \eqref{eqn:groundstate-neutral-boson-zero}]. We have also used the fact that, since we are in the untwisted sector, we can write
\begin{align}
    Z_{\mathbf{1},e} = Z_{\mathbf{1},e}^{RB} Z_{\mathbf{1},e}^{LA}
\end{align}
and, as usual, $Z_{\mathbf{1},LA}$ takes a similar form to that of $Z_{\mathbf{1},RB}$. We can express the partition function in terms of modular functions:
\begin{align}
\begin{split}
Z_{\mathbf{1},e}^{RB} &= \frac{\theta_0^0(pa^2b^2 \tau) + \theta_{1/2}^0(p a^2 b^2 \tau)}{\eta(\tau)} \times \\
& \left( \frac{1}{4} \left[ \sqrt{\frac{\theta_0^0(\tilde{\tau})}{\eta(\tilde{\tau})}} + \sqrt{\frac{\theta_{1/2}^0(\tilde{\tau})}{\eta(\tilde{\tau})}}\right]  \frac{\theta_0^0(\tau_n)^{2r} + \theta_{1/2}^0(\tau_n)^{2r}}{\eta(\tau_n)^{2r}} \right. \\
& \left. + \frac{1}{4} \left[ \sqrt{\frac{\theta_0^0(\tilde{\tau})}{\eta(\tilde{\tau})}} - \sqrt{\frac{\theta_{1/2}^0(\tilde{\tau})}{\eta(\tilde{\tau})}}\right] \frac{\theta_0^0(\tau_n)^{2r} - \theta_{1/2}^0(\tau_n)^{2r}}{\eta(\tau_n)^{2r}} \right).
\end{split}
\end{align}
Applying the usual modular transformations and taking the large length limit, we find
\begin{align}
    S_{\mathbf{1}} = -2\ln(4 \sqrt{pa^2 b^2}) + \frac{\pi L}{3} \left( \frac{2r}{v_{e,n}} + \frac{1}{v_e} + \frac{1}{2 \tilde{v}_e} \right).
\end{align}
Hence, the TEE for this nonuniform interface on the torus (in the vacuum sector) is given by
\begin{align}
\gamma_{\mathbf{1}} = 2\ln(4 \sqrt{pa^2 b^2}),
\end{align}
which is another of the main results of this paper. Note that this differs from that of the same-parity interface [cf. Eq. \eqref{eqn:tee-opp-parity}].

\subsection{Relation to Parent Topological Phase}

We now provide a physical interpretation for the values of the TEE associated with the non-uniform interface between $\mathcal{A}$ and $\mathcal{B}$, which is based on determining whether a gapped interface can be formed between phases $\mathcal{A}$ and $\mathcal{B}$ using anyon condensation. 
This approach has been fruitful in classifying gapped interfaces of 2D Abelian phases~\cite{SantosHughes-2017,Santos2018} as well as the case where the bulk topological order
is non-Abelian \cite{Bais2009a,Lou2019}.

Suppose $\mathcal{A}$ and $\mathcal{B}$ share a common parent phase $\mathcal{C}$ -- that is to say, a phase in which condensing one set of anyons yields $\mathcal{A}$ and condensing a different set of anyons yields phase $\mathcal{B}$. Then, one can form an interface between $\mathcal{A}$ and $\mathcal{B}$ by starting with $\mathcal{C}$, condensing down to $\mathcal{A}$ in one region, and then condensing down to $\mathcal{B}$ in another region, yielding a configuration which is gapped everywhere as follows:
\begin{align}
\left. (\mathcal{A}) \quad \right| \quad (\mathcal{C}) \quad \left| \quad (\mathcal{B}) \quad  \right. .
\end{align}
Shrinking the region containing $\mathcal{C}$ yields a gapped interface between $\mathcal{A}$ and $\mathcal{B}$. Similarly, a gapped interface can be formed if $\mathcal{C}$ is a daughter phase of $\mathcal{A}$ and $\mathcal{B}$ -- that is, $\mathcal{A}$ and $\mathcal{B}$ can be condensed to obtain $\mathcal{C}$.

The intermediate state $\mathcal{C}$ can be thought of as originating from $\mathcal{A}$ or $\mathcal{B}$ by gauging of an appropriate discrete symmetry, insofar as anyon condensation can be viewed as the inverse operation of gauging an anyonic symmetry \cite{Teo2015,Barkeshli2019} (related observations of the connection between boundary physics and bulk physics have been made in Ref. \cite{Lichtman2020}). Consequently, the local interactions that gap the interface manifest this symmetry, which, in the Abelian case can be precisely shown to contribute to a correction to the TEE \cite{Santos2018}.
Furthermore, in Ref. \cite{Lou2019}, it was argued that the choice of $\mathcal{C}$ determines the ground state of the interface to be a particular Ishibashi state, from which the interface TEE was calculated to be $\ln \mathcal{D}_{\mathcal{C}}$, where $\mathcal{D}_{\mathcal{C}}$ is the total quantum dimension of $\mathcal{C}$. In this subsection, after first reviewing this construction for interfaces of Laughlin states, we identify the appropriate parent phases for the two classes of MR interfaces identified above, as determined by the choice of gapping interaction, and verify this relation with the TEE.

\subsubsection{Review of Laughlin Interfaces}

Let us again consider an interface between Laughlin $\nu^{-1} = pb^2$ and $\nu^{-1} = pa^2$ states, where $a$ and $b$ are co-prime \cite{SantosHughes-2017}. In this case the parent topological phase is a Laughlin state at inverse filling $\nu^{-1}=pa^2b^2$:
\begin{align}
\left. (\mathcal{A}) \quad \nu = \frac{1}{pb^2} \quad \right| \quad (\mathcal{C}) \quad \nu = \frac{1}{pa^2b^2} \quad \left| \quad (\mathcal{B}) \quad \nu = \frac{1}{pa^2}, \right.
\end{align}
The state $\mathcal{C}$ originates from $\mathcal{A}$ and $\mathcal{B}$ by gauging discrete $\mathbb{Z}_{a}$ and $\mathbb{Z}_{b}$ symmetries, respectively. As such, the local gapping interaction of the $\mathcal{A}-\mathcal{C}$ interface displays a discrete $\mathbb{Z}_a$ symmetry associated with the pairing of $a$ local quasiparticles of $\mathcal{A}$ with one local quasiparticle of $\mathcal{A}$. Similarly, on the $\mathcal{B}-\mathcal{C}$ interface, the local interaction displays a $\mathbb{Z}_{b}$ symmetry. Consequently, as the phase ``thins out," one is left with the $\mathcal{A}-\mathcal{B}$ interface
where $a$ local quasiparticles of $\mathcal{A}$ bind to $b$ local quasiparticles of $\mathcal{B}$.

Now, the anyon content of $\mathcal{C}$ is given by
\begin{align}
\mathcal{C} = \{ e^{ir\phi} \}_{r=1,\dots , pa^2b^2}.
\end{align}
These anyons have spin
\begin{align}
h_r = \frac{1}{2} \frac{r^2}{pa^2 b^2}.
\end{align}
Consider the anyon labelled by $r_0 = pa^2 b$. It has the same spin, $h_{r_0} = \frac{1}{2} pa^2$, as the electron operator in the $\nu^{-1} = pa^2$ Laughlin state. The mutual statistics between $r_0$ and all other anyons is given by
\begin{align}
\theta_{r_0,r} = \exp\left(2\pi i \frac{r}{b} \right).
\end{align}
So, if we condense $r_0$, only anyons of the form $r = bl$ will remain deconfined. These remaining anyons have mutual statistics
\begin{align}
\theta_{l,l'} = \exp\left(2\pi i \frac{ll'}{pa^2} \right).
\end{align}
This precisely describes the topological order of $\mathcal{B}$. It is easy to see that condensing $r=pab^2$ would instead give $\mathcal{A}$. Thus, $\mathcal{C}$ is indeed the parent state of $\mathcal{A}$ and $\mathcal{B}$.

Now, the total quantum dimension of a Laughlin $\nu^{-1} = pa^2b^2$ state is $\mathcal{D} = \sqrt{pa^2 b^2}$, which agrees with the value of the TEE for an entanglement cut lying along the physical interface, $\gamma = \ln \sqrt{pa^2 b^2}$, as computed in Ref. \cite{Cano2015}.

\subsubsection{Extension to Moore-Read Interfaces}

Let us now consider the interface between $\nu^{-1} = pb^2$ and $\nu^{-1} = pa^2$ MR states, with $a$ and $b$ both odd. We calculated the TEE in this scenario to be given by $\gamma = \ln (2\sqrt{pa^2 b^2})$. This is precisely the TEE for a uniform $\nu^{-1} = pa^2 b^2$ MR state. We thus claim that the parent phase for the $\nu^{-1} = pb^2$ and $\nu^{-1} = pa^2$ MR states, with $a$ and $b$ both odd, is the $\nu^{-1} = pa^2 b^2$ MR state:
\begin{align}
\left. (\mathcal{A}) \quad \mathrm{MR}_{pb^2} \quad \right| \quad (\mathcal{C}) \quad\mathrm{MR}_{pa^2b^2} \quad \left| \quad (\mathcal{B}) \quad \mathrm{MR}_{pa^2}, \right.
\end{align}
where we have introduced the shorthand $\mathrm{MR}_{\nu^{-1}}$ to denote the MR state at filling $\nu$. Now, $\mathcal{C}$ has the anyon content
\begin{align}
\mathcal{C} = \{ e^{ir\phi}, \chi e^{ir\phi}, \sigma e^{i(r+1/2)\phi} \}_{r=1,\dots , pa^2 b^2}.
\end{align}
In order to obtain, say, phase $\mathcal{B}$, we must condense an anyon of the form $\chi e^{ir\phi}$, since this will serve as the new electron operator and we wish to obtain another MR state. From the discussion of the Laughlin interface, it is straightforward to see that condensing $\psi_{\mathcal{B}} = \chi e^{i pa^2 b \phi}$ will yield the correct Laughlin quasiparticle content, as well as Majorana content (since $\chi$ has trivial braiding with itself, under a full $2\pi$ rotation). 

As for the non-Abelian anyons, $\sigma e^{i(r+1/2)\phi}$, their braiding with $\psi_{\mathcal{B}}$ is given by,
\begin{align}
\theta_{\sigma e^{i(r+1/2)\phi},\psi_{\mathcal{B}}} = \exp\left[2\pi i \left(\frac{b + 2r + 1}{2b} \right) \right].
\end{align}
In order for this phase to be trivial, we require
\begin{align}
2r+1 = b(2m+1), \quad m \in \mathbb{Z}.
\end{align}
Both the LHS and $2m+1$ are odd, and so a solution exists if and only if \emph{$b$ is also odd}. If this is the case, we find that the non-Abelian anyons parameterized as
\begin{align}
r = b(m+1/2) - 1/2 \implies \sigma e^{i(r+1/2)\phi} = \sigma e^{ib(m+1/2)\phi}
\end{align}
remain deconfined. These anyons have spin
\begin{align}
h_r = \frac{1}{16} + \frac{1}{2} \frac{(m+1/2)^2}{pa^2},
\end{align}
which are precisely the spins of the non-Abelian anyons in the $\mathrm{MR}_{pb^2}$ state. We can also compute the braiding of these anyons and the deconfined Abelian anyons, $e^{ilb\phi}$, to be
\begin{align}
\theta_{\sigma e^{i(r+1/2)\phi},e^{ilb\phi}} = \exp\left[2\pi i \frac{l(m+1/2)}{pa^2} \right] .
\end{align}
This is the expected phase for braiding of the corresponding anyons in the $\mathrm{MR}_{pb^2}$ state. It is straightforward to see that the correct braiding statistics between the remaining non-Abelian anyons and Majoranas will also be obtained. We thus conclude that by condensing $\chi e^{i pa^2 b \phi}$ in phase $\mathcal{C}$, we obtain phase $\mathcal{B}$. \emph{Provided $a$ is odd}, it follows immediately that condensing $\chi e^{i pa b^2 \phi}$ in phase $\mathcal{C}$ will yield phase $\mathcal{A}$. We thus conclude that if both $a$ and $b$ are odd, we can obtain a GI between $\mathrm{MR}_{pb^2}$ and $\mathrm{MR}_{pa^2}$ states which is characterized by an intervening $\mathrm{MR}_{pa^2b^2}$ state, consistent with the fact that the TEE for this interface is $\gamma = \ln (2\sqrt{pa^2 b^2})$. 

Let us now consider the case where one of $a$ and $b$, say $a$, is even and the other odd. Our claim is that the parent phase in this case is given not by a MR state, but by an $\mathrm{Ising}\times U(1)_{4pa^2b^2}$ theory:
\begin{align}
\left. (\mathcal{A}) \,\, \mathrm{MR}_{pb^2} \quad \right| \quad (\mathcal{C}) \,\, \mathrm{Ising}\times U(1)_{4pa^2b^2}\quad \left| \quad (\mathcal{B}) \,\, \mathrm{MR}_{pa^2} \right.
\end{align}
The anyon content of $\mathcal{C}$ is given by
\begin{align}
\mathcal{C} = \{1, \chi, \sigma \} \times \{ e^{il\phi} \}_{l=1,\dots, 4pa^2 b^2} .
\end{align}
It is readily seen that $\mathcal{C}$ has the correct total quantum dimension,
$\mathcal{D}_{\mathcal{C}} =4\sqrt{pa^2b^2}$, given that the TEE for this interface is given by $\gamma = \ln (4\sqrt{pa^2b^2})$.

Suppose we condense
\begin{align}
\psi_{\mathcal{B}} = \chi e^{i2pa^2b \phi}.
\end{align}
This quasiparticle has spin
\begin{align}
h_{\mathcal{B}} = \frac{1}{2} + \frac{1}{2} \frac{4 p^2 a^4 b^2}{4pa^2 b^2} = \frac{1}{2} + \frac{1}{2} pa^2 ,
\end{align}
which matches that of the electron operator in phase $\mathcal{B}$. Now, the braiding of a Laughlin quasiparticle $e^{il\phi}$ with $\psi_{\mathcal{B}}$ is given by
\begin{align}
\theta_{e^{il\phi},\psi_\mathcal{B}} = \exp \left[ 2\pi i\frac{(2pa^2 b) l}{4pa^2 b} \right] = \exp \left[ 2\pi i \frac{l}{2b} \right] ,
\end{align}
which is trivial when $l = 2b m$, $m \in \mathbb{Z}$. So, all Laughlin quasiparticles except those of the form $e^{i2bm \phi}$ are confined. The remaining Laughlin quasiparticles have mutual statistics
\begin{align}
\theta_{e^{im\phi},e^{im'\phi}} = \exp \left[ 2\pi i \frac{(2b m)(2 b m')}{4pa^2 b}\right] = \exp\left[ 2\pi i \frac{m m '}{pa^2}\right],
\end{align}
which are precisely the mutual statistics of the Laughlin anyons in phase $\mathcal{B}$. It immediately follows that anyons of the form $\chi e^{i2bm \phi}$ are also deconfined and reproduce the Majorana sectors of phase $\mathcal{B}$. The braiding statistics of the non-Abelian anyons, $\sigma e^{it\phi}$ with $\psi_{\mathcal{B}}$ is given by
\begin{align}
\frac{1}{2\pi} \theta_{\sigma e^{it\phi},\psi_\mathcal{B}} = \frac{1}{2} + \frac{(2pa^2b)t}{4pa^2 b^2} = \frac{1}{2} + \frac{t}{2b} = \frac{b + t}{2b} .
\end{align}
The deconfined non-Abelian anyons thus satisfy
\begin{align}
b + t = 2b (r+1) \implies t = b(2r + 1)
\end{align}
with $r \in \mathbb{Z}$. These deconfined anyons have spin
\begin{align}
h_r = \frac{1}{16} + \frac{b^2(2r+1)^2}{4pa^2b^2} = \frac{1}{16} + \frac{(r+1/2)^2}{pa^2},
\end{align}
which matches that of the non-Abelian anyons in phase $\mathcal{B}$. Hence, condensing $\psi_{\mathcal{B}}$ in $\mathcal{C}$ correctly reproduces phase $\mathcal{B}$. It follows, of course, that by instead condensing $\psi_{\mathcal{A}} = \chi e^{i2pab^2 \phi}$, we would have obtained phase $\mathcal{A}$.  Hence, $\mathcal{C} = \mathrm{Ising}\times U(1)_{4pa^2b^2}$ appears to be the correct intermediate phase to describe the $a$ even, $b$ odd interface. Note that, however, at no point was it necessary to impose that one of $a$ and $b$ was even and the other odd; indeed, both could have been odd as well. This is consistent with the fact that the $a,b$ odd interface could, in principle, also be gapped using the tunneling terms of Eq. (\ref{eqn:charge-neutral-gapping-term}).

\section{Discussion and Conclusion \label{sec:discussion}}

In this paper we extended the cut-and-glue approach to calculating entanglement entropy of two-dimensional topologically ordered phases to interfaces of the simplest non-Abelian fractional quantum Hall states, namely the generalized Moore-Read states. By carefully taking into account the Hilbert space structure of the MR CFT, as reviewed in Section \ref{sec:mr-review}, we first demonstrated, in Section \ref{sec:uniform}, that we can reproduce the entanglement spectrum and hence the topological entanglement entropy for each of the topological sectors of the MR state on a torus. In Section \ref{sec:nonuniform-gapping-terms}, we investigated interfaces of distinct generalized MR states, identifying when and how they can be gapped out. In particular, we looked at interfaces of MR states at inverse fillings $\nu^{-1} = pa^2$ and $\nu^{-1} = pb^2$, with $a$ and $b$ coprime, finding that they can always be gapped, but also that the form of the gapping interaction depends on whether $a$ and $b$ are both odd or if one is even. We then found that this distinction manifests itself in the TEE when the entanglement cut is placed along the interface. Indeed, we found in Section \ref{sec:non-uniform-EE} that, in the trivial sector, the TEE is given by $\gamma_{\mathbf{1}} = 2\ln (2\sqrt{pa^2b^2})$ when $a$ and $b$ are both odd and by $\gamma_{\mathbf{1}} = 2\ln (4\sqrt{pa^2b^2})$ when one of $a$ and $b$ is even. Finally, we demonstrated how this value of the TEE is connected to the existence of a parent topological phase from which both the $\nu^{-1} = pa^2$ and $\nu^{-1} = pb^2$ MR states descend.

Although we focused on the generalized MR states, in principle, the cut-and-glue approach could, in principle, be extended to other non-Abelian topological orders whose edge CFTs possess a free-field representation. Following our prescription, one can approximate a gapping term to quadratic order and then project the resulting ground state to the appropriate topological sector of the physical Hilbert space. It should be possible, for instance, to repeat our calculation for states in the Bonderson-Slingerland hierarchy \cite{Bonderson2008} and for the orbifold FQH states of Barkeshli and Wen \cite{Barkeshli2011}. It would also be interesting to see whether our methodology could be used to investigate interfaces of Abelian and non-Abelian states.

Aside from calculations of the entanglement entropy in other systems, another open question is the extent to which the anyon condensation picture of gapped interfaces of non-Abelian states is connected to the existence of explicit gapping interactions for such interfaces. In the examples we considered, we found that there did indeed appear to be a close correspondence between the two. For an interface of MR states at inverse fillings $\nu^{-1} = pb^2$ and $\nu^{-1} = pa^2$ with $a$ and $b$ both odd, we were able to write down a gapping term which simply corresponded to a local operator constructed by fusing together elements of the set of anyons to be condensed. In contrast, when one of $a$ and $b$ was even, we found it useful to resort to a topologically equivalent description of the MR edge theory to be able to write down an interaction which fully gapped the interface. Nevertheless, this interaction was still closely connected to the set of condensed anyons characterizing the interface. Now, for interfaces of \emph{Abelian} states, it is known that there is a one-to-one correspondence between Lagrangian subgroups and gapping interactions, provided one allows for the introduction of additional topologically trivial edge states (physically, this corresponds to edge reconstruction). Two Abelian theories differing from one another only by the addition of such trivial edge states are said to be \emph{stably equivalent} theories \cite{Levin2013,Barkeshli2013}. At a superficial level, our construction mirrors this notion of stable equivalence, in that we write down a theory with the same topological content, but with additional degrees of freedom. However, the additional fields which are added in our case are \emph{not} local, in contrast to the Abelian case. It is not clear how general this coset construction of topologically equivalent CFTs is, but it could perhaps be used as a basis to write down general gapping interactions for interfaces of arbitrary non-Abelian orders -- or at least those with free field representations. Such a scheme could potentially be used to derive the different sets of tunneling interactions that can be used to gap out an interface between two given non-Abelian topological orders.

Lastly, as noted in the introduction, gapped interfaces of Abelian topological phases have attracted much interest in recent years, due to the possibility of realizing non-Abelian defects at terminations of said interfaces \cite{BarkeshliQi-2012,Lindner-2012,Clarke-2013,Cheng-2012,Vaezi-2013,BarkeshliJianQi-2013-a,BarkeshliJianQi-2013-b,Mong-2014,khanteohughes-2014,SantosHughes-2017,santos2019}. In fact, as also noted in the introduction, the value of the TEE of an entanglement cut along an interface between Abelian topological phases has been connected to the emergence of a one-dimensional symmetry protected topological phase (SPT) along the interface \cite{Santos2018}. The endpoints of these SPTs support parafermions, in contrast to purely one-dimensional SPTs which can only host Majorana zero modes. It would be interesting to see whether an analogous statement holds for interfaces of generalized MR states and if one can obtain bound states more exotic than parafermions.

After the initial posting of this work, Ref. \cite{Fliss2020} appeared, which examines gapped interfaces between distinct non-Abelian Chern-Simons theories.

\begin{acknowledgments}

We thank J. Cano, C. Chamon, T. Iadecola, R. Leigh, T. Zhou, and especially E. Fradkin for helpful discussions. R.S. acknowledges the support of the Natural Sciences and Engineering Research Council
of Canada (NSERC) [funding reference number 6799-516762-2018]. RS was also supported in part by the US National Science Foundation under grant No. DMR-1725401 at the University of Illinois. BH was partially supported by the ERC Starting Grant No.~678795 TopInSy. LHS is supported by a faculty startup at Emory University. JCYT is supported by the National Science Foundation under Grant No. DMR-1653535.

\end{acknowledgments}

\appendix

\section{Modular Functions \label{sec:appendix-modular-functions}}

In this appendix we collect the definitions and basic properties of the $\theta$ and $\eta$ functions. First, we introduce the notation
\begin{align}
    q = e^{2\pi i \tau},
\end{align}
where $\tau \in \mathbb{C}$ is the modular parameter. The Dedekind $\eta$ function is defined as
\begin{align}
    \eta(\tau) = q^{1/24} \prod_{n=1}^\infty (1- q^n).
\end{align}
Under modular transformations, the $\eta$ function satisfies
\begin{align}
    \eta(\tau + 1) &= e^{\pi i /12} \eta(\tau),
    \\ \quad \eta(-1/\tau) &= \sqrt{-i\tau}\eta(\tau). \label{eqn:eta-S-transformation}
\end{align}
We also make use of the $\theta$ functions,
\begin{align}
    \theta^\alpha_\beta (\tau) = \sum_{n \in \mathbb{Z}} q^{\frac{1}{2}(n+\alpha)^2} e^{2\pi i (n+\alpha)\beta}.
\end{align}
Under modular transformations, these functions satisfy
\begin{align}
    \theta^\alpha_\beta (\tau + 1) &= e^{-\pi i \alpha (\alpha-1)} \theta^\alpha_{\alpha+\beta-\frac{1}{2}}(\tau), \\
    \theta^\alpha_\beta(-1/\tau) &= \sqrt{-i\tau}e^{2\pi i \alpha \beta} \theta^\beta_{-\alpha} (\tau). \label{eqn:theta-S-transformation}
\end{align}
The standard Jacobi $\theta$ functions (see, for instance, Ref. \cite{DiFrancescoBook}) can be expressed in terms of these more general functions:
\begin{align}
    \theta_2(\tau) &= \sum_{n \in \mathbb{Z}} q^{(n+1/2)^2 / 2} = \theta^{1/2}_0 (\tau) , \\
    \theta_3(\tau) &= \sum_{n \in \mathbb{Z}} q^{n^2 / 2} = \theta^{0}_0 (\tau) , \\
    \theta_4(\tau) &= \sum_{n \in \mathbb{Z}} (-1)^n q^{n^2 / 2} = \theta^{0}_{1/2} (\tau) . 
\end{align}
Lastly, we note that for $\tau = i \tau_2$, with $\tau_2 \in \mathbb{R}^+$, we have that
\begin{align}
    \lim_{\tau_2 \to \infty } \eta(\tau) &= q^{1/24} \label{eqn:eta-asymptotic} \, . \\
    \lim_{\tau_2 \to \infty }\theta^\alpha_\beta (\tau) &= \delta_{\alpha,0} \, . \label{eqn:theta-asymptotic}
\end{align}

\section{Details of Projected Ground States}

\subsection{Untwisted Sectors \label{sec:appendix-untwisted-GS}}

For completeness, we can write down the explicit form of the ground state in, say, the $e^{ir\phi}$ sector:
\begin{align}
\ket{\psi_{e^{ir\phi}}} = P_{e^{ir\phi},B}\ket{\hat{\psi}_{e^{ir\phi}}} = \ket{\psi_{e^{ir\phi},1}} \otimes \ket{\psi_{e^{ir\phi},2}},
\end{align}
where we made use of Eq. (\ref{eqn:untwisted-P=PA=PB}). Following Section \ref{sec:projection}, we further rewrite $ P_{e^{ir\phi},B} = P_{e^{ir\phi},RB} P_{e^{ir\phi},LB}$ so that we can express the exact interface ground states as
\begin{align}
    \begin{split}
    \ket{\psi_{e^{ir\phi},1}} &= P_{e^{ir\phi},RB} \ket{\hat{\psi}_{e^{ir\phi},1}} \\
    &= \frac{1}{2}(1 + (-1)^{F_{RB}}(-1)^{N_{RB}+r/n}) \ket{\hat{\psi}_{e^{ir\phi},1}},
    \end{split} \\
    \begin{split}
    \ket{\psi_{e^{ir\phi},2}} &= P_{e^{ir\phi},LB} \ket{\hat{\psi}_{e^{ir\phi},2}} \\
    &= \frac{1}{2}(1 + (-1)^{F_{LB}}(-1)^{N_{LB}-r/n}) \ket{\hat{\psi}_{e^{ir\phi},2}}.
    \end{split}
\end{align}
As is evident from the above expression, the effect of the projection on, say, $\ket{\hat{\psi}_{e^{ir\phi},1}}$, is to annihilate all states not satisfying $(-1)^{F_{RB}}(-1)^{N_{RB}+r/n} =1$. As discussed in the main text, the form of $\ket{\hat{\psi}_{e^{ir\phi},1}}$ is such that the remainig states will also satisfy $(-1)^{F_{LA}}(-1)^{N_{LA}-r/n} =1$. Analogous statements hold for the action of the projection on $\ket{\hat{\psi}_{e^{ir\phi},2}}$. Note that these expressions for the projections require that $\ket{\hat{\psi}_{e^{ir\phi},1/2}}$ already obey the correct quantization of the winding numbers, $N_{\mu B}$ for sector $e^{ir\phi}$.  Explicitly, we can write
\begin{widetext}
\begin{align}
\begin{split}
 \ket{\psi_{e^{ir\phi},1}} &= \frac{1}{2} \left[\ket{G_{b,\mathrm{osc},1}} \otimes  \sum_{N \in \mathbb{Z} - \frac{r}{n}} e^{-\frac{v_e \pi n N^2}{2L}} \ket{N_{RB}=N}\ket{N_{LA}=-N} \otimes \prod_{k>0} \left( 1 + i e^{-\tilde{v}_e k / 2} d_{-k}^\dagger c_k^\dagger \right) \ket{0} \right. \\
 	&\quad  \left. + \ket{G_{b,\mathrm{osc},1}} \otimes  \sum_{N \in \mathbb{Z} - \frac{r}{n}} (-1)^{N+\frac{r}{n}} e^{-\frac{v_e \pi n N^2}{2L}} \ket{N_{RB}=N}\ket{N_{LA}=-N} \otimes \prod_{k>0} \left( 1 - i e^{-\tilde{v}_e k / 2} d_{-k}^\dagger c_k^\dagger \right) \ket{0} \right],
\end{split} \label{eqn:psi-eirphi-1-state}
\end{align}
with $\ket{\psi_{e^{ir\phi},2}}$ taking a similar form. Focusing on the explicit expression for $\ket{\psi_{e^{ir\phi},1}}$, we see that every state appearing in the second line of Eq. (\ref{eqn:psi-eirphi-1-state}) with $(-1)^{F_{RB}}(-1)^{N_{RB}+r/n} = -1$ will indeed cancel with a corresponding state in the first line. 

\subsection{Twisted Sectors \label{sec:appendix-twisted-GS}}
\subsubsection{Ground State}
We present here the explicit form of the approximated ground state in the $a = \sigma e^{i(r+1/2)\phi}$ sector:
\begin{align}
    \ket{\psi_a} = P_B\ket{\hat{\psi}_a} = \frac{1}{2}(1 + (-1)^{F_B}(-1)^{N_{RB}+N_{LB}}) \ket{\hat{\psi}_a},
\end{align}
where we used Eq. (\ref{eqn:twisted-P=PA=PB}) to write $P\ket{\hat{\psi}_a} = P_B\ket{\hat{\psi}_a}$. As in the untwisted sector case of the previous subsection, we made use of the fact that all the states appearing in $\ket{\hat{\psi}_a}$ obey the correct quantization of the winding modes, $N_{\mu\alpha}$, appropriate to the $a= \sigma e^{i(r+1/2)\phi}$ sector to write down a closed form expression for $P_B\ket{\hat{\psi}_a}$. The effect of the projection is to annihilate all states not satisfying $(-1)^{F_B}(-1)^{N_{RB}+N_{LB}} = 1$. Again, as discussed in the main text, the remaining states will also automatically satisfy $(-1)^{F_A}(-1)^{N_{RA}+N_{LA}} = 1$. Explicitly evaluating the above expression for $\ket{\psi_a}$, we can write
\begin{align}
\begin{split}
\ket{\psi_a} 	&= \frac{1}{2}\left[\ket{G_{b,\mathrm{osc},1}} \otimes \sum_{N \in   \mathbb{Z}-\frac{r+1/2}{n}} e^{-\frac{v_e \pi n N^2}{2L}} \ket{N_{RB}=N}\ket{N_{LA}=-N} \otimes \prod_{k>0} \left( 1 + i e^{-\tilde{v}_e k / 2} d_{-k}^\dagger c_k^\dagger \right) \ket{0} \right] \\
				&\,\otimes \left[\ket{G_{b,\mathrm{osc},2}} \otimes \sum_{N \in \mathbb{Z}-\frac{r+1/2}{n}} e^{-\frac{v_e \pi n N^2}{2L}} \ket{N_{LB}=-N}\ket{N_{RA}=N} \otimes \prod_{k>0} \left( 1 + i e^{-\tilde{v}_e k / 2} \tilde{c}_{-k}^\dagger \tilde{d}_k^\dagger \right) \ket{0} \right] \\
				&\,\otimes \frac{1}{\sqrt{2}}(\ket{0_A,0_B} + i \ket{1_A,1_B} ) \\
				&+ \frac{1}{2}\left[\ket{G_{b,\mathrm{osc},1}} \otimes \sum_{N \in \mathbb{Z}-\frac{r+1/2}{n}} (-1)^{N+\frac{r+1/2}{n}} e^{-\frac{v_e \pi n N^2}{2L}} \ket{N_{RB}=N}\ket{N_{LA}=-N} \otimes \prod_{k>0} \left( 1 - i e^{-\tilde{v}_e k / 2} d_{-k}^\dagger c_k^\dagger \right) \ket{0} \right] \\
				&\,\otimes \left[\ket{G_{b,\mathrm{osc},2}} \otimes \sum_{N \in \mathbb{Z}-\frac{r+1/2}{n}} (-1)^{-N-\frac{r+1/2}{n}} e^{-\frac{v_e \pi n N^2}{2L}} \ket{N_{LB}=-N}\ket{N_{RA}=N} \otimes \prod_{k>0} \left( 1 - i e^{-\tilde{v}_e k / 2} \tilde{c}_{-k}^\dagger \tilde{d}_k^\dagger \right) \ket{0} \right] \\
				&\,\otimes \frac{1}{\sqrt{2}}(\ket{0_A,0_B} - i \ket{1_A,1_B} ).
\end{split}
\end{align}
\end{widetext}
Although this is a rather cumbersome expression, we can parse its meaning as follows. The first three lines are simply a reexpression of $\ket{\hat{\psi}_a}$. The last three lines correspond to the state obtained by acting on $\ket{\hat{\psi}_a}$ with $(-1)^{F_B}(-1)^{N_{RB}+N_{LB}}$. Every state appearing in the last three lines for which $(-1)^{F_B}(-1)^{N_{RB}+N_{LB}} = -1$ will thus cancel with a state in the first three lines, leaving only states with $(-1)^{F_B}(-1)^{N_{RB}+N_{LB}} = 1$, as desired.

\subsubsection{Zero Mode Fermion Parity}
Now, as alluded to in the main text, there is a subtlety regarding how to interpret the fermion parity of the zero mode. We constructed the fermion $f_A$ from the MZMs $d_0$ and $\tilde{d}_0$ as $f_A = (d_0 + i\tilde{d}_0)/\sqrt{2}$. However, we can also define $f'_A = (\tilde{d}_0 + id_0)/\sqrt{2}$, so that 
\begin{align}
\ket{0,\tilde{0}} &= \frac{1}{\sqrt{2}}(\ket{0_A,0_B} + i \ket{1_A,1_B} ) \\
                    &= \frac{1}{\sqrt{2}}(\ket{1_A',0_B} - \ket{0_A',1_B} ) .
\end{align}
In other words, $f_A$ being occupied is equivalent to saying that $f_A'$ is unoccupied and vice versa. The point at issue is that the $\mathbb{Z}_2$ symmetry operators, $G_\alpha$ [see Eq. (\ref{eqn:g_alpha})], are defined in terms of the total fermion parities, $(-1)^{F_\alpha}$, and one must decide whether this parity is measured relative to the occupation of $f_A$ or $f_A'$. Indeed, if we measured it with respect to $f_A'$, one would find that $P_a\ket{\hat{\psi}_a} = 0$ since, for each state appearing in $\ket{\hat{\psi}_a}$, the total fermion parities of $A$ and $B$ would be opposite to one another.

In order to remove this ambiguity in the definition of the fermion parity, it is necessary to resort to physical arguments which can provide additional input, which we now provide. Before physically cutting the torus into the cylinders $A$ and $B$, the torus starts in the ground state with a $\sigma e^{i(r+1/2)\phi}$ Wilson loop wrapping around the $y$-cycle (i.e. the cycle perpendicular to the entanglement cut). On performing the physical cut of the torus into two cylinders, the Wilson loop is cut into two Wilon lines with endpoints at the edges of the cylinders. Physically, this configuration corresponds to having a $\sigma e^{i(r+1/2)\phi}$ anyon on one end of each cylinder and the corrsponding conjugate anyon on the other end of each cylinder.

Let us label these anyons as $\sigma_{A,r} \equiv \sigma_A e^{i(r+1/2)\phi_A}$, $\bar{\sigma}_{A,r} \equiv \bar{\sigma}_A e^{i(r+1/2)\bar{\phi}_A}$, $\sigma_{B,r} \equiv \sigma_B e^{i(r+1/2)\phi_B}$, and $\bar{\sigma}_{B,r} \equiv \bar{\sigma}_B e^{i(r+1/2)\bar{\phi}_B}$. We claim that the pairs of anyons at each interface must fuse to the identity, and not to a neutral Majorana: 
\begin{align}
\sigma_{A,r} \times \bar{\sigma}_{B,r} = \sigma_{B,r} \times \bar{\sigma}_{A,r} = 1.
\end{align}
Physically, we can think of the electron tunneling terms which glue the edges together as hybridizations of $\sigma_{A,r}$ with $\bar{\sigma}_{B,r}$ and $\sigma_{B,r}$ with $\bar{\sigma}_{A,r}$. This would make it energetically preferable for each pair of these anyons to fuse into the identity, as opposed to a Majorana fermion. In particular, when we expanded the tunneling term about one of its minima, we did so assuming that this corresponded to the ground state, which one should interpret as the vacuum. 

Now, we wish to identify what the allowed fusion possibilities for $\sigma_A \times \bar{\sigma}_A$ and $\sigma_B \times \bar{\sigma}_B$ should be. From the previous paragraph, we see that fusing all four of the twist anyons should yield the vacuum. This requires that either $\sigma_A \times \bar{\sigma}_A = \sigma_B \times \bar{\sigma}_B = 1$ \emph{or} $\sigma_A \times \bar{\sigma}_A = \sigma_B \times \bar{\sigma}_B = \chi$. This suggests that we should define the complex fermions $f_{1,2}$ on cylinders $A$ and $B$ to be such that we can express the ground state as 
\begin{align}
\ket{0,\tilde{0}} &= \frac{1}{\sqrt{2}}(\ket{0_1,0_2} + \alpha \ket{1_1,1_2}),
\end{align}
where $\alpha$ is some unimportant phase. Hence, our choice of measuring the fermion parity relative to $f_A$ and $f_B$ is consistent with this physical picture. We note that this line of reasoning is similar to more carefully constructed arguments for determining the ground state degeneracy of the Moore-Read state on the torus -- see, for instance, Refs. \cite{Oshikawa2007,Iadecola2019}.

\section{Alternative Representation of the Ising CFT \label{sec:appendix-ising}}

In this section, we consider the edge theory of the Ising topological order which, conventionally, is described by the Ising CFT. We first write down a CFT description of the edge which is topologically equivalent to the Ising CFT, in a sense to be made more precise shortly. We then show how we can write down an explicit gapping interaction for the interface of two Ising edges using this alternative CFT description, which does not appear possible (at least based on a superficial analysis) in the standard Ising CFT description of the edge.

\subsection{Coset Construction and Hilbert Space Structure}

In the usual free-field representation, the Ising edge theory contains a single chiral Majorana:
\begin{align}
    \cL' = \psi \frac{i}{2} (\partial_t - \partial_x) \psi .
\end{align}
The three topological sectors in the theory are $1$, $\sigma$, and $\psi$ -- the vacuum, twist operator, and Majorana sectors, respectively. Gapping an interface between two Ising topological orders thus appears difficult, as any local tunneling operator would have to involve terms quadratic in both the left- and right-moving Majoranas, which na\"ively would square to unity.

We instead make use of the coset representation,
\begin{align}
\mathrm{Ising} = \frac{SO(N+1)_1}{SO(N)} \sim SO(N+1)_1 \boxtimes \overline{SO(N)}_1.
\end{align}
Here we take $N=2r$, $1 < r \in \mathbb{Z}$. On the left hand side of the equivalence, we have a theory of $N+1$ chiral Majoranas in which we gap out $N$ of them. On the right hand side, we have a theory of $N+1$ chiral Majoranas and $N$ anti-chiral Majoranas in which we have condensed a certain set of bosonic anyons so as to identify certain topological sectors. Now, the Ising CFT is identical to the coset $SO(N+1)_1/SO(N)$, in that they have the same primary operator content as well as total and chiral central charges. In contrast,we will say the Ising CFT and the $SO(N+1)_1 \boxtimes \overline{SO(N)}_1$ CFT are \emph{topologically} equivalent, in that they possess the same primary operator content (i.e. topological sectors) and chiral central charge, but not the same total central charge \cite{Bais2009,Moore1989}.

Let us now outline in detail the structure of the $SO(N+1)_1 \boxtimes \overline{SO(N)}_1$ theory. Placing the Ising topological order on a cylinder, as in Fig. \ref{fig:cylinder}, the $\mu=L,R=+,-$ edges are described by the Lagrangians
\begin{align}
    \cL_\mu = \sum_{\alpha=0}^N \psi^\alpha_\mu \frac{i}{2} (\partial_t - \mu \partial_x) \psi^\alpha_\mu + \sum_{a=1}^N \bar{\psi}^a_\mu \frac{i}{2} (\partial_t + \mu \partial_x) \bar{\psi}^a_\mu. \label{eqn:ising-coset-fermionic-Lagrangian}
\end{align}
So, on edge $L$ ($R$), there are $N+1$ chiral (anti-chiral) Majoranas, $\psi_\mu^\alpha$ and $N$ anti-chiral (chiral) Majoranas, $\bar{\psi}^a_\mu$. For simplicity, we have set all velocities to unity. Additionally, we adopt the convention that Greek indices $\alpha,\beta$ run from $0$ to $N$ and the Latin indices $a,b$ from $1$ to $N$. This theory possesses the currents
\begin{align}
J^{\alpha\beta}_\mu = i \psi^\alpha_\mu \psi^\beta_\mu, \quad \overline{J}^{ab}_\mu = i \overline{\psi}^a_\mu \overline{\psi}^b_\mu, \label{eqn:ising-coset-currents}
\end{align}
which generate the $SO(N+1)_1$ and $\overline{SO(N)}_1$ Kac-Moody algebras of the two edges, respectively. Additionally, the operators
\begin{align}
    \quad M^{\alpha a}_\mu = i \psi^\alpha_\mu \overline{\psi}^a_\mu  \label{eqn:ising-coset-local-electronic-operators}
\end{align}
correspond to the condensed bosons encoded in the tensor product, $\boxtimes$, and hence, like the currents, are local-electronic objects. 
Using these expressions, we can see that this theory is in fact topologically equivalent to the Ising CFT. For instance, starting with one Majorana fermion, say, $\psi^\alpha$, we can obtain any other Majorana $\psi^\beta$ or $\bar{\psi}^a$ by fusing it with $J^{\alpha\beta}$ or $M^{\alpha a}$. Hence, there is only \emph{one} distinct Majorana fermion sector, as in the Ising theory. 

Although not strictly necessary, it will prove convenient for our purposes to bosonize as many of the fermions as possible. Since we have taken $N=2r$ to be even, we can pair up all the Majoranas in the $\overline{SO(N)}_1$ factor into Dirac fermions and bosonize them: 
\begin{align}
\overline{c}^j_\mu = (\overline{\psi}^{2j-1}_\mu + i \overline{\psi}^{2j}_\mu)/\sqrt{2} \sim e^{i \overline{\phi}^j_\mu}, \quad j=1,\dots,r.
\end{align}
Hence,
\begin{align}
\overline{\psi}^{2j-1}_\mu \sim \cos(\overline{\phi}^j_\mu), \quad \overline{\psi}^{2j}_\mu \sim \sin(\overline{\phi}^j_\mu).
\end{align}
As for the $SO(N+1)$ factor, we can bosonize all but one of the Majoranas, say the $\mu=0$ one:
\begin{align}
c^j_\mu = (\psi^{2j-1}_\mu + i \psi^{2j}_\mu) /\sqrt{2} \sim e^{i \phi^j_\mu}, \quad j=1,\dots,r.
\end{align}
Hence
\begin{align}
\psi^{2j-1}_\mu \sim \cos(\phi^j_\mu), \quad \psi^{2j}_\mu \sim \sin(\phi^j_\mu).
\end{align}
The $\mu=L,R=+,-$ edges are then described by the Lagrangians
\begin{align}
\begin{split}
\cL_\mu = &\frac{1}{4\pi} \sum_{j=1}^r \left[  \partial_x \phi^j_\mu (\mu\partial_t - \partial_x) \phi^j_\mu +  \partial_x \overline{\phi}^j_\mu (-\mu\partial_t - \partial_x) \overline{\phi}^j_\mu \right] \\
&+ \psi_\mu \frac{i}{2} (\partial_t - \mu\partial_x)\psi_\mu,
\end{split} 
\end{align}
where we have relabelled $\psi^{\alpha=0}_\mu \equiv \psi_\mu$. In this partially bosonized language, the currents are given by
\begin{align}
J^{\alpha\beta} &\sim \begin{cases}
 \cos(\phi^{(\alpha+1)/2})\cos(\phi^{(\beta+1)/2}), \, \alpha\text{ odd},\,\beta\text{ odd} \\
 \cos(\phi^{(\alpha+1)/2})\sin(\phi^{\beta/2}), \, \alpha\text{ odd},\,\beta\text{ even} \\
 \sin(\phi^{\alpha/2})\sin(\phi^{\beta/2}), \, \alpha\text{ even},\,\beta\text{ even}
\end{cases}  \label{eqn:SO(N+1)-currents-1}
\end{align}
for $\alpha,\beta \neq 0$, while,
\begin{align}
J^{0\beta} &\sim \begin{cases}
 \psi \cos(\phi^{(\beta+1)/2}),\, \beta\text{ odd} \\
 \psi \sin(\phi^{\beta/2}),\, \beta\text{ even}
\end{cases}, \label{eqn:SO(N+1)-currents-2}
\end{align}
for $\beta\neq 0$, and
\begin{align}
\overline{J}^{ab} &\sim \begin{cases}
 \cos(\overline{\phi}^{(a+1)/2})\cos(\overline{\phi}^{(b+1)/2}), \, a\text{ odd},\,b\text{ odd} \\
 \cos(\overline{\phi}^{(a+1)/2})\sin(\overline{\phi}^{b/2}), \, a\text{ odd},\,b\text{ even} \\
 \sin(\overline{\phi}^{a/2})\sin(\overline{\phi}^{b/2}), \, a\text{ even},\,b\text{ even}.
\end{cases} \label{eqn:SO(N)-currents}
\end{align}
The local-electronic operators, $M^{\alpha a}$, are likewise given by
\begin{align}
M^{\alpha a} &\sim \begin{cases}
 \cos(\phi^{(\alpha+1)/2})\cos(\overline{\phi}^{(a+1)/2}), \, \alpha\text{ odd},\,a\text{ odd} \\
 \cos(\phi^{(\alpha+1)/2})\sin(\overline{\phi}^{a/2}), \, \alpha\text{ odd},\,a\text{ even} \\
 \sin(\phi^{\alpha/2})\sin(\overline{\phi}^{a/2}), \, \alpha\text{ even},\,a\text{ even}
\end{cases} \label{eqn:condensed-currents-1}
\end{align}
for $\alpha \neq 0$, and by
\begin{align}
M^{0a} &\sim \begin{cases}
 \psi\cos(\overline{\phi}^{(a+1)/2}), \, a\text{ odd} \\
 \psi\sin(\overline{\phi}^{a/2}), \, a\text{ even}.
\end{cases}
 \label{eqn:condensed-currents-2}
\end{align}
We have suppressed the $\mu=L,R$ edge subscript for compactness in the above expressions.

Now, as discussed in Section \ref{sec:mr-review} for the MR theory, it is important that we understand the organization of the Hilbert space as dictated by the currents. Let us first work in the fermionic language of Eq. (\ref{eqn:ising-coset-fermionic-Lagrangian}) and Eq. (\ref{eqn:ising-coset-currents}). As described above, there are three topological sectors: $1$, $\psi$, and $\sigma$. Since the currents are all bilinears in the Majorana fields, it immediately follows that all states within a topological sector have the same total fermion parity, $(-1)^F$, where $(-1)^F$ anti-commutes with \emph{all} the Majorana fields.

Similar statements hold in the (partially) bosonized language. From Equations (\ref{eqn:SO(N+1)-currents-1})-(\ref{eqn:condensed-currents-2}) we see that the current operators either change the bosonic winding number parity of two bosonic fields, or change the bosonic winding number parity of one field and the Majorana fermion parity. In other words, in the identity sector, the total bosonic winding number parity (of both the barred and unbarred fields) must much that of the fermion parity -- note the similarity with the ``gluing" constraint in the Moore-Read CFT.

In order to express this Hilbert space organization more formally, let us identify the operator which generates the underlying $\mathbb{Z}_2$ gauge symmetry. As usual, we write the bosonic winding numbers as
\begin{align}
N^j_{\mu} = \int_0^L \frac{\partial_x \phi^j_\mu}{2\pi} dx, \quad \overline{N}^j_{\mu} = \int_0^L \frac{\partial_x \overline{\phi}^j_\mu}{2\pi} dx,
\end{align}
which have \emph{integer} eigenvalues. We then define the operator 
\begin{align}
I = (-1)^F (-1)^{\sum_j (N_R^j + N_L^j)} (-1)^{\sum_j (\overline{N}_R^j + \overline{N}_L^j)},
\end{align}
where $(-1)^F$ anti-commutes with the Majorana fields $\psi_L$ and $\psi_R$. This generates the $\mathbb{Z}_2$ transformation,
\begin{align}
\psi^\mu \to - \psi_\mu , \quad \phi^j_\mu \to \phi^j_\mu + \mu \pi , \quad \overline{\phi}^j_\mu \to \overline{\phi}^j_\mu - \mu\pi,
\end{align}
under which the currents are manifestly invariant. The physical Hilbert space is defined by the constraint $I=1$, which simply states the total number of fermionic excitations (recalling that the vertex operators $e^{i\phi^j_\mu}$ and $e^{i\bar{\phi}^j_\mu}$ obey fermionic statistics) is even. 

\subsection{Gapping Term}

Let us now return to the question which motivated the search for an alternative representation of the Ising edge theory, namely, how to gap out an interface of Ising edges. For instance, suppose we would like to glue the two edges of the cylinder in Fig. \ref{fig:cylinder} together by bringing them close together and adding an interaction to gap them out. To do so, we can simply write down a current-current interaction, which is local by definition and takes the form of a Gross-Neveu interaction \cite{Gross1974}:
\begin{align}
H_{\mathrm{gap}} &= u \sum_{\alpha,\beta} J^{\alpha\beta}_R J^{\alpha\beta}_L + u \sum_{a,b} J^{ab}_R J^{ab}_L \\ 
    &= -u (\bm \psi_R \cdot \bm \psi_L)^2 - u (\overline{\bm \psi}_R \cdot \overline{\bm \psi}_L)^2. \label{eqn:ising-fermionized-gapping-int}
\end{align}
In the language of the standard Ising edge theory (i.e. the usual Ising CFT), this interaction heuristically corresponds to $(\psi_L\psi_R)^2$, as one would expect on the basis of an anyon condensation picture of the gapped interface. Indeed, condensing $\psi_L\psi_R$ in the $\mathrm{Ising}\times \bar{\mathrm{Ising}}$ theory yields the Toric code topological order, which can further be condensed to a trivial order. In the partially bosonized language, this Gross-Neveu interaction becomes (dropping terms which only renormalize velocities)
\begin{align}
\begin{split}
H_{\mathrm{gap}} = &-u \sum_{j_1 \neq j_2}\left[ \cos(2\Theta^{j_1}) \cos(2\Theta^{j_2}) +  \cos(2\overline{\Theta}^{j_1}) \cos(2\overline{\Theta}^{j_2})\right] \\
        &- u \sum_{j=1}^{r} \cos(2\Theta^j) i \psi_L \psi_R \label{eqn:ising-bosonized-gapping-int}
\end{split}
\end{align}
where we have defined
\begin{align}
2 \Theta^j \equiv \phi^j_R - \phi^j_L , \quad 2\overline{\Theta}^j = \overline{\phi}^j_R - \overline{\phi}^j_L .
\end{align}
It is straightforward to see that Eq. (\ref{eqn:ising-bosonized-gapping-int}) will gap out the interface -- the sine-Gordon terms will pin the angle variables, which in turn will result in a mass term for the remaining Majoranas.

\bibliography{references}

\end{document}